\documentclass{aa}%[referee]
\usepackage[varg]{txfonts}
\usepackage{graphicx}
\usepackage{xcolor}
\usepackage{hyperref}
\begin{document}
\title{S2D2: Small-scale Significant substructure DBSCAN Detection}
\subtitle{I. NESTs detection in 2D star-forming regions}
\author{Marta Gonz\'{a}lez \inst{\ref{inst1}, \ref{inst7}} \and Isabelle Joncour \inst{\ref{inst1}} \and Anne S. M. Buckner \inst{\ref{inst2}} \and Zeinhab Khorrami \inst{\ref{inst3}}   \and Estelle Moraux \inst{\ref{inst1}} \and Stuart L. Lumsden \inst{\ref{inst4}}  \and Paul Clark  \inst{\ref{inst3}} \and Ren\'{e} D. Oudmaijer  \inst{\ref{inst4}} \and Jos\'{e} Manuel Blanco \inst{\ref{inst5}}  \and Ignacio de la Calle \inst{\ref{inst5}}  \and Jos\'{e} Mar\'{i}a  Herrera-Fernandez \inst{\ref{inst5}} \and Jes\'{u}s J. Salgado \inst{\ref{inst5}}  \and Luis Valero-Mart\'{i}n \inst{\ref{inst5}}  \and Zoe Torres \inst{\ref{inst1}} \and \'{A}lvaro Hacar \inst{\ref{inst6}} \and Ana Ulla\inst{\ref{inst7}}}

\institute{Univ. Grenoble Alpes, CNRS, IPAG, 38000 Grenoble, France\label{inst1} \and University of Exeter, Stocker Road, Exeter EX4 4PY \label{inst2}  \and School of Physics and Astronomy, Cardiff University, The Parade, CF24 3AA, U.K. \label{inst3} \and School of Physics and Astronomy, University of Leeds, Leeds LS2 9JT, U.K.\label{inst4} \and Quasar Science Resources, S.L. Camino de las Ceudas 2, 28232, Las Rozas de Madrid, Madrid, Spain \label{inst5} \and Leiden Observatory, Niels Bohrweg 2, 2333 CA Leiden, The Netherlands \label{inst6} \and Universidade de Vigo, Campus Universitario Lagoas-Marcosende, 36210, Vigo, Spain \label{inst7}  } 

\bibpunct{(}{)}{;}{a}{}{,} % to follow the A&A style

\abstract {The spatial and dynamical structure of star-forming regions can help provide insights on stellar formation patterns. The amount of data from current and upcoming surveys calls for robust and objective procedures to detect structure, so the results can be statistically analysed and different regions compared.}  {We aim to provide the community with a tool able to detect the small scale significant structure, above random expectation, in star-forming regions, which could be the imprint of the stellar formation process. The tool makes use of the one point correlation function to determine an appropriate length scale $\epsilon$ and of nearest neighbour statistics to determine a minimum number of points $N_{min}$ for the DBSCAN algorithm in the $\epsilon$ neighbourhood.} {We implement the procedure and apply it to synthetic star-forming regions of different nature and characteristics to obtain its applicability range. We also apply the method to observed star-forming regions to demonstrate its performance in realistic circumstances and analyse its results.} {The procedure successfully detects significant small scale substructures in heterogeneous regions, fulfilling the goals it was designed for, and providing very reliable structures. The analysis of regions close to complete spatial randomness ($Q \in [0.7,0.87]$) shows that, even when some structure is present and recovered, it is hardly distinguishable from spurious detection in homogeneous regions due to projection effects. Interpretation should thus be done with care. For concentrated regions, we detect a main structure surrounded by smaller ones, corresponding to the core plus some Poisson fluctuations around it. We argue that these structures do not correspond to the small compact regions we are looking for. In some realistic cases, a more complete hierarchical, multi-scale analysis would be needed to capture the complexity of the region.} {We have developed implementations of our procedure, and a catalogue of the NESTs (Nested Elementary STructures) detected by it in four star-forming regions (Taurus, IC 348, Upper Scorpius, and Carina), which will be publicly available to the community. Implementations of the 3D, and up to 6D versions of the procedure including proper movements are in progress, and will be provided as future work.}

\maketitle
\section{Introduction}

Although some of the processes are well established, a coherent and detailed portrait of stellar formation is still not complete. Dynamical, thermal, magnetic and gravitational effects may appear at all scales producing an exceedingly complex and chaotic process \citep[see e.g.][for a review]{Larson07}. In particular, the specific relations between the geometry of parent clouds, prestellar cores, and young stellar objects (YSOs) are currently subject of very active research.
Our current understanding depicts molecular clouds with intricate structure which undergo very complex fragmentation, where dense filamentary structures appear. These filaments, and particularly their intersections, host dense molecular cores, and star formation \citep[see][and references therein]{Robitailleetal19, Hacaretal17}. 

In such scenarios, star formation is not expected to occur in isolation, justifying the importance of environmental effects on the whole process. The different effects that forming and young stellar objects can have on their environment (particularly if massive) suggests that the clustered environment of forming stars will have significant effects on a variety of observable phenomena, such as: high-mass star formation, protoplanetary disk survival, binary ratio, or the H$_\alpha$ cut-off observed in disc galaxies \citep[see e.g.][]{Pfalzneretal12, Larson07, ReiterParker19, PflammAltenburgKroupa08}. The european project StarFormMapper \footnote{https://starformmapper.org/} was born to study the influence of the natal environment on star formation.

Star-forming regions, henceforth SFRs, are the perfect observational laboratory to evaluate stellar formation and evolution theories. In addition, large surveys like Gaia and Herschel \citep[][]{Gaia2016, Herschel} provide the scientific community with unprecedented quality and volume of data both on the gas and stellar components. These observations, coupled with simulations and the development of appropriate analytical and statistical techniques, will allow us to characterise the process leading from gas and molecular clouds to stellar clusters. In order to take full advantage of the data,  the development of statistical methods, amongst other efforts, is required \citep{Siemiginowskaetal19}.  In particular, we are interested in the development of robust, statistical procedures for the objective detection of significant spatial and spatio-kinematical structure. We should be able to ensure that the structures detected have the same observational properties, if not the same physical origin, to allow comparison amongst SFRs. The method must also be readily applicable in different regions while guaranteeing the reliability of the structure detections. 

We are particularly interested in small spatial structures in young, pristine SFRs, as in the research by \citet{Joncouretal17, Joncouretal18} (henceforth J17 and J18) which lays the theoretical foundation of this work. J17 and J18 are the only studies, to our knowledge, focusing specifically on the small, local scale, and comparing the samples to a complete random distribution to ensure the significance of the structure retrieved. Traditionally, non-parametric clustering methods discard clusters with low number of members due to the difficulty of distinguishing them from random fluctuations \citep[see e.g.][]{KirkMyers11, Gutermuthetal09}, while parametric methods choose parameters like the number of clusters based on criteria associated to the likelihood of the underlying model \citep[see e.g.][]{Kuhnetal14, FeigelsonBabu12}. 
%Traditionally, clusters with less than 10 members are discarded due to the difficulty of distinguishing it from random fluctuations \citep[see e.g.][and references therein]{KirkMyers11, Gutermuthetal09}. 
% Parametric approaches, like the one used by \citet{Kuhnetal14}, decide on the appropriate model of the subclusters using criteria based on likelihood of the fit and penalizing the number of parameters, like the Akaike Information Criterion \citep[see e.g.][for a description]{FeigelsonBabu12}. 
Through meticulous analysis of the small-scale structure obtained in Taurus, J17 and J18 showed that these compact, local structures can be the imprints of the fragmentation of massive dense cores or clustering of cores.

The primordial nature and evolution of spatio-kinematical structure in SFRs, clusters and associations are still active questions. The complexity involved in the dynamical evolution of such systems is huge, represented by highly non-linear models strongly dependent on the specific initial conditions \citep[see e.g.][]{Aarsethetal08, Clarke10}.
Spatial analysis of observations and simulations has been applied to density and radius estimation, membership and multiplicity determination, or mass segregation  \citep[see e.g. ][]{CasertanoHut85, Gomezetal93, Larson95, MaizApellanizetal04, CartwrightWhitworth04, Allisonetal09, ParkerGoodwin15, MaschbergerClarke11, Buckneretal19}. 
In recent years, analyses have been extended to the spatio-kinematical phase space, allowing to estimate the kinematical state of clusters and associations, generate catalogues, or distinguish between different populations within the Milky Way  \citep[see e.g.][]{Fureszetal06,  Parkeretal14, Wrightetal14, AlfaroGonzalez16, GonzalezAlfaro17, ParkerWright18, CantatGaudinetal18}.

Studies on the correlation of the degree of structure with age in open clusters \citep{SanchezAlfaro09,Dibetal18}, indicate that if such a correlation exists, it is weak. Simulations from \citet{ParkerMeyer12} point to a rapid erasure of primordial structure, in agreement with \citet{FujiiZwart16}, who propose that clumpy SFRs lose their structure through the expulsion of residual gas and two body relaxation.  

However, \citet{HetemGregorioHetem19} also analysed a large sample of clusters, finding that according to their mass segregation indicators $\Lambda_{\rm{MSR}}$ and $\Sigma_{\rm{MST}}$ \citep[as introduced in][]{Allisonetal09, MaschbergerClarke11} their structural characteristics did not change within their first 10 Myrs of age, although it must be noted that clusters in their sample have a relatively low degree of structure. \citet{Pfalzneretal12} performed simulations of single and multimodal clusters (formed by two or more single clusters combined in one sample) and evaluated local surface and nearest neighbour methods to find the theoretical density profile and modes. They found that incompleteness and resolution can prevent from distinguishing modes, and also that cluster age is not a reliable indicator of dynamical state, at least for embedded clusters. In addition, they warn about the reliability of age estimates when combining several low mass clusters into a single sample. These subsamples are not usually at the same evolutionary stage, and some of them may still be forming stars. Including these very young subclusters in the average calculation, will keep the global age estimates low for a variety of clusters with different ages for the older subsamples.

Considering all this, there are still significant chances of the structure in young, clumpy SFRs being primordial and reflecting the nature of the star-forming process in a particular region. In that case, analysing the reliable, significant small scale structure will help better understand the characteristics of the process. Even if the structure were not primordial, the objective method of structure retrieval presented in this work makes them relevant and robust, allowing for statistical comparison of the characteristics of different SFRs.

In this work, we analyse the behaviour and define the range of applicability of S2D2, a procedure based on the method in J17 and J18 which was successfully applied in Taurus to retrieve significant small scale structure. To that end, as described in section \ref{method}, we have developed an automatic tool, S2D2, and tested it in simulated SFRs of different nature. We have also applied the procedure to four observed SFRs (Taurus, IC 348, Upper Scorpius and Carina), evaluating its performance in realistic situations. In section \ref{results} we show the results of the procedure on synthetic SFRs, that will allow us to define the range of applicability of S2D2. Section \ref{real} shows the results of further testing of the procedure on observed SFRs. Finally, in section \ref{conclusions} we will summarise the main results and conclusions from this work. We note that even though in this paper we will treat only two dimensions, the procedure can be readily extended to 3D and up to 3+3D, and these versions will be also made available for the community. Appendix \ref{implementations} contains information on the different implementations of the procedure that are publicly available.

\section{Description of S2D2}\label{method}

%\subsection{Development of automated tool for the scientific community}
S2D2 combines a classic data mining algorithm as DBSCAN with a statistically sound, theoretically rooted method to choose its parameters so that it searches for the smallest scale significant structures in a sample. 
Its basis was developed and extensively discussed in J17 and J18, where it was successfully applied to Taurus, a characteristic example of a very young and structured SFR. There, the small scale significant structures (that we will also call NESTs, or Nested Elementary Structure, following their nomenclature) had a high likelihood of being the pristine imprints of the stellar formation process. In the following, we briefly explain the procedure and the slight modifications and additions that allow its robust application in a general case. 
\subsection{DBSCAN}\label{DBSCANsection}
We now present the commonly used clustering algorithm DBSCAN, introduced in \citet{Esteretal96}. DBSCAN is a general purpose data mining algorithm that has been widely used in a variety of contexts, including structure detection in spatio-kinematical domains \citep[see e.g][and references therein]{Costadoetal17, Canovasetal19, KounkelCovey19}. We refer the reader to the appendix in J17 or the report \citet{JoncourReportClustering} for an in-depth review of DBSCAN and other clustering methods.

As its name indicates, DBSCAN (Density Based Spatial Clustering of Applications with Noise) detects clusters or structures in a specific domain according to density criteria, introducing an associated concept of reachability that characterises all the points in a cluster. This density requirement is based on two parameters: a length scale, $\epsilon$, and a minimum number of points, $N_{min}$. A point p in a point pattern is a core point if there are at least $N_{min}-1$ different points of the pattern within distance $\epsilon$ of such point (also called $\epsilon$-neighbourhood or vicinity). All points in an $\epsilon$-neighbourhood of a core point are said to be directly ($\epsilon$-)reachable from that core point and are assigned to the same cluster. The reachable points that fulfil the $N_{min}$ condition are also core points of the cluster, while those reachable points that do not satisfy the condition become border points. The points that are not directly reachable from a core point and do not meet the $N_{min}$ requirement are labeled as noise. 
\begin{figure*}[h!]
\begin{center}
\includegraphics[width=17cm]{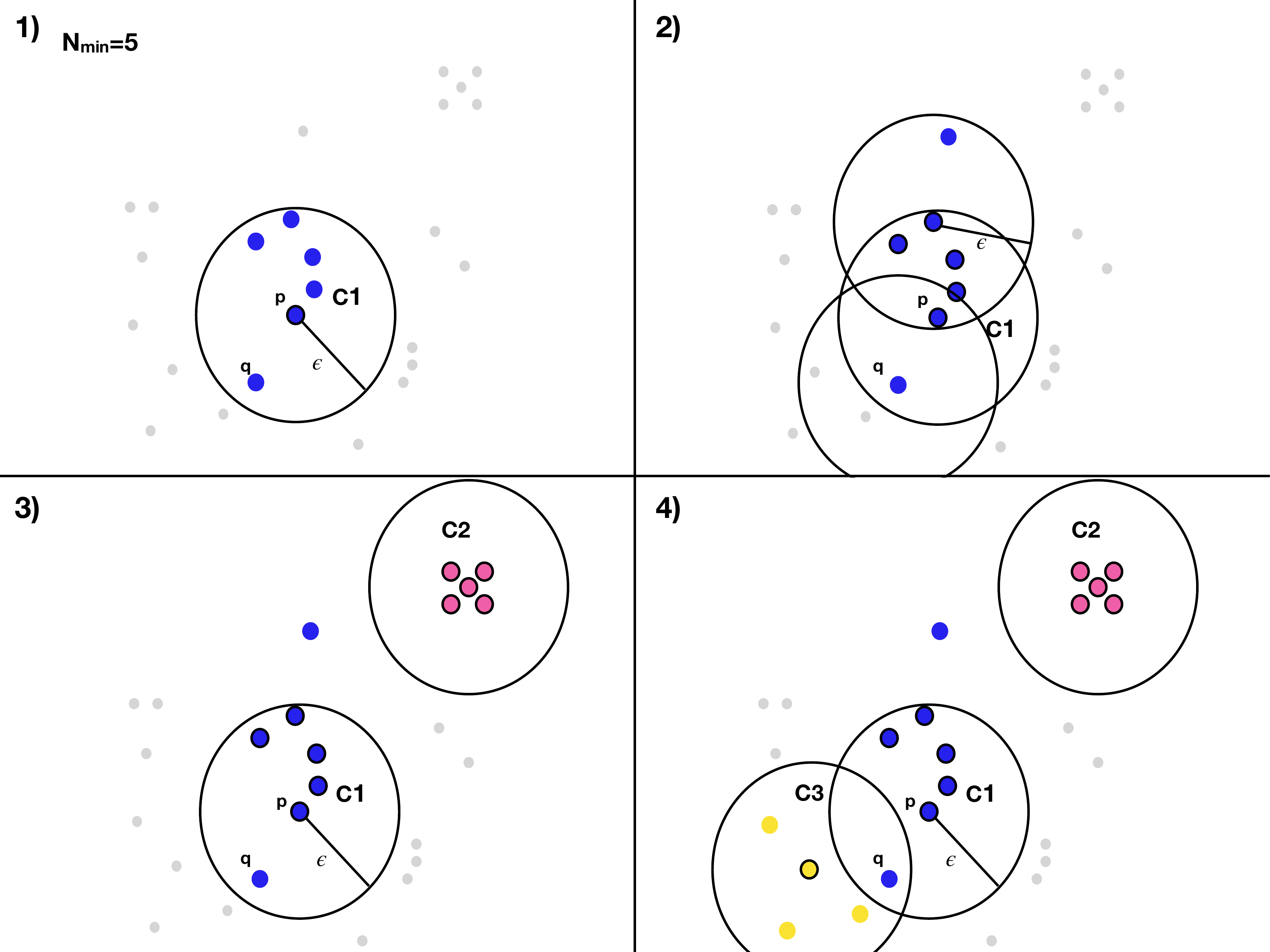}

\end{center}

\caption{Application of DBSCAN algorithm to an example sample. Each panel represents one step in the process, with panel 4 showing the final result of the algorithm. Grey dots represent points that are not assigned to any cluster (noise points in panel 4), coloured, big dots depict points assigned to clusters (C1, C2, and C3), and each cluster is represented with a different colour. Coloured points with black border are core points in a cluster, while those without represent border points. Black circles show the $\epsilon$ environment of some representative points, which are also labeled. The first panel depicts the detection of a cluster within the sample. The $\epsilon$-vicinity of the point labeled as $p$ has more than $N_{min}=5$ sample members inside, so $p$ is identified as a core point (represented as a black-edged, coloured point) of the first identified cluster, C1, and the points within its $\epsilon$-neighbourhood are marked as blue, since they are directly reachable from $p$ and, thus, belong to the cluster C1. A subsequent step is to check the $\epsilon$-vicinities of these points belonging to C1, which leads to the discovery of new members of C1, and the classification of some of them as core points. The bottom left panel shows a latter epoch in the algorithm application, where a second cluster C2 composed only of core points is identified. The last panel in Fig. \ref{DBSCAN} shows the final results of the algorithm, where a third cluster C3 with a single core point is identified. The point $q$ is in the $\epsilon$-vicinities of core points in C1 and C3, so it will be assigned to either cluster depending on the processing order of the points. }
\label{DBSCAN}
\end{figure*}

Figure \ref{DBSCAN} shows four stages of the application of DBSCAN algorithm to a sample, illustrating several possibilities that will  be mentioned in this section. The first panel shows the detection of the first core point, $p$, defining cluster C1. In the second panel more points of C1 are detected, studying the environment of the already known members of C1. Panel 3 shows a second identified cluster, C2, and panel 4 shows the final stage of the algorithm, where a third cluster, C3 has also been identified.

The parameters $\epsilon$ and $N_{min}$ define a minimum local density requirement for core points, that we will call nominal density $\rho_{nom}=\frac{N_{min}}{\pi \epsilon^2}$. This density $\rho_{nom}$  is not necessarily the minimum local density of the clusters, since by definition the border points belonging to a cluster do not have to reach it. This means that the DBSCAN algorithm is not completely equivalent to a cut in local density since the minimum local density of the members of a cluster will be that of its border points.  
Also, the parameter $N_{min}$ is not necessarily either the minimum number of points in each cluster. Even though DBSCAN is a hard clustering method (meaning that each point will belong to only one cluster, if any) in some situations border points can be assigned to different clusters depending on the order of processing by the algorithm. This is illustrated in Fig. \ref{DBSCAN}, where the point labeled as $q$ does not reach the minimum number of points in its $\epsilon$-vicinity, but it belongs to the $\epsilon$-vicinity of a core point in C1 and also  a core point in C3. The point $q$ is assigned to cluster C1 because it was identified first, but it is located in the boundary of both C1 and C3, so if the core point of C3 had been processed before $p$, $q$ would be assigned to C3. As a result of this configuration and the point processing order, C3 has only $4=N_{min}-1$ members.

DBSCAN is one of the few classic algorithms where noise is handled, meaning that not all the points in the sample are assigned to a cluster. This is a necessity in problems like ours, where just the relevant substructures must be retrieved. Another advantage of the algorithm is that it can detect clusters of arbitrary shape and does not impose any a priori number of substructures in the sample. 
The main drawback of the algorithm is the single scale $\epsilon$ used to search for structures, which may not reflect realistically the complex nature of all SFRs. 
 To overcome this issue, a multi-scale, hierarchical extension of this method, following the philosophy of choosing relevant length scales and number of points guaranteeing reliable structures, is in development and has been presented in \citet{Joncour19}.

\subsection{Selection of the size scale $\epsilon$: The one point correlation function} \label{epsSelect}

In this section we will use some definitions from the spatial statistics field, which studies the statistical distribution of objects within a spatial domain (point patterns). In this work, we will focus on point distributions given by the positions of stellar objects in SFRs, but more general domains can be considered. 
The most basic point distribution is complete spatial randomness, where the probability of having $n$ objects within a specific region is only dependent on the volume of the region. We will refer to complete spatial randomness by its acronym CSR or as homogeneous distributions. CSR distributions are described by a Poisson distribution with constant density (or intensity), and represent distrtibutions of points without mutual interaction. In spatial statistics CSR serves as reference to compare the characteristics of more complex and real life derived distributions, where interactions and other phenomena are expected to occur.  
We refer the reader to \citet{Diggle03,Illianetal08} for a comprehensive introduction to spatial statistics and point patterns; and to \citet{FeigelsonBabu12} for its astronomy applications.

There are several tools for comparing a specific point distribution to CSR, and evaluating whether it is compatible with CSR, clustered, or inhibited. Throughout this text, clustering (resp. inhibition) means that the points are closer (resp. more separated) than random expectation.
%The two point correlation function ($\xi(r)$ or TPCF), is a tool used traditionally in astronomy to quantify the spatial distribution of galaxies \citep{Peebles80}. For each distance $r>0$, $\xi(r)$ compares the observed distribution of distances in the sample with a random distribution, measuring the excess probability of two points of being at distance R. This means that if $\xi(r)>0$ (resp. <0) the frequency of pairs of at distance $r$ is larger (resp. smaller) than randomly expected. In terms of the TPCF, clustering is characterised by an excess of pairs at close distances ($\xi(r)>0$ for small $r$) and a deficit at large $r$. Conversely, inhibition is characterised by a deficit of pairs distances, and an excess at large $r$ with respect to random.  
The pair correlation function $g(r)$ (PCF) compares the distribution of distances in the sample with that of a random distribution, and is  %is related to the TPCF by the formula $g(r) = 1+\xi(r)$ and can be
 described by the ratio of the observed distribution and CSR, as a function of distance $r$. Distances where $g(r)>1$ indicate an excess of pairs with distance $r$ compared to random and, conversely $g(r)<1$ indicates a deficit.  The slopes of $g$ at small scales have been used to distinguish stellar binaries, clusters, and associations \citep[e.g][]{Gomezetal93,Larson95,KrausHillenbrand08}. The % use of the 
 PCF (or equivalent functions like the two point correlation function) has been used traditionally  in astronomy to quantify the spatial distribution of galaxies \citep{Peebles80} and is indicated by \citet{Retteretal19} for the testing of CSR in the context of SFRs. It is, however, not exempt of problems, such as its dependence on the geometry and size of the window where the spatial distribution is defined. The PCF measures second order statistics, variance and covariance, associated to interactions between points such as attraction or repulsion.

J17 introduced the One Point Correlation Function (OPCF) $\Psi(r)$, which is given by the ratio of the distribution of the first nearest neighbour in the sample and that of a homogeneous distribution. As the ratio of the sample and CSR distributions at different scales, the OPCF resembles the PCF, but it measures first order effects associated to density variations in the study area, and is also less sensitive to edge and size effects from the window, particularly at short size scales (as was shown in J17). The choice of the first nearest neighbour ensures that we are considering the closest environment of each star, treating the smallest possible scales. Analogously to the pair correlation function, distances, $r$, where $\Psi(r)>1$ indicate an excess of stars with nearest neighbour at distance $r$ compared to random and, conversely $\Psi(r)<1$ indicates a deficit.

The size scale $\epsilon$ for DBSCAN is chosen by S2D2 following J17 and J18, as the smallest transition distance between excess and deficit of stars with respect to CSR in terms of the OPCF, where $\Psi(\epsilon)=1$. Given the OPCF definition, stars whose nearest neighbour is at a distance smaller than $\epsilon$ are clustered, compared to the theoretical expectation for CSR.

As functions based on local properties, both the one point and the pair correlation functions are evaluated at different scales $r$ within the sample. In practice, this is usually done with histograms, that discretise the range of densities involved (e.g \citealt{Larson95}, J17, J18, \citealt{KrausHillenbrand08}). Histograms pose an issue, namely that the size (or number) of bins needs to be carefully chosen depending on the sample to avoid empty bins and also reflect variations.  
In this work, we avoid this issue by using a Gaussian kernel representation for the first nearest neighbour density where the bandwidth $h$ is computed using Silverman's formula \citep{Silverman86}, which is given by:
\begin{equation}
h=0.9\cdot \min \left\lbrace \hat{\sigma}, \frac{IQR}{1.34} \right\rbrace \cdot n^{-1/5}
\label{silverman}
\end{equation}

\noindent where $n$ is the size, $\hat{\sigma}$ the standard deviation, and $IQR$ the inter-quartile range of the sample. We performed additional tests with the more complex Botev's algorithm \citep{Botev10} to calculate the bandwidth, but it did not improve the results. In fact, the $\epsilon$ estimates obtained from the OPCF calculated using both bandwidths are very close, as shown in appendix \ref{AppendixOPCF}. As a consequence, we keep using Silverman's formula, which is simpler and much more popular. 
We note that the OPCF values are dependent on the density of the region, which is used to derive the first nearest neighbour distribution for CSR, which in turn is associated to the window chosen for the study. In order to obtain a robust procedure that can be applied without the need to explicitly study each region to choose an appropriate window, we rely on nearest neighbour statistics.
The probability density function of the $k$-th nearest neighbour in a 2D CSR distribution is given by J17:

\begin{equation}
\mathcal{P}_{k}(r)=\frac{2(\pi \rho_{CSR})^{k}}{\Gamma(k)}\cdot r^{2k -1}\cdot \rm{exp}(-\pi \rho_{CSR} r^2)
\label{pdfNNk}
\end{equation}

\noindent{where} $\rho_{CSR}$ is the density or intensity of the CSR point distribution, and $\Gamma$ represents the $\Gamma$ (gamma) function \footnote{The $\Gamma$ function is a special mathematical function that extends the factorial function for complex numbers, and as such, arises frequently in statistical calculations. For a complex number $z$ with positive real part, $\Gamma(z)=\int_{0}^{\infty} t^{z-1}e^{-t}dt$.}.%  The incomplete upper (resp. lower) $\Gamma$ functions are a generalization from the complete $\Gamma$ function, including a new variable $x$ that substitutes the lower (resp. upper) integration limit. }. 
The expected value of the $k$-th nearest neighbour distance is then:

\begin{equation}
{\bar{r}}_{k}=\frac{\Gamma(k+1/2)}{\Gamma(k)}\cdot (\pi \rho_{CSR})^{-\frac{1}{2}}
\label{r6Mean}
\end{equation}

If we solve Eq. \ref{r6Mean} for the intensity of the process $\rho_{CSR}$ we obtain:

\begin{equation}
\rho_{CSR}=\frac{\Gamma(6.5)^2}{\Gamma(6)^2}\frac{1}{\pi \bar{r_{6}^{2}}}=\frac{5.755412}{\pi \bar{r_{6}^{2}}}
\label{eqRho}
\end{equation}

\noindent where we have substituted for $k=6$, an intermediate value which balances the locality of the estimate and the smoothing of random fluctuations for a variety of sample sizes according to \citet{CasertanoHut85}. They argue that lower values of $j$ (like $j=3$) are well suited for regions with a very small number of particles, $N_{star}<30$, but in the case of larger simulations could include noise, biasing the estimates. Later work using local density estimates \citep[such as, e.g.][amongst others]{ParkerGoodwin15,MaschbergerClarke11,GonzalezAlfaro17, Buckneretal19} has confirmed that values for $j$ between 5 and 7 are appropriate for 2 and 3D distributions. 
The expression we propose for the density of the region, $\rho_{CSR}$ in Eq. \ref{eqRho} is formally equal to the unbiased estimator for the local density proposed by \citet{CasertanoHut85} based on the $6^{th}$ nearest neighbour $\Sigma_{6}$, differing from it only through a constant factor $\sim 1.15$. Our choice of a representative density for the region to make a fair comparison with CSR ($\rho_{CSR}$) is thus coherent with our method and with classical estimators of local density.

\subsection{Selection of the minimum number of points $N_{min}$: significance}\label{nMinSelect}

As in all statistical distributions, CSR patterns show noise and deviations from their theoretical distributions, due to finite sampling effects. This implies that, when generating a point pattern with constant, uniform density the points are not evenly distributed within the volume. Such fluctuations in local density for CSR can be detected and interpreted as significant structure instead of noise.

 The nearest neighbour distribution can help us evaluate the theoretical probability of finding fluctuations with a particular density, as developed in J18.
 
For a CSR distribution, the probability $\alpha$ of having $k$ companions within radius $r$ is given by the integral between 0 and $r$ of the $k$-th nearest neighbour probability density function, shown in Eq. \ref{pdfNNk}. This value $\alpha$ is the statistical significance level of a structure of $k+1$ members in a $r$ neighbourhood, and $1-\alpha$ is its confidence level. If we fix $r$ and $\rho_{CSR}$, the confidence value of a structure with respect to random fluctuations increases as we increase the number of neighbours $k$ requirement. It is also interesting to note, that for fixed $r$ and $k$, the confidence value decreases as we increase the density of the region. 

Taking all this into account, we will require a confidence level larger than $3\sigma$, $(1-\alpha)= 0.9985$, to consider the structures significant, using the scale $\epsilon$ and the density $\rho_{CSR}$ chosen by our procedure as previously described. In other words, we choose $N_{min}=k_{0.0015}+1$, with $k_{0.0015}$ being the smallest number of neighbours $k$ for which the confidence level of the structures found is above 0.9985.

The $N_{min}$ obtained and the strict level of confidence imposed by S2D2 implies that we may be losing some real, significant structure that is not statistically distinguishable from random fluctuations in a clear way. We have made a conservative choice between completeness and reliability of the detections for the sake of obtaining structures that can be interpreted and compared across different clusters. 

\subsection{Summary of the algorithm}
We now summarise the steps of S2D2:
\begin{enumerate}

\item{Choose $\epsilon$ scale (section \ref{epsSelect})}
\begin{itemize}
\item{Calculate the representative density of the region, that will be used to compare with random, $\rho_{CSR}$.}
\item{Calculate the first nearest neighbour distribution and the OPCF $\Psi(r)$}
\item{Use the OPCF to obtain $\epsilon$ scale separating the smallest scale at which there is a transition from excess to deficit of stars with respect to CSR.}
\end{itemize}
\item{Choose $N_{min}$ to guarantee significance(section \ref{nMinSelect}): For the $\epsilon$ value previously obtained, iteratively increase the $N_{min}$ until a fixed significance value $1-\alpha$ is reached.}

\item{Apply DBSCAN with the obtained $\epsilon$ and $N_{min}$}

\end{enumerate}

\section{Results}\label{results}

\subsection{Tests of S2D2 in synthetic clusters}
We start the analysis of the behaviour of the procedure using simulated test clusters (that represent the stellar content of SFRs), where the underlying distribution is, by construction, known.

The test clusters display a variety of characteristics, allowing us to determine the range of applicability of the method. This way, we can ensure that the structure found is significant and that the comparison and analysis of structure across different regions is coherent and robust. 

\subsubsection{Synthetic cluster generation and treatment}

\begin{figure*}[h!]
\begin{center}
\includegraphics[width=17cm]{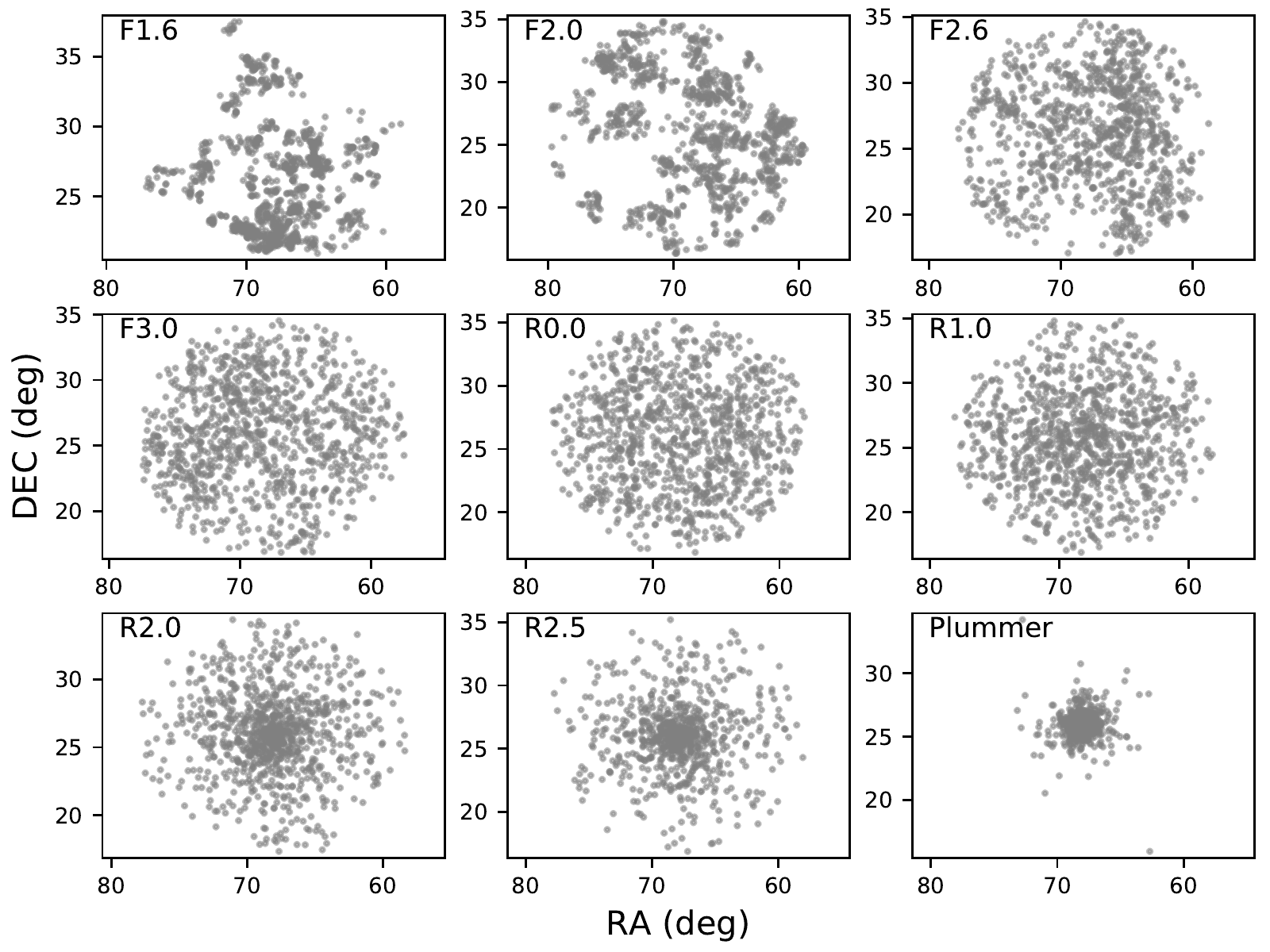}

\end{center}
\caption{Example realisation of each of the distributions in Table \ref{tablaDist}.}
\label{Distributions}
\end{figure*}

\begin{table}
\caption{Spatial distributions, with parameters and references, used to generate synthetic clusters}
\label{tablaDist}      
\centering
                                  
\begin{tabular}{l | c | c | c | c | c |c|c|c|c|c|}  

Distribution&Parameter&Reference\\
\hline
Fractal & D=1.6& \citet{Kuepperetal11}\\
Fractal & D=2.0& \citet{Kuepperetal11}\\
Fractal & D=2.6& \citet{Kuepperetal11}\\
Fractal & D=3.0& \citet{Kuepperetal11}\\
Radial & E=0&\citet{CartwrightWhitworth04}\\
Radial &E=1&\citet{CartwrightWhitworth04}\\
Radial & E=2&\citet{CartwrightWhitworth04}\\
Radial & E=2.5&\citet{CartwrightWhitworth04}\\
Plummer &$\frac{r_{cut}}{r_{Plum}}=5$ &\citet{Aarsethetal08}\\
\end{tabular}
\end{table}

For each of the distributions in Table \ref{tablaDist}, we have simulated 10 3D clusters (i.e. 10 different realisations of each distribution) with $N_{star}=1000$ points each.  The number of bona-fide members of observed clusters  is between  is typically between 100 and 10000 \citep[see e.g. the survey by][]{CantatGaudinetal18} so studies with synthetic clusters also use those ranges for $N_{star}$ \citep[see e.g.][]{Jaffaetal17,Lomaxetal18, Parker18}. Even though in this work we do not specifically explore the effect of $N_{star}$ in our procedure, larger (resp. smaller) values of  $N_{star}$ will sample better (resp. worse) the theoretical underlying distributions, making it  easier (resp. harder) to distinguish them. In Figure \ref{Distributions} we show an example realisation of each of the distributions in Table \ref{tablaDist}. 

Fractal distributions of all dimensions were generated with McLuster \citep{Kuepperetal11} without imposing any radial gradient, and the radial ones according to the recipe in \citet{CartwrightWhitworth04} (henceforth CW04). These distributions produce regions with varying levels of substructure and concentration. We have also included a Plummer distribution, using the generating function in \citet{Aarsethetal08}, to account for concentrations of different nature. The regions sampled from fractal distributions range from highly structured (fractal dimension D=1.6) up to homogeneous regions (fractal dimension D=3.0). Similarly, radial distributions show different levels of concentration, according to the exponent of their density. They range from homogeneous (exponent E=0) to highly concentrated regions (exponent E=2.5).  We note that in the case of concentrated regions, by construction, the density has a gradient, larger for larger concentration exponent. The fractal distribution with dimension with D=3.0 and the radial distribution with E=0 produce homogeneous distributions statistically equal to CSR, since they are different ways to generate the same theoretical Poisson homogeneous point distribution with constant density. 

As for the Plummer models, the specific scale radius $r_{Plum}$ defining the core size is not important in itself, since, as we explain later, the clusters are rescaled afterwards. However, in practice, the generating function is not bounded in terms of radius, so a cutoff radius is usually enforced to avoid the appearance of very extreme outliers. We have chosen $\frac{r_{cut}}{r_{Plum}}=5$, since the theoretical Plummer model has $\sim 95\%$ of its mass within 5 $r_{Plum}$.

In all cases, for the sake of easing comparison amongst regions, simulations were translated and rescaled to the approximate position and size of the Taurus SFR (allowing comparison with J17 and J18), setting the units so the radius of the cluster is $\sim 9 \deg$ at Taurus' distance of 145 pc without modifying the relative sizes of the axes. 
Then, the 3D clusters were projected into (RA, DEC) coordinates, to mimic the 2D data available in observations of young stellar objects in SFRs. Finally, the sample was treated for binaries and chance alignments in projection, merging multiple systems (considering as part of a multiple system objects at separations below 1000 AU) into one single object, as done in J17 and J18. The limit of 1000 AU was chosen in these works for two reasons: it is close to the resolution limit in the regions, and it also is within the lower separation estimates for wide binaries. After merging multiple systems in the initially generated sample of 1000 stars, the final number of objects in each synthetic star-forming region is between 966 and 1000. 

\subsubsection{Synthetic cluster description}

\begin{figure}[h!]
\begin{center}
\resizebox{\hsize}{!}{\includegraphics{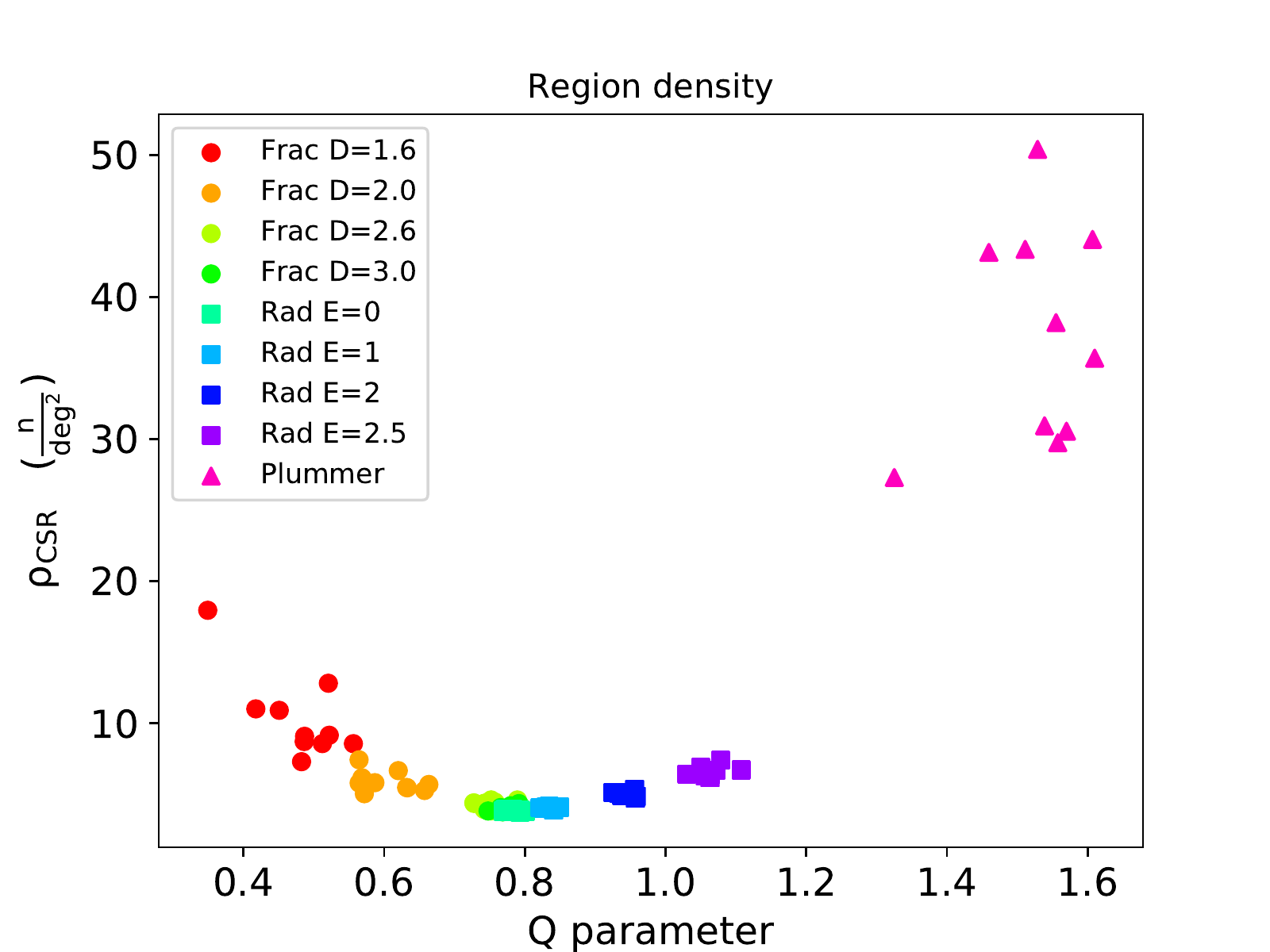}}
\end{center}
\caption{Density against the $Q$ structure parameter for all synthetic regions. Different symbols indicate whether the parent distribution is fractal or radial, while colour indicates a different parent distribution.}
\label{DensVsQ}
\end{figure}

For all synthetic star-forming regions, we have calculated the structure parameter $Q$ introduced in CW04. $Q=\bar{m}/\bar{s}$ is the ratio between the normalised mean branch length of the minimum spanning tree $\bar{m}$, and the normalized mean distance between points $\bar{s}$. The scale-free $Q$ structure parameter has been widely used to quantify the structure of stellar clusters, star-forming regions, and even dense cores \citep{Alfaroetal18,Parker18} since fractals have low $Q$ values, lower for lower fractal dimension, while radial distributions have large $Q$ values, increasing with the concentration. There is even a threshold of $Q=0.8$ that can separate concentrated from structured regions, corresponding to homogeneous distributions. 
The limit values $D=3$ for fractal distributions and $E=0$ for radials both correspond to a homogeneous distribution, with constant density and the $Q$ values obtained from approaching homogeneity from both perspectives converge. 

 Figure \ref{DensVsQ} shows the typical local density of a region $\rho_{CSR}$ (used for comparison with random and calculated using eq. \ref{eqRho}) against the $Q$ parameter for all the simulations, with colours and shapes marking the type of distribution for each synthetic cluster. Both the strength and limitations of the $Q$ parameter are clear from the plot: in effect, each distribution has a specific range of $Q$ values, that can be associated to their substructure and/or concentration, and in addition, the $Q$ values for the homogeneous distributions, approached both from a fractal and radial density, are coherent. The drawback of the $Q$ parameter is that the dispersion amongst realisations of a single distribution, larger for fractals, causes an overlap that makes it difficult to distinguish between distributions, particularly when they are close to homogeneous. Despite its limitations, which are discussed in appendix \ref{AppendixQ}, we have used the $Q$ structure parameter to graphically separate the synthetic distributions in the plots, and as a global indicator of the presence of substructure, as recommended by \cite{DaffernPowellParker20}. 

From Figure \ref{DensVsQ} it is clear that the density of the region is associated to its level of structure, given that we have rescaled all the synthetic regions to guarantee that their sizes are comparable. The density of a region $\rho_{CSR}$ increases with both substructure and concentration. This is an expected and desirable behaviour, since both structured and concentrated simulations are examples of clustered patterns, characterised precisely for their excess of stars at small distances, which decreases the average $6^{th}$ nearest neighbour distance with respect to random. In other words, their average local density $\Sigma_{6}$ is larger.
 Analogously to the $Q$ parameter, the density of the region $\rho_{CSR}$ is more disperse in the case of structured regions, and the density values of both the fractal and radial approximation to an homogeneous distribution are similar. 
 The Plummer distribution, being outside the Box-Fractal/radial model paradigm, shows a behaviour that globally corresponds to a concentrated distribution (large $Q$ and $\rho_{CSR}$ values) but nevertheless different from the radials.

\begin{figure}[h!]
\begin{center}
\resizebox{\hsize}{!}{\includegraphics{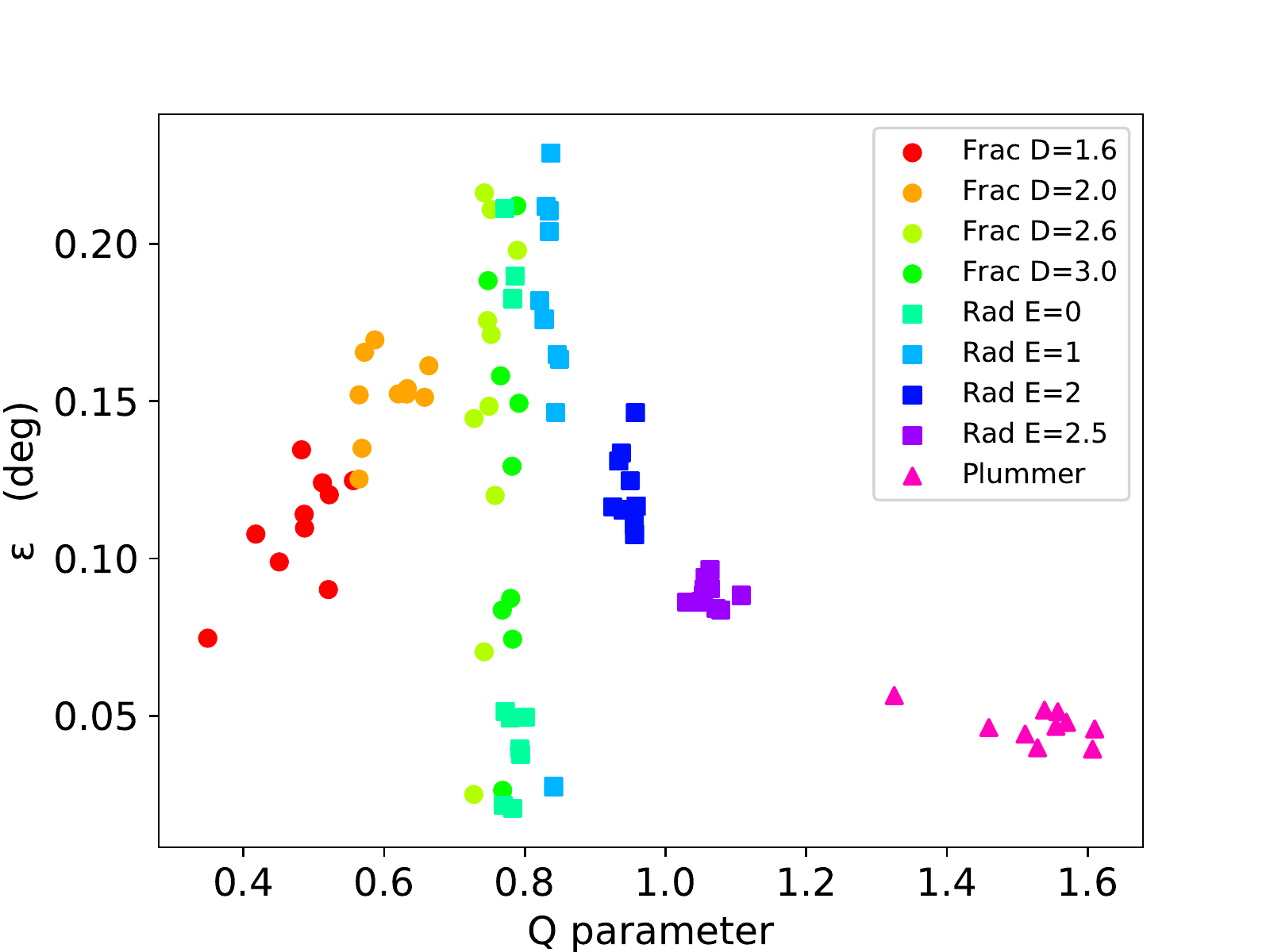}}
\resizebox{\hsize}{!}{\includegraphics{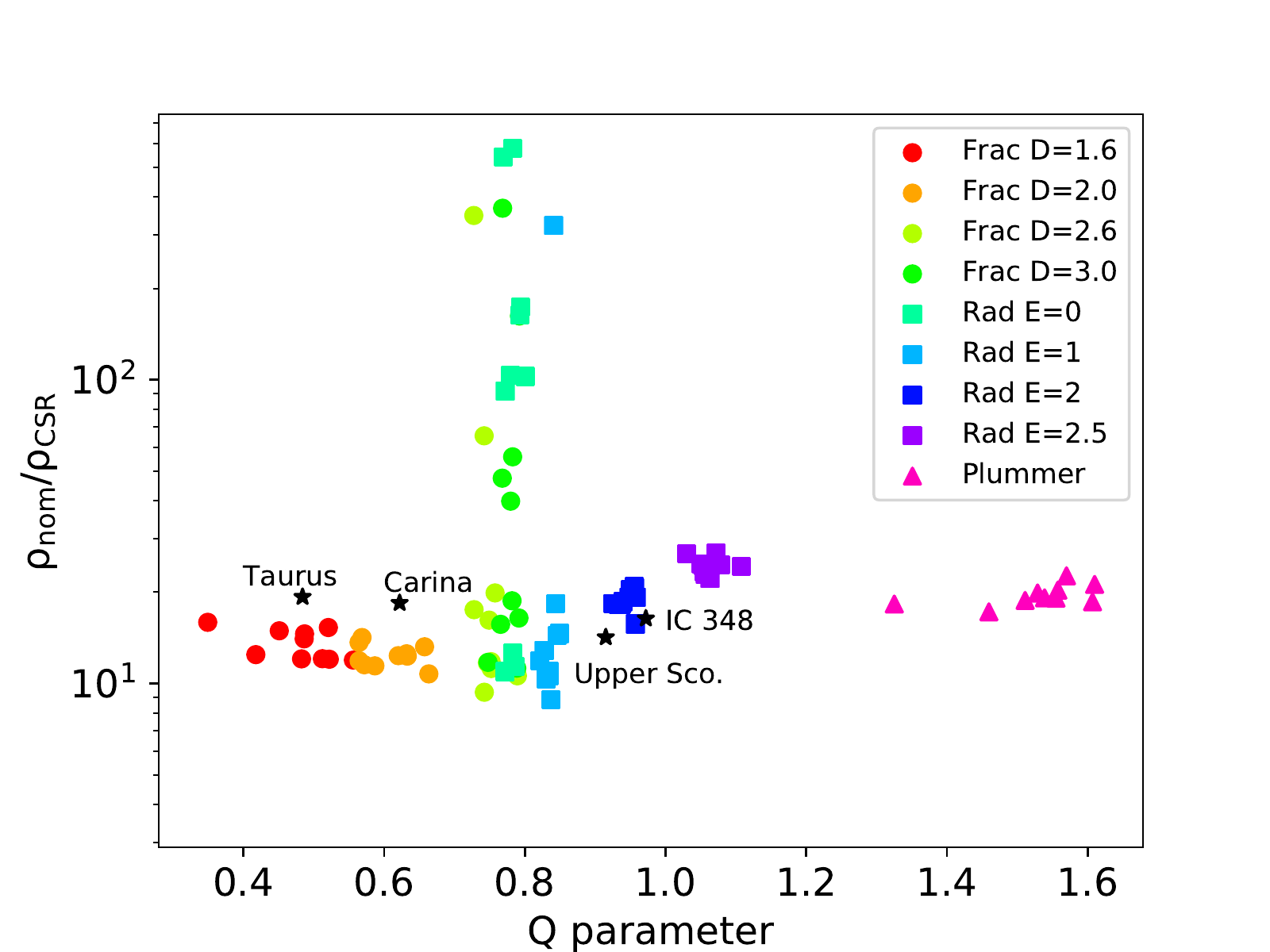}}
\resizebox{\hsize}{!}{\includegraphics{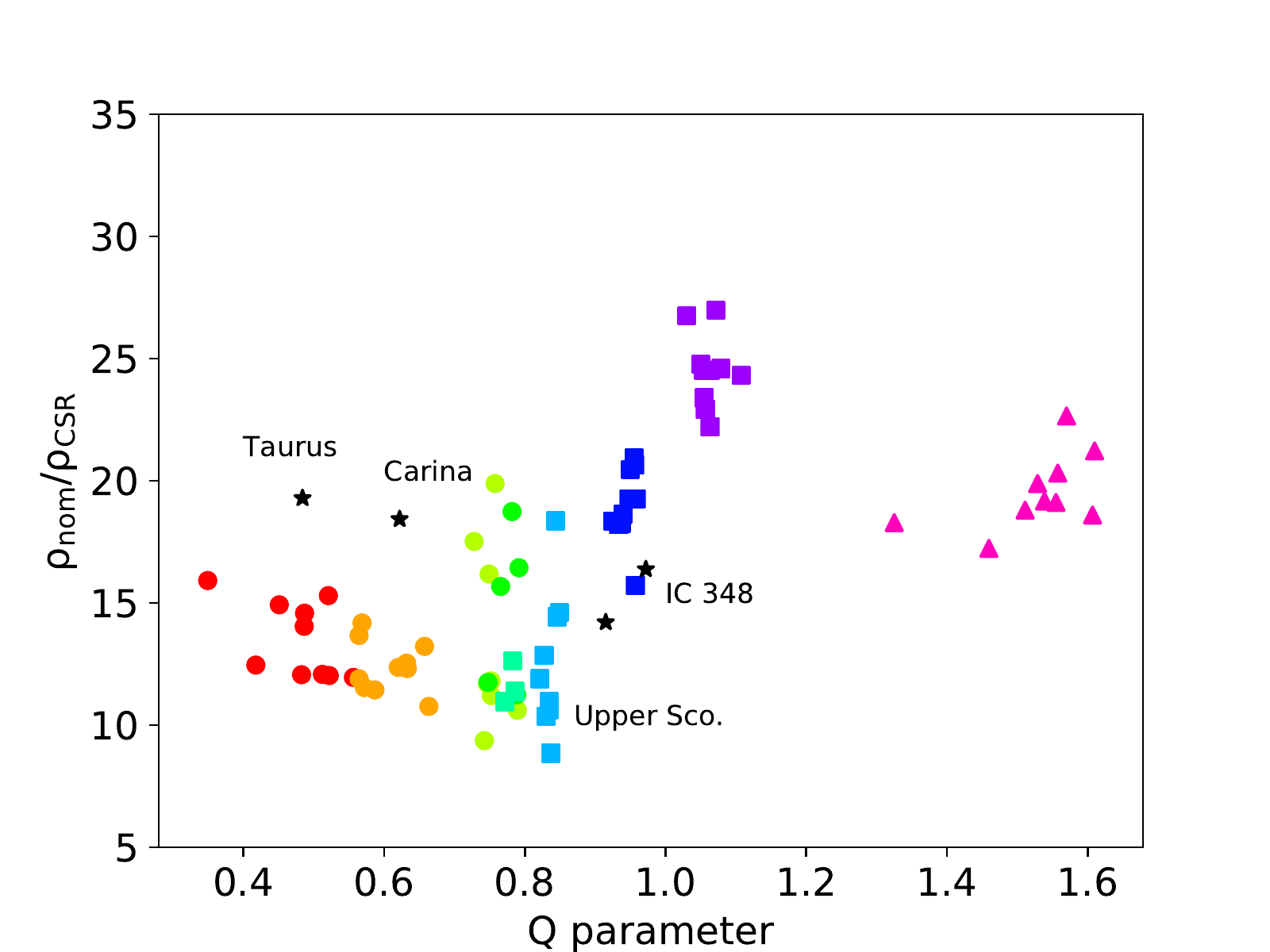}}

\end{center}

\caption{Top: $\epsilon$ scale parameter for DBSCAN calculated with the procedure presented in section \ref{epsSelect} against $Q$ structure parameter. Middle: Relative nominal density of the detected structures ($\frac{\rho_{nom}}{\rho_{CSR}}$, as described in the text) against $Q$ structure parameter. Colours and symbols represent parent distributions, with the same code as in Figure \ref{DensVsQ}. Bottom: Same plot as in the middle panel, linearly scaled and zoomed, so the general trends are clearer.}
\label{EpsNminVsQ}
\end{figure}

The top panel of Figure \ref{EpsNminVsQ} shows the parameter $\epsilon$ for each synthetic cluster calculated as described in section \ref{epsSelect} against the $Q$ structure parameter, with the same symbol and colour code as in Figure \ref{DensVsQ}. The $\epsilon$ scale parameter is, in general, smaller for both more structured and more concentrated regions, and larger for homogeneous.
This is consistent with the fact, shown in Figure \ref{DensVsQ}, that structured and concentrated clusters are also locally denser than homogeneous. 
In general, $\epsilon$ shows significant dispersion, particularly for clusters approaching homogeneity and obtained from radial distributions, where very small $\epsilon$ values can appear. This is expected from the method, since the OPCF is close to complete spatial randomness and the threshold $\epsilon$ such that $\Psi(\epsilon)=1$ ( separating excess from defect of stars with nearest neighbour at distance $\epsilon$) can be crossed by fluctuations at a variety of distances. For an illustration of the summarised behaviour of the OPCF  for the different synthetic clusters in this work, the reader is referred to Appendix \ref{AppendixOPCF}. We note that the method is devised for substructured regions, characterised by star distributions where some pristine substructure associated to the cloud fragmentation might be retained.  

To clarify the relationships between structure, scales, and densities, we have weighted the nominal density $\rho_{nom}$ of the structures detected by DBSCAN (as described in section \ref{DBSCANsection}) with the density of the regions $\rho_{CSR}$ (henceforth relative nominal density of the structures or $\rho_{nom}/\rho_{CSR}$) and plotted it against the $Q$ structure parameter, as shown in the middle and bottom plots of Figure \ref{EpsNminVsQ}. The middle plot is in logarithmic scale, to show the complete span of values reached, and the bottom plot is zoomed and in linear scales. This panel shows a clear trend of larger relative nominal density required for concentrated regions. A similar, slight increasing trend of relative nominal density with fractality is also present, but it is not as obvious as with concentration. This is partly due to the fact that regions close to CSR show very large dispersion, as was the case with $\epsilon$ in the previous panel. We note that the nominal density required for substructure detection is in all cases larger than 8.8 times the density of the region, confirming the strict criteria for significance in S2D2.

In addition to the results in synthetic clusters, Figure \ref{EpsNminVsQ} and all the following figures in this section also contain the values of the observed clusters analysed in section \ref{real}. As will be shown in more detail in section \ref{real} and appendix \ref{AppendixQ}, the simple Box-Fractal/radial and Plummer models do not capture all the features in real clusters, with the exception of IC348.

\begin{figure}[h!]
\begin{center}
\resizebox{\hsize}{!}{\includegraphics{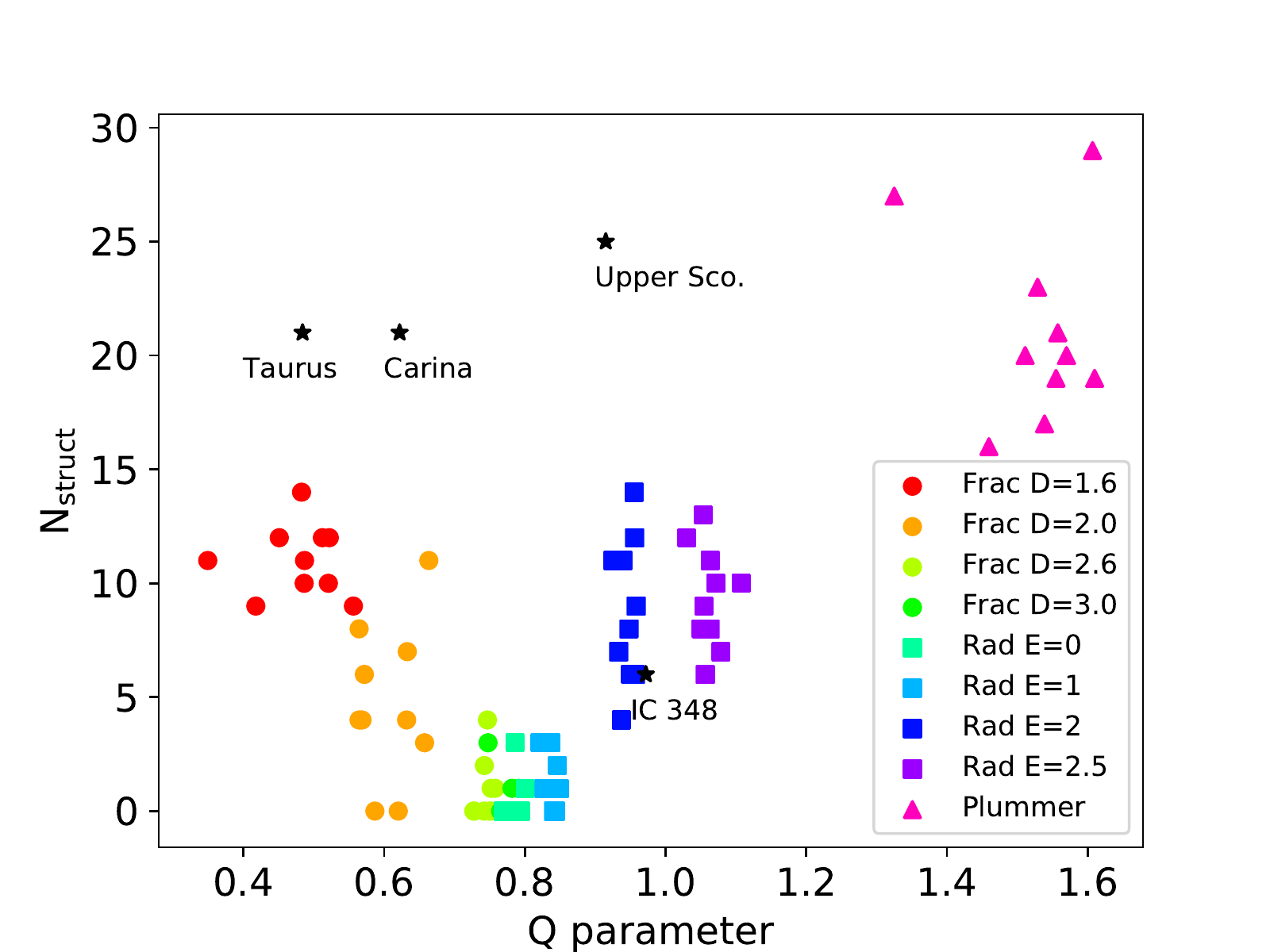}}
\end{center}
\caption{Number of structures detected with the procedure against $Q$ structure parameter. Colours and symbols represent parent distributions, with the same colour and symbol code as in Figure \ref{DensVsQ}.}
\label{NstructVsQ}
\end{figure}

Figure \ref{NstructVsQ} shows the number of structures detected in each region against the $Q$ parameter. Despite the dispersion, there is a clear trend coherent with the expectations and objectives from the method proposed. The more structured a region, the larger the amount of small scale significant substructure detected, with homogeneous regions showing almost no traces of significant structures. 

It is important to mention that there are cases where some structure can still be detected by the procedure in homogeneous regions, due to the projection of 3D structures. The retrieval of this spurious structure in S2D2 is not frequent, but projection effects should not be confused with substructure to avoid the over-interpretation of results. We note that the $Q$ values and thresholds used in this work correspond to the 2D analysis of 3D data and, thus, take into account projection effects. A cut in the $Q$ parameter of 0.7 will discard some quasi-homogeneous structured regions, but gives more than 2$\sigma$ certainty that we are dealing with a structured region, considering CW04,\citet{Cartwright09,SanchezAlfaro09}, covering an ample range of sample sizes.  Analogously, regions with $Q$ between 0.8 and 0.87 are within the 2$\sigma$ dispersion range of homogeneous regions, and contain also regions of light concentration.
In clusters with $Q$ values within these ranges, special care must be taken to try and distinguish whether they are projected CSR.

It is also clear from Fig.~\ref{NstructVsQ}, in general, that the number of structures detected also increases with concentration. This is an effect associated with the density distribution of these samples that, by construction, have a density gradient. In subsequent sections (specifically in sections \ref{subConcentrated} and \ref{real}) we will explain this effect in more detail. 
The fact that real observed regions, shown as black stars in Figure \ref{NstructVsQ}, show in general a larger amount of structures than simulations with a close $Q$ parameter value will be discussed in section \ref{real}, where we will analyse the results of S2D2 applied to real data and the differences with simulations. 

\begin{figure}[h!]
\begin{center}
\resizebox{\hsize}{!}{\includegraphics{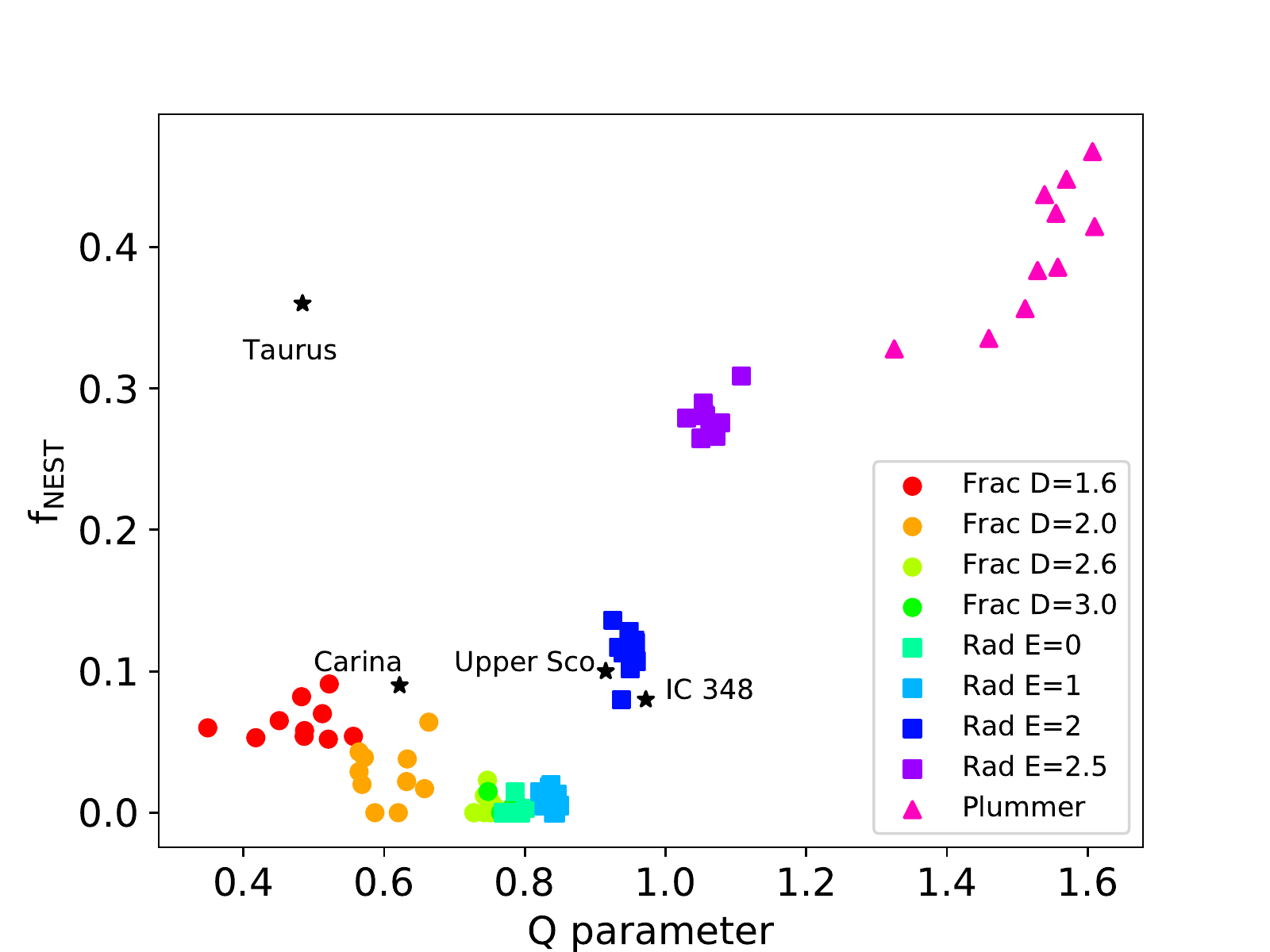}}
\resizebox{\hsize}{!}{\includegraphics{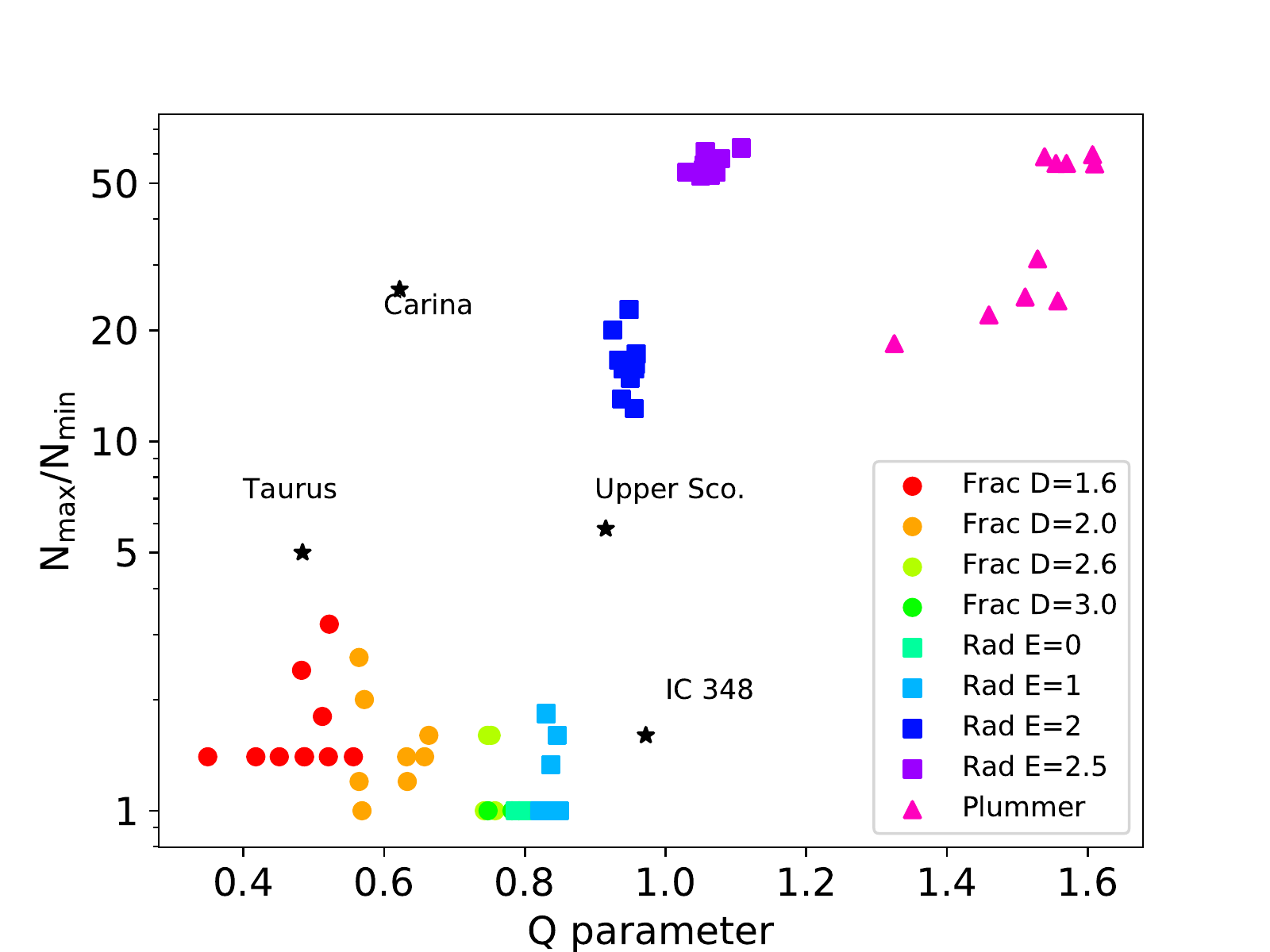}}
\end{center}

\caption{Top: Fraction of stars in NESTs against $Q$ structure parameter. Bottom: Relative maximum size of structure detected $N_{max}$ weighted by the significant number of points $N_{min}$ against $Q$ structure parameter.
Colours and symbols represent parent distributions, with the same code as in Figure \ref{DensVsQ}.}
\label{nInNestsVsQ}
\end{figure}

Figure \ref{nInNestsVsQ} shows two additional results from the application of the procedure. The top plot in Fig. \ref{nInNestsVsQ} shows the fraction of stars belonging to NESTs against the $Q$ structure parameter. There is a clear trend for concentrated regions, where the fraction of stars within structures reaches almost 30\% for radial distributions, and 50\% for Plummer. There is also a slight tendency of more fractal distributions to have a larger amount of stars belonging to NESTs. These trends are partly due to the larger amount of structures detected in concentrated and structured regions (as shown in Figure \ref{NstructVsQ}), but also to the fact that the structures themselves are larger for concentrated regions. To explore this last effect, in the bottom plot of Figure \ref{nInNestsVsQ} we show the relative maximum population of NESTs, given by the ratio of the maximum number of stars of a structure $N_{max}$ and the minimum number of points required by the procedure $N_{min}$. For highly concentrated regions, this ratio is larger than 10 meaning that at least one of the structures is too abundantly populated to be considered small scale. We cannot give an objective strict limit for large scale structures, but given the results for synthetic regions, Taurus, and Upper Scorpius in Fig. \ref{nInNestsVsQ}, it is reasonable to individually study all the characteristics of a region before deciding whether it is concentrated, particularly for values of $\frac{N_{max}}{N_{min}}$ between 5 and 10 .

\begin{table*}[h!]
\caption{Sample mean $\pm$ sample standard deviation of the procedure results in synthetic clusters, grouped by distribution.}              %
\label{tableSynth}      
\centering
                                  
\begin{tabular}{l | c|c | c | c | c |c|c|c|c|}          
                    
Distrib.& Param.& $Q$ & $\rho_{CSR}$ ($\rm{n/deg^2}$) & $\epsilon$ (deg) & $N_{min}$ & $N_{struct}$ & $f_{NEST}$ & $N_{max}/N_{min}$ & $\rho_{nom}/\rho_{CSR} $ \\    % table heading
\hline                                   
    Fractal& D=1.6 &0.48 $\pm$	0.06 &10.40 $\pm$	3.09& 0.11 $\pm$	0.02	& 5.0 $\pm$	0.0 & 11.0 $\pm$1.5& 0.06 $\pm$0.01& 1.7$\pm$0.6& 13.5 $\pm$	1.6\\     
    Fractal &D=2.0 &0.61 $\pm$	0.04 &5.88 $\pm$	0.71& 0.15 $\pm$	0.01	& 5.2 $\pm$	0.4& 4.7$\pm$	3.4& 0.03 $\pm$0.02& 1.6 $\pm$0.5& 12.4 $\pm$	1.1\\
    Fractal &D=2.6 &0.75 $\pm$	0.02 &4.34 $\pm$	0.26&0.15 $\pm$	0.06	& 4.9 $\pm$	1.0& 0.8 $\pm$	1.3& 0.00 $\pm$0.01& 1.3 $\pm$0.3& 52.1 $\pm$	105.2\\     
        \hline

    Fractal &D=3.0 &0.78 $\pm$	0.01 &3.98 $\pm$	0.20&0.11 $\pm$	0.06	& 4.3 $\pm$	0.9& 0.5 $\pm$	1.0& 0.00 $\pm$0.00& 1.0 $\pm$0.0& 74.6 $\pm$	112.4\\
    Radial &E=0.0 &0.78 $\pm$	0.11 &3.84 $\pm$	0.07&0.08$\pm$	0.08	& 3.7 $\pm$	1.1& 0.4 $\pm$	1.0& 0.00 $\pm$0.00& 1.0 $\pm$0.0& 178.9 $\pm$	209.2\\     
        \hline

    Radial &E=1.0 &0.84 $\pm$	0.01 &4.06 $\pm$	0.07&0.17 $\pm$	0.06	& 5.2 $\pm$	0.9& 1.5 $\pm$	1.2& 0.01 $\pm$0.01& 1.2 $\pm$0.3& 43.5 $\pm$	98.0\\
    Radial &E=2.0 &0.95 $\pm$	0.01 &5.01 $\pm$	0.17&0.12 $\pm$	0.01	& 4.4 $\pm$	0.5& 8.8 $\pm$	3.2& 0.11 $\pm$0.02& 16.4 $\pm$3.1& 18.9 $\pm$	1.5\\     
    Radial &E=2.5 &1.06$\pm$	0.02 &6.62 $\pm$	0.36&0.09 $\pm$	0.00	& 4.0$\pm$	0.0& 9.4 $\pm$2.2& 0.28 $\pm$0.01& 55.8 $\pm$3.5& 24.5 $\pm$	1.5\\   
    Plummer&--          &1.52$\pm$ 0.08 &37.34$\pm$.    7.70&0.05$\pm$       0.00  &4.9$\pm$     0.3 &21.1$\pm$4.1&0.40$\pm$0.04&40.8$\pm$18.0&19.5$\pm$1.6\\
    \hline                                             
\end{tabular}

\end{table*}

Table \ref{tableSynth} shows a summary of the general results of the simulations. The sample mean and standard deviation give us a central and dispersion measurement of the magnitudes calculated in this work across all the realisations of each distribution.

\subsection{Fractal clusters}
\begin{figure}[h!]
\begin{center}
\resizebox{\hsize}{!}{\includegraphics{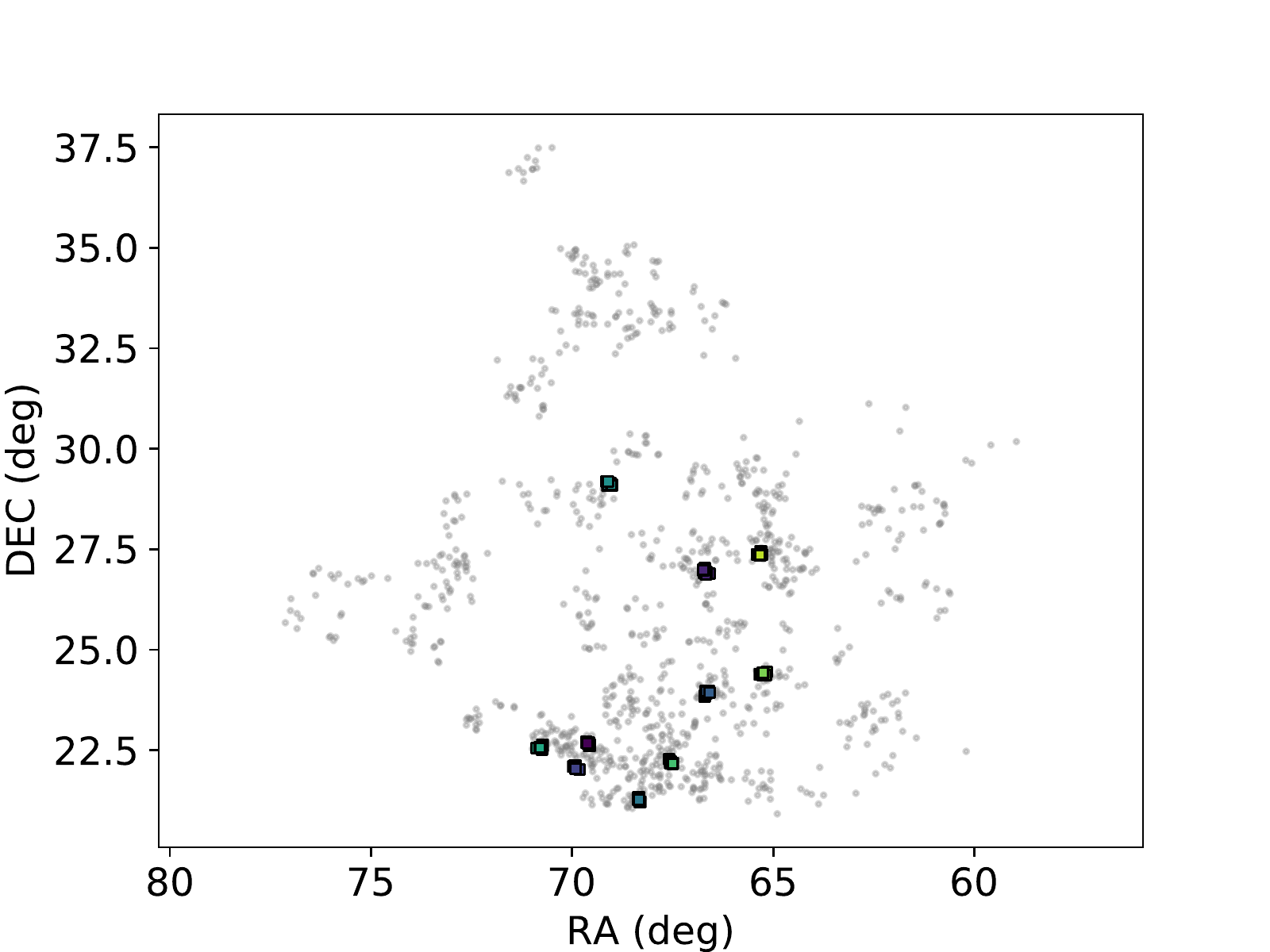}}
\resizebox{\hsize}{!}{\includegraphics{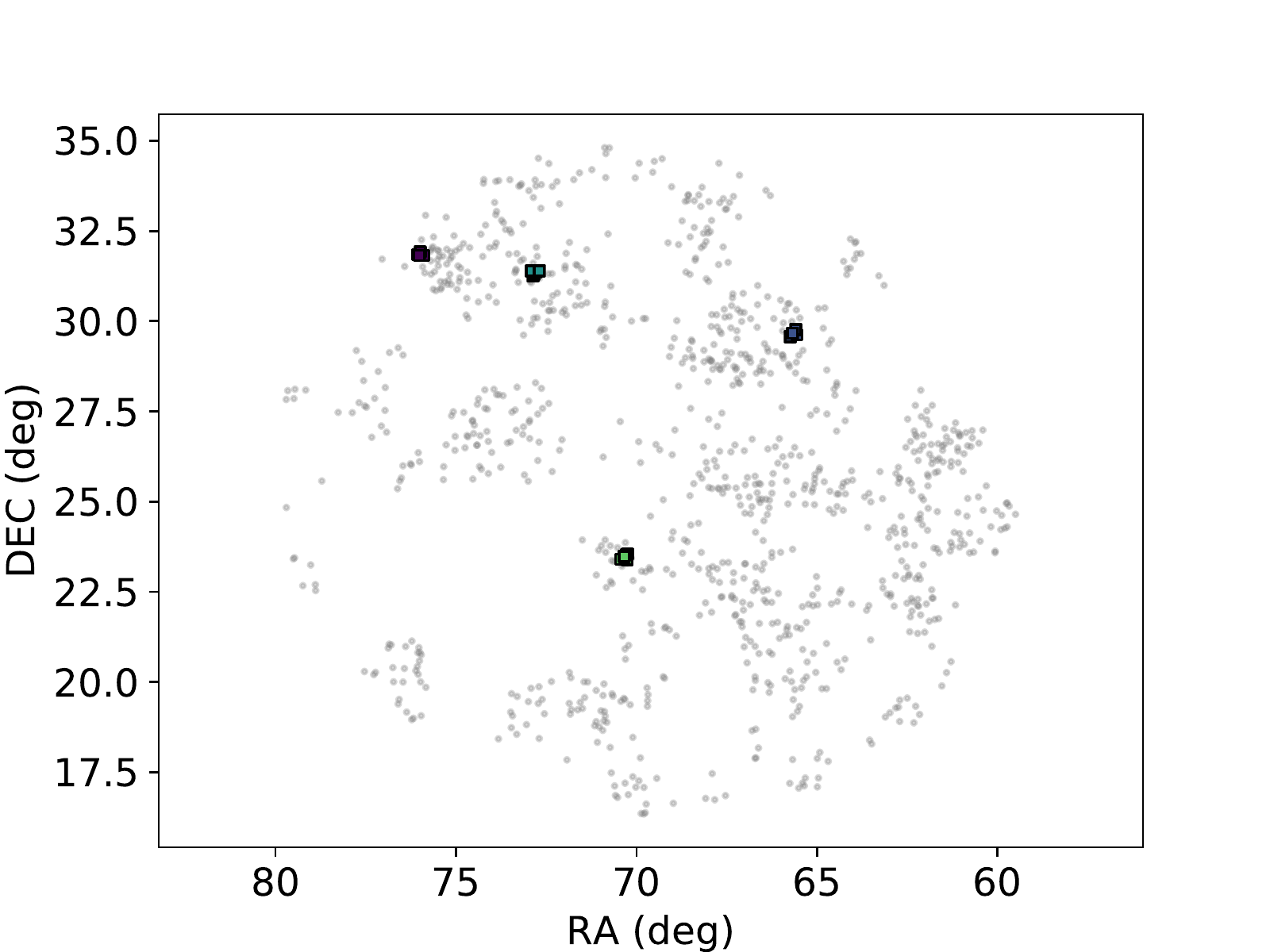}}
\resizebox{\hsize}{!}{\includegraphics{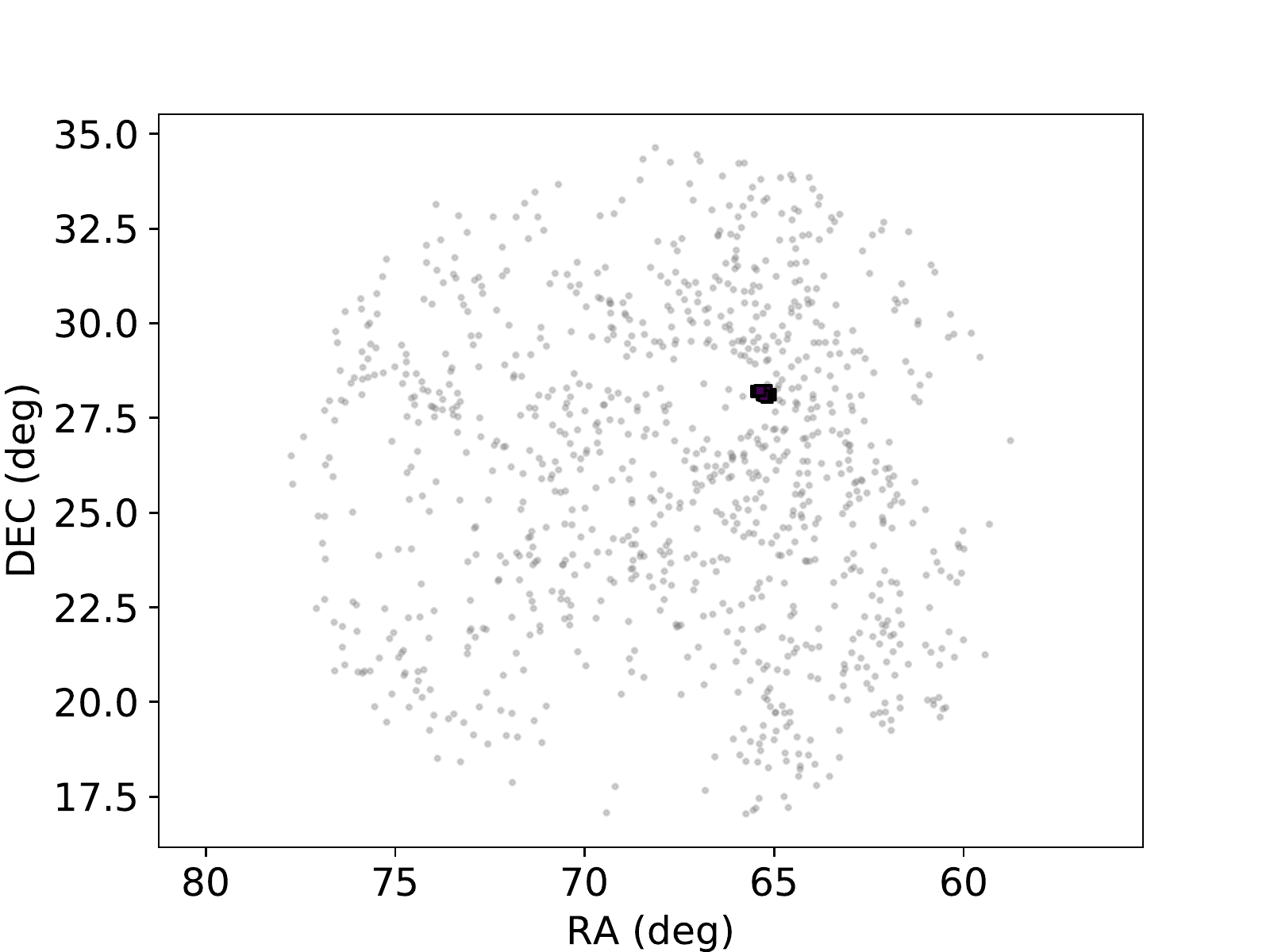}}

\end{center}

\caption{Top: Map of structures in a fractal with D=1.6. Middle: Same for a fractal with D=2.0. Bottom: Same for a fractal with D=2.6. Noise stars are in grey, while stars in structures are shown as coloured squares. Each shade of purple, blue, green and yellow represents a different small scale structure retrieved by S2D2.}
\label{MapsStructured}
\end{figure}

Figure \ref{MapsStructured} shows examples of synthetic structured clusters, where the complete sample is depicted by grey dots and the stars in the detected substructures are shown as coloured squares.
The upper plot in Fig. \ref{MapsStructured} shows an example of a fractal cluster with fractal dimension D=1.6, the smallest used in this work. The general sample is highly structured, with obvious clumps of stars of different size. The procedure detects several significant structures in the southern and central part of the sample, while the structures that can be identified by visual inspection in the northern areas are deemed not significant. The compactness and small scale of the structures found is evident.
The middle panel of Fig. \ref{MapsStructured} shows the significant structure in a fractal synthetic cluster with dimension 2. The region is clearly structured, although the clumps present are less dense and clear than in the example of fractal dimension D=1.6. Four small compact significant structures are detected by the procedure, marked in colour.
The lower plot in Fig. \ref{MapsStructured} shows a synthetic fractal cluster with dimension 2.6. This distribution is relatively close to CSR, and the structured nature of the distribution is not clear at all. Our procedure only detects one structure, small and compact, fulfilling the requirements of significance.

We can say that, in structured regions, the structures found by S2D2 are small, compact, and very reliable. There is a trade-off between reliability and retrieval of structures, where we have obviously favoured reliability, even at the risk of losing some of the structure. We believe that this is important in a systematic procedure for statistical comparisons amongst regions, minimising the chance of including artefacts in such comparisons. In the implementations available for the community, the user will be able to manually introduce a value of $\epsilon$ and $N_{min}$ for DBSCAN, relaxing the $\rho_{nom}$ required for detection.

\subsection{Homogeneous clusters}

\begin{figure}[h!]
\begin{center}
\resizebox{\hsize}{!}{\includegraphics{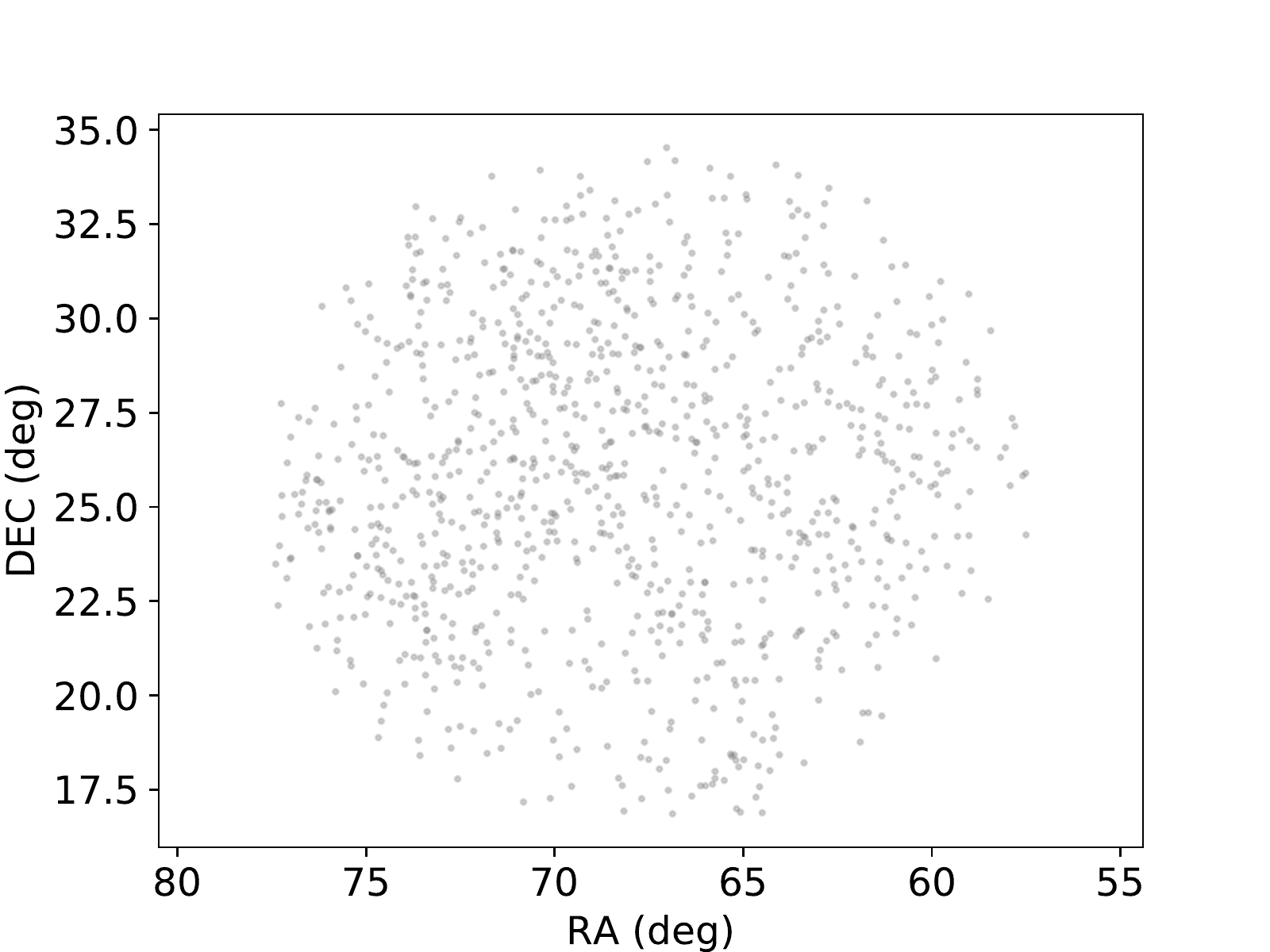}}
\resizebox{\hsize}{!}{\includegraphics{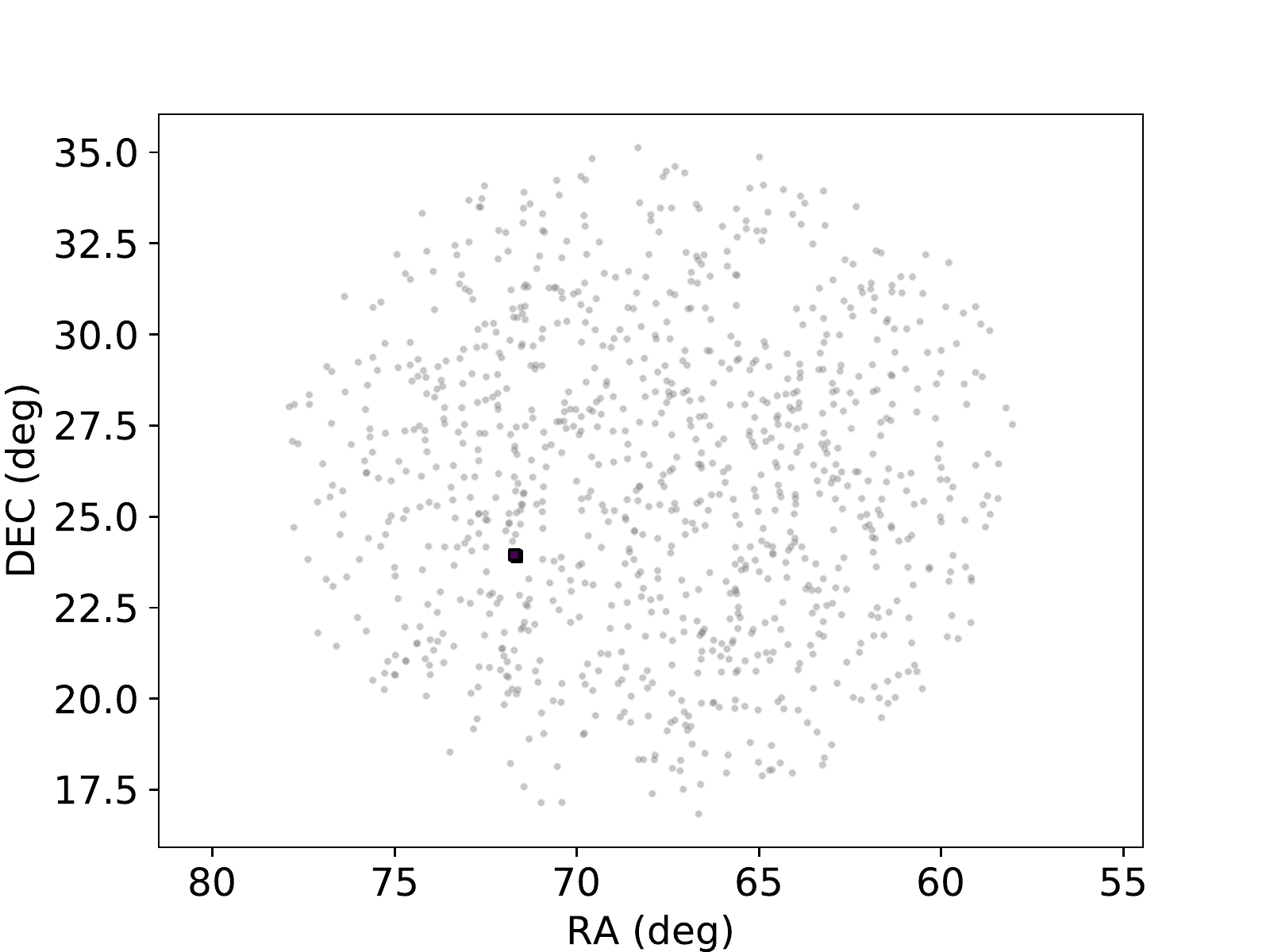}}

\end{center}

\caption{Top: Map of structures in a fractal with D=3.0. Bottom: Same for a Radial with E=0.0. Noise stars are in grey, while the purple squares in the lower plot belong to a significant structure retrieved by the procedure.}
\label{MapsHomog}
\end{figure}

Figure \ref{MapsHomog} shows maps of synthetic regions drawn from both the fractal and radial recipes to obtain a CSR distribution, the fractal with dimension D=3 and the radial with exponent E=0. It is clear that the stars in these CSR samples are not evenly distributed in the spatial domain and show density variations. These are a combination of statistical fluctuations and projection effects.  

The upper plot shows the fractal with D=3, where S2D2 finds no significant structure, and the bottom panel shows a region with radial distribution, where despite the strict level of significance (in this particular case, the nominal density $\rho_{nom}$ required for detection is more than a factor 100 larger than the density of the region), some of this structure is retrieved by the procedure and shown in purple.

In any case, as previously explained, we are warned by the $Q$ values associated to these regions (0.79 and 0.8, respectively) to carefully analyse and decide whether the structure retrieved could be significant or an artefact due to projection effects.

\subsection{Concentrated clusters} \label{subConcentrated}

\begin{figure*}[h!]
\begin{center}
\includegraphics[width=17cm]{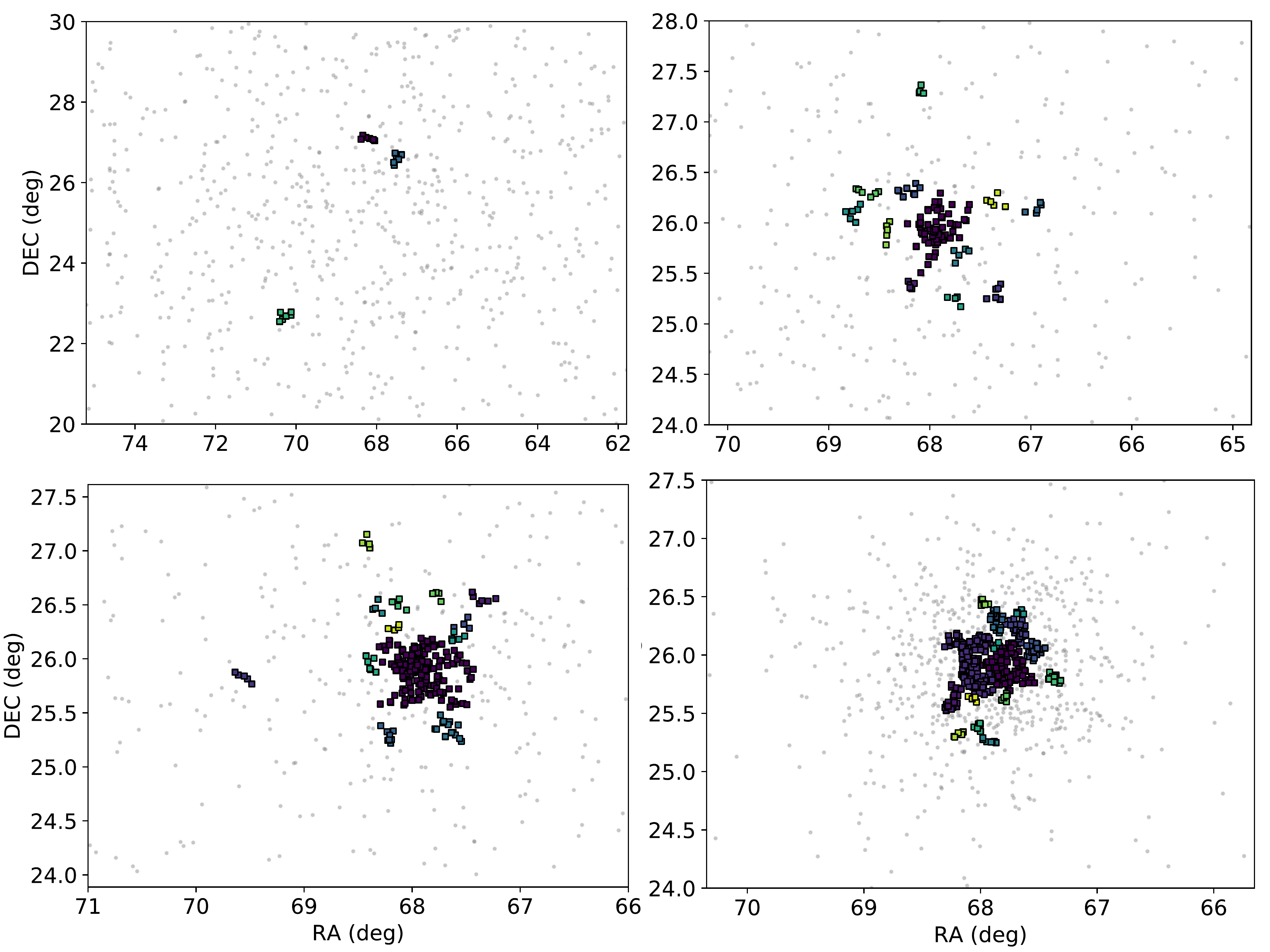}

\end{center}

\caption{Maps with the structures found by the procedure for example simulations drawn from concentrated distributions. From left to right and top to bottom, radial with exponent E=1.0, radial with E=2.0, radial with E=2.5, and Plummer. Noise stars are represented as grey dots, while the stars belonging to substructureare squares coloured in different shades of blue, purple, green and yellow according to the small scale structure that they belong to. We note that only the central area of the clusters is shown, so the retrieved substructure is clearer.}
\label{MapsConcentrated}
\end{figure*}

Figure \ref{MapsConcentrated} shows examples of synthetic clusters drawn from concentrated distributions (radials with exponents E > 0 and Plummer). From top to bottom and left to right, panels show radials with exponent 1, 2, and 2.5, and a Plummer distribution. In all four cases, the plot is restricted to the central part of the cluster. The increase of concentration with the exponent for radials is clear from Fig. \ref{MapsConcentrated}, as is the different nature of the concentration in the Plummer case. The increasing amount of structures detected with concentration that we mentioned in the previous section is clear, and its pattern can now be seen: the central concentration characterising these clusters is retrieved as a large structure, surrounded by smaller secondary structures, and, in general, the larger the concentration, the larger the central structure and the larger amount of secondary structures. For the Plummer example several large structures can be distinguished next to each other in the central area. In the case of low central concentration (as with E=1 in the top left panel of Fig. \ref{MapsConcentrated}), the central structure is not significantly retrieved, and only some secondary structures are there. Despite being concentrated, examples of this distribution are close to CSR, as indicated by the $Q$ values close to 0.84, so the central increase in density is not clear.

The secondary structures are a consequence of the density gradients present in these samples, which are by construction proportional to their degree of concentration. A large density gradient implies that the range of densities present in the sample is also large, so the choice of a single density (no matter how carefully done) does not represent the complexity of the sample. Secondary structures represent Poisson fluctuations at densities larger than $\rho_{CSR}$, the region density. 

Concentrated regions are beyond the scope for which S2D2 was originally designed, since they do not present the kind of local small-scale structures that we search for. In addition, simulations by 
\citet{ParkerMeyer12, DaffernPowellParker20} to study the persistence of substructure indicate that it is quickly erased through dynamical interactions, with the $Q$ parameter increasing rapidly. Thus, the concentrated regions could be dynamically evolved, devoid of spatio-kinematical traces of stellar formation. This is in agreement with \citet{Sillsetal18}, where they used the extension of $Q$, $Q^{+}$ to quantify the structure. We refer the reader to appendix \ref{AppendixQ} for more information on these tools.
The fact that even in these cases the behaviour of S2D2 is consistent, finding the large main structure with Poisson fluctuations around, is a sign of the robustness of our procedure. For cases like this, a multi-scale approach of the structure analysis, as in \citet{Joncour19}, Joncour, in prep., will allow to distinguish these fluctuations, capturing the complexity of the density pattern in the region.

\section{Results on observed clusters}\label{real}
In this section we show the results of further testing of the procedure, this time in observed clusters, that will allow us to calibrate its limits in realistic samples, beyond the idealised nature of synthetic clusters. The catalogues of substructure found in these four regions can be accessed from the StarformMapper web.

Table \ref{tableReal} shows a summary of the results of applying S2D2 to four test cases: Taurus, IC 348, Upper Scorpius and Carina, which constitute a varied sample of initial conditions. We have processed the YSO samples in all regions to merge multiple systems using the same distance limit of 1000 AU used in the simulations and in J17, J18 for Taurus, despite the different distances at which they are located. This does not hinder the calibration objective we pursue in this section, but we recommend careful consideration of the spatial resolution and method used to obtain the sample members of a cluster before studying its substructure with S2D2, or any other tool.

In the following, we will describe the specific results in each region. 

\begin{table*}[h!]
\caption{Results of the procedure in real clusters}              %
\label{tableReal}      
\centering
                                  
\begin{tabular}{l | c | c | c | c | c |c|c|c|c|c|}          
                    
Region&$N_{stars}$ & $Q$ & $\rho_{CSR}\, (\rm{deg^{-2}}) $& $\epsilon\, (\rm{deg}$) & $N_{min}$ & $N_{struct}$ & significance &$f_{NEST}$ & $N_{max}/N_{min}$ & $\rho_{nom}/\rho_{CSR} $ \\    % table heading
\hline                                   
    Taurus &405& 0.48 & 7.47 & 0.094 & 4 &  21 &99.87&0.36 &5.0&19.29\\     
    IC 348 &465& 0.97 & 2148.98 &0.007 & 5 &  6 & 99.97&0.08&1.6&16.38\\
    Upper Scorpius &1611& 0.92 &  10.96 &  0.101 & 5 &25&99.95&0.10&5.8&14.21\\
    Carina &2787& 0.62 &  9578.83 & 0.003 & 5 &21 &99.98&0.09&25.8&18.43\\
    \hline                                             
\end{tabular}

\end{table*}

\begin{figure*}[h!]
\begin{center}
\includegraphics[width=17cm]{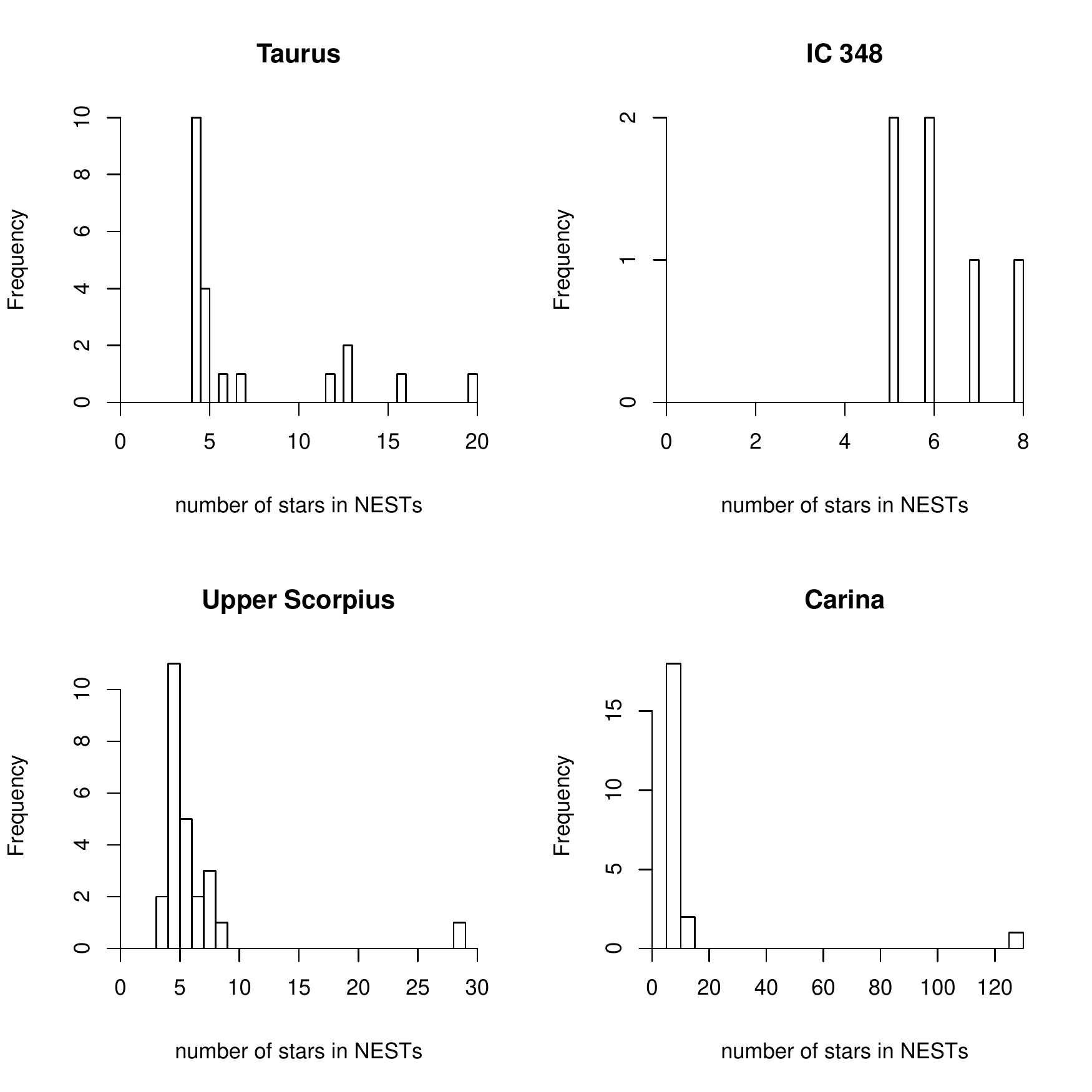}

\end{center}
\caption{Histogram of the number of members of the NESTs found in the observed regions.}
\label{HistoNests}
\end{figure*}

\subsection{Taurus}
We first test the proposed method in the Taurus region, which, as a young (1-2 Myr) and close (140 pc) SFR has been the object of numerous studies \citep[see e.g.][and references therein]{Kenyonetal08,Luhmanetal10}. Taurus is a typical example of a structured region, making it perfect to check the correct behaviour of the procedure.
J17 and J18 found and studied in detail ultra-wide pairs and significant small scale structures of higher multiplicity, NESTs, proposing that they are pristine and reflect the characteristics of the cloud fragmentation process. These works set the basis of the method presented in section \ref{method}.
In this work we use the updated catalogue from \citet{Luhman18}, which has $\sim 100$ more members than the sample by \citet{Luhmanetal10} used in J17 and J18. The sample in \citet{Luhman18} has also data from Gaia DR2, a significant portion of which includes parallax and proper motion information, so this analysis will be extended into more dimensions in future work.

\begin{figure*}[h!]
\begin{center}
\includegraphics[width=14cm]{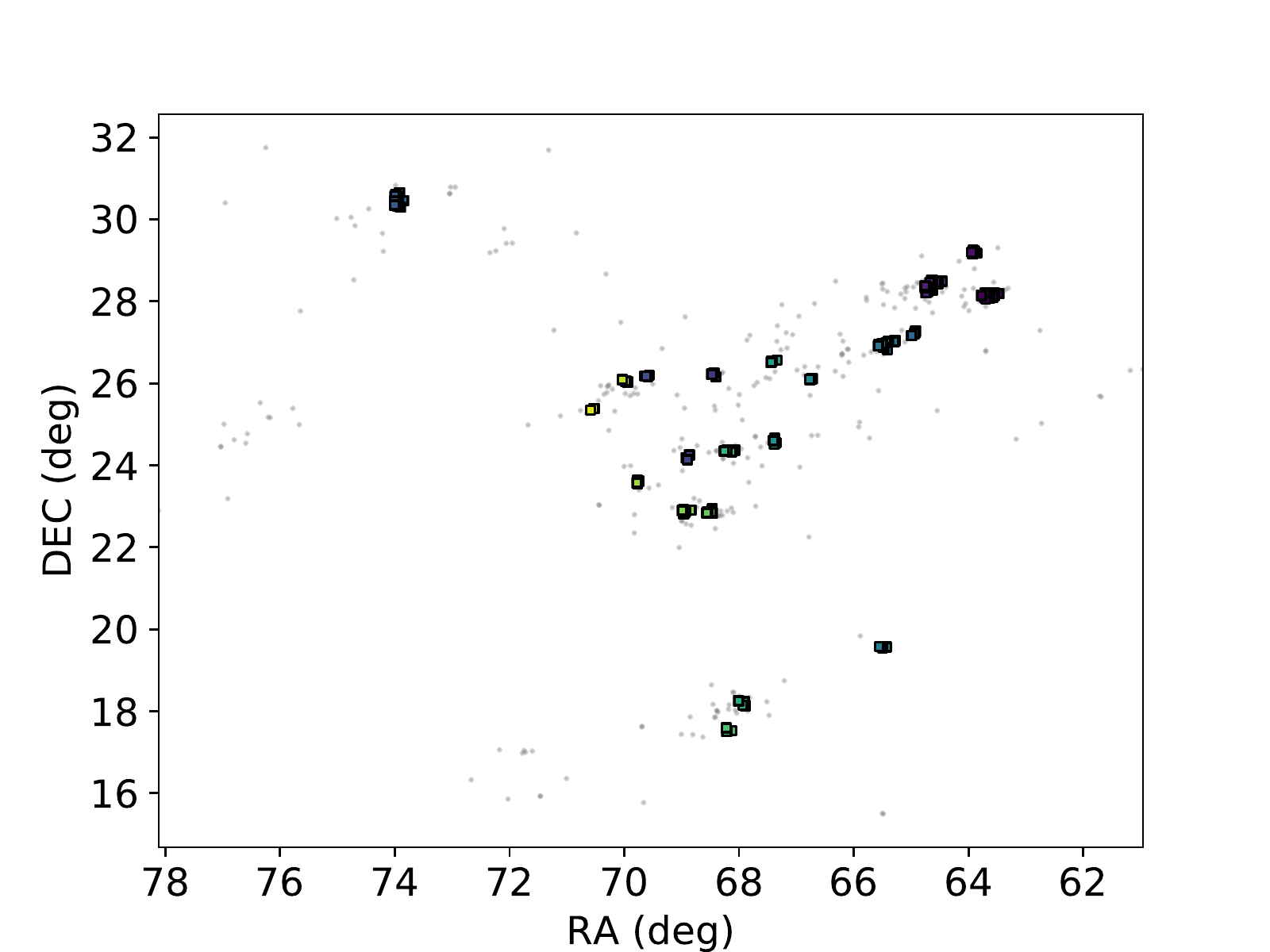}
\includegraphics[width=14cm]{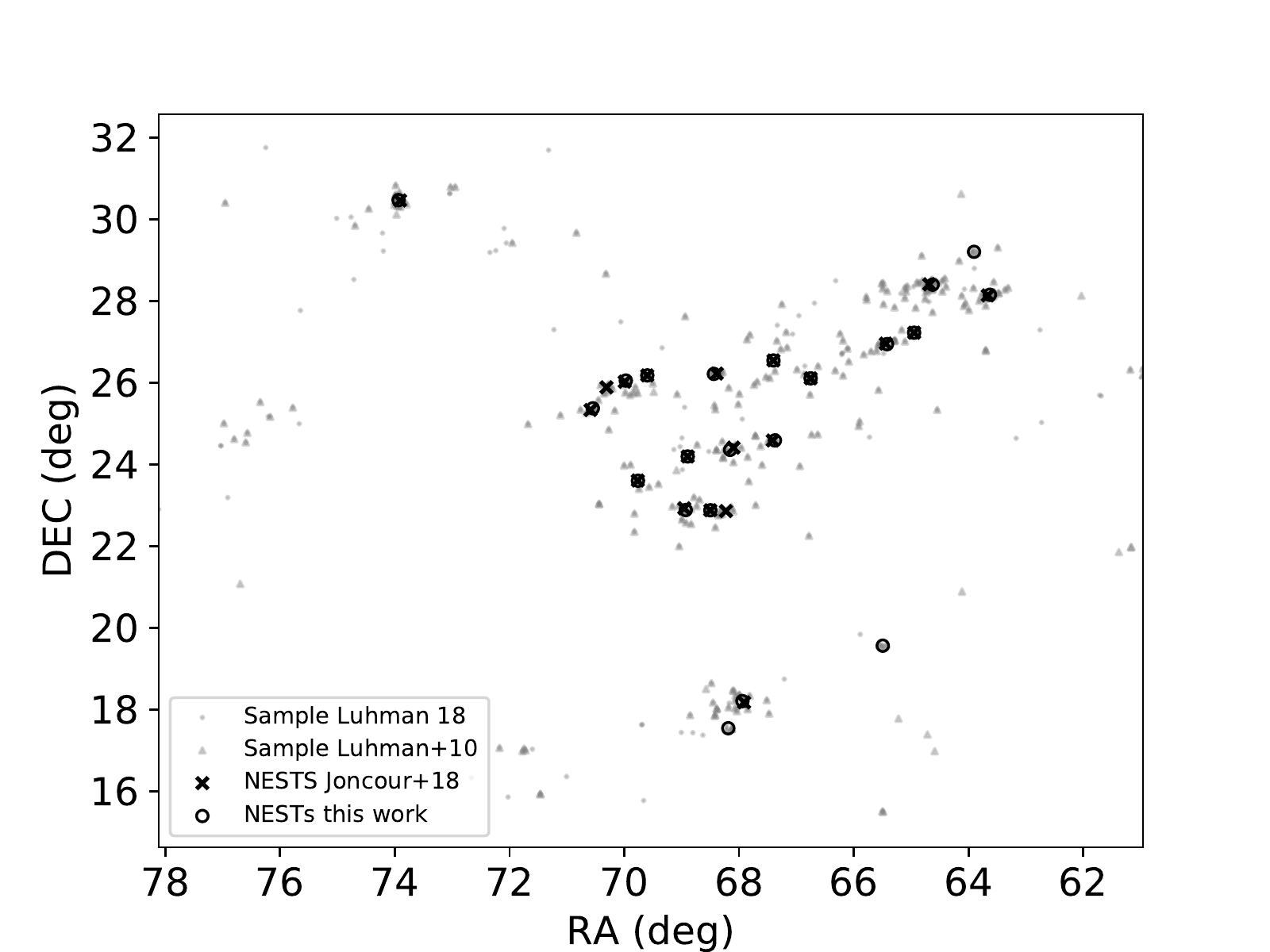}

\end{center}
\caption{Top: Position map (RA,DEC) of the Taurus sample. Noise stars are depicted as grey small dots, while stars belonging to the structures found are shown as coloured squares, with different shades of purple, blue, yellow and green representing different small scale structures. Bottom: Comparison of the position of the NESTs found in J18 and those detected in this work. Grey triangles show the stars from the analysis of \citet{Luhmanetal10} , while grey circles show stars in the updated samples from \citep{Luhman18}. Black symbols mark the centroids of the NESTs detected: crosses the NESTs found in J18, and empty circles the NESTs found in this work. }
\label{TaurusNests}
\end{figure*}

The original updated sample of \citet{Luhman18} has 438 stars. During preprocessing of the sample for multiple systems (collapsing multiples with distances lower than 1000 AU)  we have found 32 such systems, 31 double and 1 triple, yielding a final sample of 405 members. The summary of the results of the analysis of the region is shown in Table \ref{tableReal}, while Fig. \ref{HistoNests} shows the number of members within NESTs. Taurus has a density $\rho_{CSR}=7.469\, \rm{deg^{-2}}$ and a structure parameter $Q=0.484$, consistent with the simulations of a highly structured region. S2D2 obtained a relevent scale $\epsilon=0.094\, \rm{deg}$ and minimum number of members in structre $N_{min}=4$. Figure~\ref{TaurusNests} shows a map of our results in Taurus, where the grey circles represent noise and coloured squares are the stars within structures. Despite the differences in sample size (which is ~30\% larger in this work), and method (we now automatically calculate $\rho_{CSR}$ independently of the window, and use a kernel density approximation for the calculation of the OPCF) our results are consistent with those of J18 in terms of number, position and size of the NESTs detected. The bottom plot in Figure \ref{TaurusNests} is a map of the Taurus SFR comparing the positions of the structures detected in J18 and those in this work.  

When we compare the results of the analysis of Taurus region (in Table \ref{tableReal}) with those of synthetic regions (in Table \ref{tableSynth}), we observe general qualitative compatibility with significantly structured regions, that is, a large number of very  compact ($\rho_{nom}/\rho_{CSR}>15$) structures with a relatively low number of members ($N_{max}/N_{min}=5$) is retrieved.
Hovever, the numerical indicators are different from the simple box fractal models studied in section \ref{results}. The amount of NESTs, relative nominal density, the fraction of stars within them and the size of the most populated NEST compared to $N_{min}$ are larger in Taurus than in the fractal simulations. Our results indicate that even for an archetypically structured region, the nature of the structure is more complex than reflected by simple box-fractal models. This is consistent with the results by \citet{Lomaxetal18}, and will be throroughly analysed in Appendix \ref{AppendixQ}.

\subsection{IC 348}

\begin{figure*}[h!]
\begin{center}
\includegraphics[width=17cm]{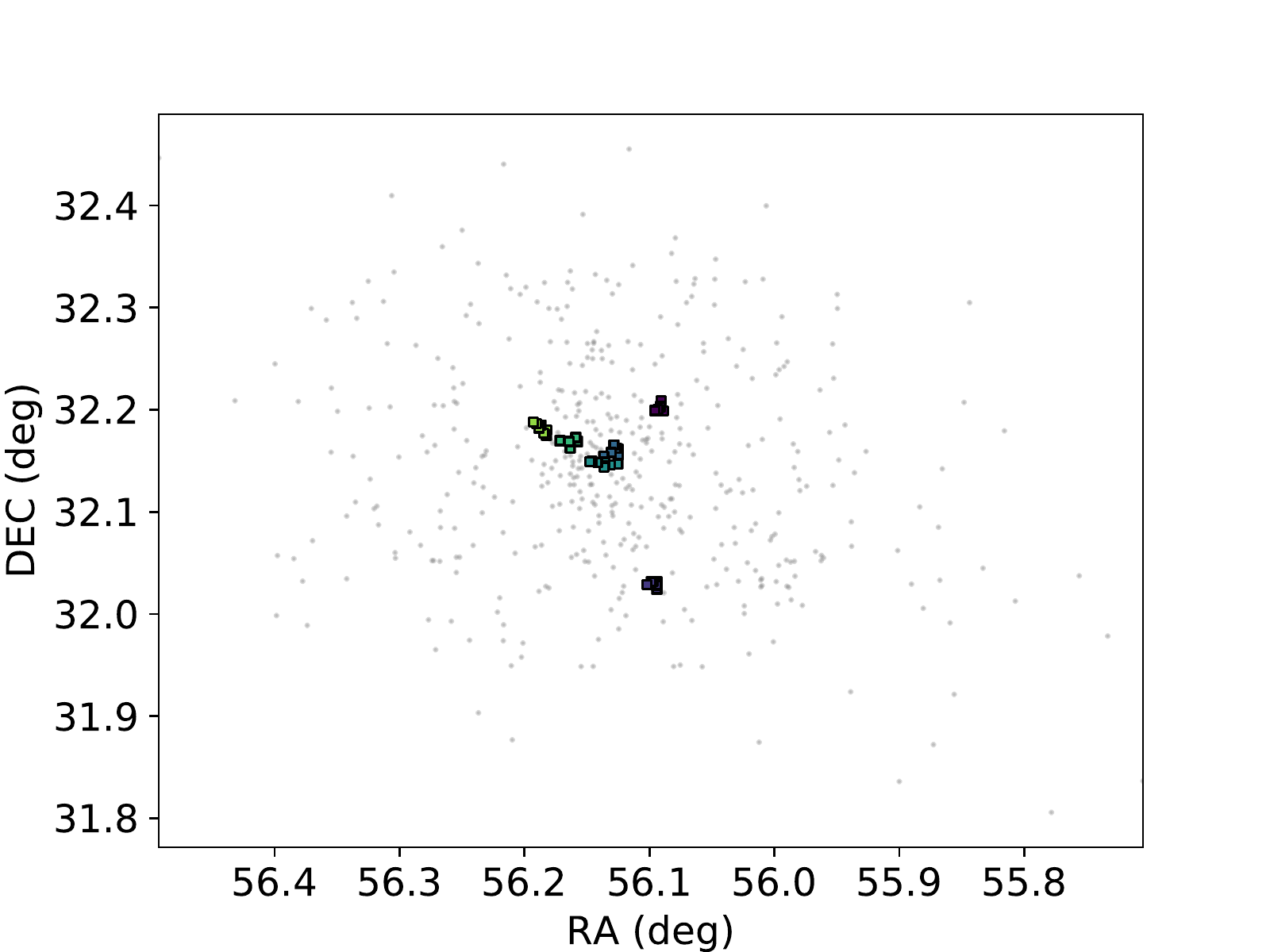}
\end{center}
\caption{Position map (RA,DEC) of the IC 348 sample. Noise stars are depicted as grey dots, while stars belonging to the structures found are shown in colour. Each shade of purple, blue and green represents a different NEST retrieved by S2D2.}
\label{ic348Nests}
\end{figure*}

IC 348 is a SFR in the Perseus cloud,  with YSOs of ages between ~2 and 6 Myr, and at a distance of ~315 pc, according to \citet{Luhmanetal16}, which provided an updated sample of its members. The general spatial trend of the YSO sample is clearly centrally concentrated, as can be seen in Fig. \ref{ic348Nests}.
 
The initial sample contains 478 stars, and preprocessing (as explained in section \ref{method}) found 13 binary systems, so the procedure was applied to the final sample of 465 stars. For this region (and also for Carina), densities and distances cannot be directly compared to other regions or to the simulated clusters due to the different physical size of the clusters.

Figure \ref{ic348Nests} shows a map with the detected structures, shown as coloured dots over the grey population. The general concentrated character of the region is in agreement with the value $Q=0.97$, consistent with a radial distribution with an exponent E=2.
We find 6 structures with $\epsilon=0.0067, \rm{deg}$ and $N_{min}=5$. Figure \ref{ic348Nests} shows a map of the YSOs in the sample, where coloured squares indicate belonging to a particular structure, and Figure \ref{HistoNests} shows a histogram of the number of members in each NEST.
The summarised results of the analysis of IC 348 are shown in Table \ref{tableReal}.  The  fraction of stars in structures, $f_{NEST}$, is close to distributions with significant central concentration (radial gradient with exponent E=2).  However, the relative maximum size of the structures, $N_{max}/N_{min}$ is much smaller, suggesting less concentrated regions, with exponent E=1. We note that in the centre of the cluster there are two substructures very close to each other, such that a small increase in the scale $\epsilon$ for DBSCAN would merge them within a larger single structure. This would decrease the significance of the NESTs, which is very high. The number of structures $N_{struct}$, or the relative nominal density for significant structures, can be coherent with both exponents of 1 or 2. 

These results indicate that the YSO distribution in IC 348 is reasonably  well represented by a radial density distribution with constant exponent between 1 and 2, and the substructures found correspond to Poisson fluctuations at densities larger than $\rho_{CSR}$, not to imprints of star formation sites.

\subsection{Upper Scorpius}

 Upper Scorpius is a region of the Scorpius-Centaurus OB association relatively close ($\sim$ 145 pc), young ($\sim$ 11 Myr), according to \citet{Luhmanetal18}, which also provided with a rich, new catalogue of YSO members. 
     
Visual inspection of the sample indicates some non spherically symmetric concentration with some structure, or at least significant density variations within. This is compatible with the $Q$ parameter of this sample $Q=0.915$, which points to clear though mild concentration.  

\begin{figure*}[h!]
\begin{center}
\includegraphics[width=17cm]{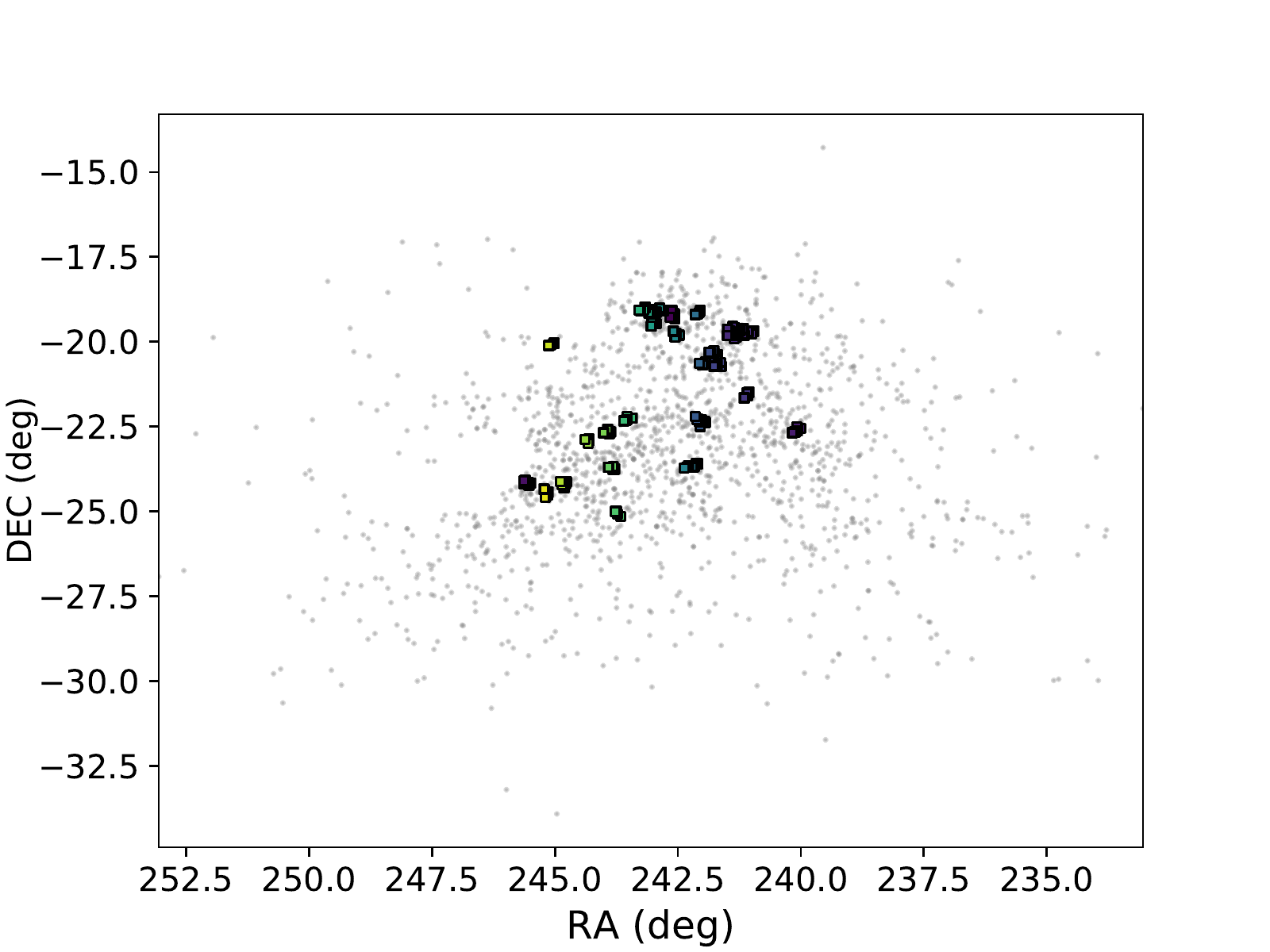}

\end{center}
\caption{Position map (RA,DEC) of the Upper Scorpius sample. Noise stars are depicted as grey dots, while stars belonging to the structures found are shown in colour. Each shade of purple, blue, green and yellow represents a different NEST retrieved by S2D2.}
\label{upperScorpiusNests}
\end{figure*}

In general, our results (shown in Figure \ref{upperScorpiusNests}) are compatible with moderate central concentration (radial with E=1), although the amount of structure detected, or the fraction of stars within NESTs are more in agreement with greater spatial concentration. There is a large structure with 29 stars towards the north, and plenty of small scale structures dispersed through the region.

Given the shape of the distribution, and the differences between radial and Plummer concentrated distributions, we believe that in this case we are dealing with a complex form of concentration with more than one slope, giving rise to a central density  'plateau'.
The situation is complex, and the small scale structures hard to interpret. There is a wide range of densities in the region, so a multi-scale analysis would shed light on the real nature and significance of these structures.

\subsection{Carina}
\begin{figure*}[h!]
\begin{center}
\includegraphics[width=17cm]{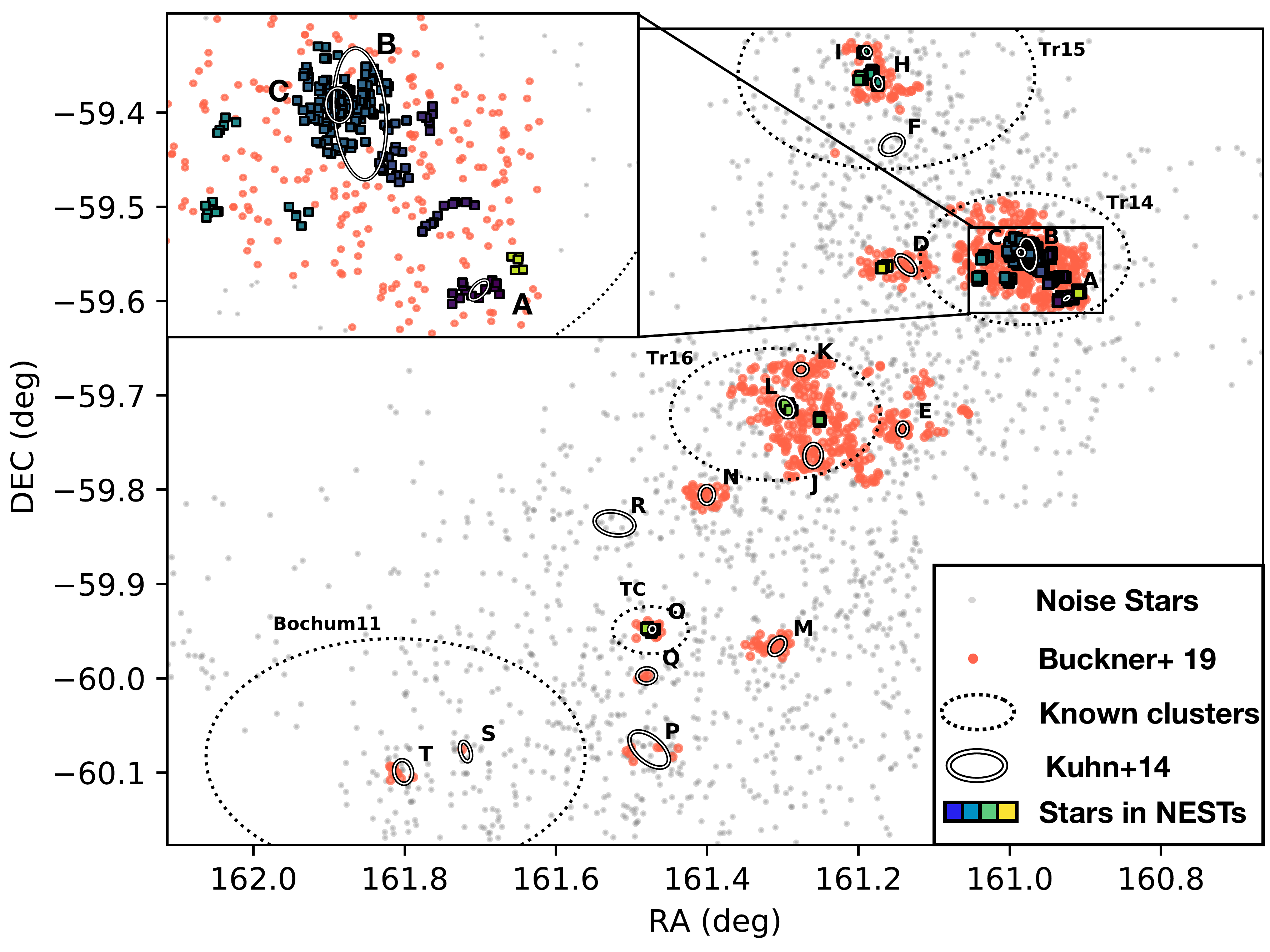}

\end{center}
\caption{Position map (RA, DEC) of the Carina Nebula sample, comparing the results of S2D2 with those of B19 and K14. Dotted ellipses show the position of known clusters in the area, while black rimmed white ellipses show the structures obtained by \citet{Kuhnetal14}, tagged with letters accordingly. Noise stars are depicted as grey dots, the stars with significant clustering index $I_{5}$ from \citet{Buckneretal19} are coloured in red. Stars belonging to structures found by S2D2 are squares coloured following the same colour scheme as in Figure \ref{TaurusNests}, with different NESTS represented by different shades of purple, blue, yellow and green.}
\label{carinaNests}
\end{figure*}

We finally apply our procedure to the YSOs in the Carina Nebula, a very large SFR with complex spatial structure (Fig. \ref{carinaNests}). It is known for hosting several open clusters (Trumpler 14, 15 and 16, Treasure Chest, and Bochum 11 in the window that we will analyse) and numerous massive stars, making it an interesting laboratory to study massive star formation and its effects on its surroundings.  

For this reason, the Carina SFR has been widely studied, allowing us to compare our results with two previous works. Carina was studied in \citet{Kuhnetal14} (henceforth K14) as part of the MYStIX survey which also provided a catalogue of members that we use in this work. The authors retrieved structure using a parametric finite mixture model fitting isothermal ellipsoids to substructures. These structures are varied in terms of scale and density, though always of the same shape. The final parameters, such as the number of structures, are decided by comparing the results with different values, to maximise the Akaike Information Criterium \citep[which is porportional to the likelihood of the model and includes a penalty for its complexity, as explained in][]{FeigelsonBabu12}. A total of 20 subclusters were identified by K14.

The same sample of Carina was also recently studied by \citet{Buckneretal19} (henceforth B19) with the tool INDICATE, which is not a structure detection tool, but a diagnostic of the local clustering trends in a sample. INDICATE assesses the clustering tendency of each star in a sample and assigns it an index, where larger indices imply stars with a higher degree of spatial association. The index values are calibrated against random distributions to define a `significance threshold' above which a star is considered to be spatially clustered.

In Fig. \ref{carinaNests}, the results from B19 and K14 are shown, and compared with those from S2D2. The stars with a clustering index from B19 above the significance level are shown in red. The structures found by K14 are depicted as black rimmed white ellipses labeled with letters, and black dotted ellipses show the known clusters in the area. The stars belonging to the different structures found by S2D2 are squares coloured following the same palette as in previous plots, with different shades of purple, green and yellow. The area of Trumpler 14 is zoomed so the retrieved NESTs are clearer.

The number of members of the structures found by S2D2 in Carina  is shown in Fig. \ref{HistoNests}.  In general, the structures retrieved by S2D2 are in agreement with previous work, with all of them in areas where the clustering index from B19 is significant, and the majority (~70 \%) of them in areas were K14 also found structures. We retrieve a large number of structures, mostly in the densest northern area, in the positions of Trumpler 14, 15 and 16. There are indications that the region is very complex in nature, with significant local radial concentrations, despite globally having a $Q$ value indicating substructure. The $\rho_{CSR}$ value of the region density is dominated by the northern area. The large value of $N_{max}/N_{min}$ indicates that Trumpler 14, where a structure with more than 100 stars is retrieved with some associated secondary structures, is concentrated and shows a gradient. This is consistent with B19, where the highest index values are found, and also with the A and B structures in K14. Some small scale structure is detected in the Trumpler 15 and 13 areas, also consistent with previous work. In the northern region of Carina, the results given by the three methods are broadly compatible, with the exception of structure F given by K14. Here, no structure is retrieved by S2D2 since its density is more than a factor of two lower than the nominal density required for detection. The probability of structure F being an artefact due to random sampling or projection effects
is not negligible. This is in agreement with B19 who found F had no stars with significant clustering tendencies and proposed it may not be real cluster but
instead fluctuations in the dispersed population field. We therefore recommend great care before interpreting this structure or including it in statistical analyses.

In the south, the only significant detection is a small structure in the Treasure Chest area, compatible with structure O from K14, and with significant clustering index according to B19. A comparison with previous work shows that we are losing some substructure due to the use of a single density (and thus, scale) for the whole region. The results given by B19 in the south also show low values for their clustering tendency compared to the northern area. K14 classified all the regions in the MYStIX catalogue according to the kind of substructure fitted by their method in four classes: \textit{simple} regions, with one single cluster; \textit{cluster- halo} regions, where a big structure surrounding a smaller, more compact one was identified; \textit{linear chain} regions, where the subclusters were organised in filaments; and \textit{clumpy} regions, where a large number of structures were found all over the area. Carina did not fit in any of these four categories, showing mixed traits, and was finally categorised as a complex region.  This is consistent with our inability to detect all the structure in the complete region simultaneously.
Additional tests with our procedure, separating the whole Carina region in South-East and North-West (using the same windows as in B19) show that separating the region before the analysis leads to similar results in the NW area while in the SE more significant structure is retrieved, still globally compatible with the results of K14 and B19. To get all the significant structures at different scales at once, we recommend a multi-scale method, like the one in Joncour, in prep., \citet{Joncour19}.

\section{Summary and Conclusions}\label{conclusions}
In this work, we present the S2D2 procedure, designed to systematically choose the parameters $\epsilon$ (length scale) and $N_{min}$ (minimum number of points) based on statistical properties of the star sample, to guarantee that DBSCAN will search, within clusters, for the smallest scale structures that are significant above random fluctuations. Different implementations of the procedure will be available for the scientific community.

We have implemented S2D2 and tested it in both synthetic and real projected young clusters, representing a range of initial conditions including and well beyond the extent where these small scale structures are present.

In substructured regions, S2D2 is very successful. The structure we retrieve is exactly what the method was designed for: very significant small scale structure, or NESTs. In young SFRs like Taurus, NESTs are the candidates to be pristine remnants of the process of fragmentation of the original gas cloud and cores.   

Even when we consider regions beyond the original scope of S2D2, (such as homogeneous clusters where the subclusters that we search for do not appear, or concentrated clusters that, in addition of not having subclusters are not  represented by a single density $\rho_{CSR}$ to define relevant scale and significance of the structures)  the methodical and robust nature of our procedure gives consistent and systematic results. 

Special care must be taken when interpreting the results of regions that are close to complete spatial randomness, characterised by values of the $Q$ structure parameter in the range $Q \in (0.7,0.87)$. In these cases, and despite retrieving only structures with more than $3-\sigma$ significance above random, we cannot guarantee that the structure retrieved, though compact and small-scale, is not associated to projection effects or the presence of a slight density gradient.

For concentrated regions, characterised by a large scale structure, the main core is detected accompanied by a halo of secondary structure. The appearance of secondary structures is one of the limitations of S2D2, as an analysis that only considers a single scale or size as relevant. These structures appear in regions characterised by stellar density gradients and are associated to random fluctuations at different local densities.
 
Tests in real regions show that YSOs in observed SFRs hardly have a structure as simple as that in idealised synthetic regions. In some cases, as shown in Carina, this is partly due to choosing very large and complex region, comprising several clusters of simpler nature and possibly different evolutionary state. However, disentangling these clusters into single samples can prove a very challenging task, and the final sample size can be too small to significantly apply statistical methods in them.

The ideal method in those cases would consider the varied and complex nature of SFRs, analysing the structure at several scales. This way, the hierarchy of structures in terms of scale and density can be completely captured. For that reason, a multi-scale version of the procedure presented in this work is in development (Joncour, in prep.), preliminary results shown in \citet{Joncour19}.

S2D2 provides consistent, robust and meaningful detection of significant substructures in star-forming regions. As such, it allows for comparison of the structure in different regions. This work will be followed by an extensive analysis of a whole set of observed SFRs to proceed to the statistical analysis of the retrieved substructures and assess their geometrical properties, evolutionary and dynamical status.

\begin{acknowledgements}
The Star Form Mapper project has received funding from the European Union's Horizon 2020 research and innovation program under
grant agreement No 687528.
\end{acknowledgements}

\bibliographystyle{aa} % style aa.bst
\bibliography{Bibliography}

\begin{thebibliography}{75}
\expandafter\ifx\csname natexlab\endcsname\relax\def\natexlab#1{#1}\fi

\bibitem[{{Aarseth} {et~al.}(2008){Aarseth}, {Tout}, \&
  {Mardling}}]{Aarsethetal08}
{Aarseth}, S.~J., {Tout}, C.~A., \& {Mardling}, R.~A. 2008, {The Cambridge
  N-Body Lectures}, Vol. 760

\bibitem[{{Alfaro} \& {Gonz{\'a}lez}(2016)}]{AlfaroGonzalez16}
{Alfaro}, E.~J. \& {Gonz{\'a}lez}, M. 2016, \mnras, 456, 2900

\bibitem[{{Alfaro} \& {Rom{\'a}n-Z{\'u}{\~n}iga}(2018)}]{Alfaroetal18}
{Alfaro}, E.~J. \& {Rom{\'a}n-Z{\'u}{\~n}iga}, C.~G. 2018, \mnras, 478, L110

\bibitem[{{Allison} {et~al.}(2009){Allison}, {Goodwin}, {Parker}, {Portegies
  Zwart}, {de Grijs}, \& {Kouwenhoven}}]{Allisonetal09}
{Allison}, R.~J., {Goodwin}, S.~P., {Parker}, R.~J., {et~al.} 2009, \mnras,
  395, 1449

\bibitem[{{Astropy Collaboration} {et~al.}(2013){Astropy Collaboration},
  {Robitaille}, {Tollerud}, {Greenfield}, {Droettboom}, {Bray}, {Aldcroft},
  {Davis}, {Ginsburg}, {Price-Whelan}, {Kerzendorf}, {Conley}, {Crighton},
  {Barbary}, {Muna}, {Ferguson}, {Grollier}, {Parikh}, {Nair}, {Unther},
  {Deil}, {Woillez}, {Conseil}, {Kramer}, {Turner}, {Singer}, {Fox}, {Weaver},
  {Zabalza}, {Edwards}, {Azalee Bostroem}, {Burke}, {Casey}, {Crawford},
  {Dencheva}, {Ely}, {Jenness}, {Labrie}, {Lim}, {Pierfederici}, {Pontzen},
  {Ptak}, {Refsdal}, {Servillat}, \& {Streicher}}]{Astropy}
{Astropy Collaboration}, {Robitaille}, T.~P., {Tollerud}, E.~J., {et~al.} 2013,
  \aap, 558, A33

\bibitem[{Baddeley {et~al.}(2015)Baddeley, Rubak, \& Turner}]{Baddeleyetal15}
Baddeley, A., Rubak, E., \& Turner, R. 2015, Spatial Point Patterns:
  Methodology and Applications with R (Chapman and Hall/CRC Press)

\bibitem[{Baines {et~al.}(2019)Baines, de~Calle, Herrera-Fernandez, Ibarra,
  Salgado, \& Valero-Martin}]{Bainesetal19}
Baines, D., de~Calle, I.~l., Herrera-Fernandez, J.~M., {et~al.} 2019, in
  Astronomical Society of the Pacific Conference Series, Vol. 523, Astronomical
  Data Analysis Software and Systems XXVIII (Astronomical Society of the
  Pacific)

\bibitem[{Botev {et~al.}(2010)Botev, Grotowski, \& Kroese}]{Botev10}
Botev, Z.~I., Grotowski, J.~F., \& Kroese, D.~P. 2010, The Annals of
  Statistics, 38, 2916

\bibitem[{{Buckner} {et~al.}(2019){Buckner}, {Khorrami}, {Khalaj}, {Lumsden},
  {Joncour}, {Moraux}, {Clark}, {Oudmaijer}, {Blanco}, {de la Calle},
  {Herrera-Fernand ez}, {Motte}, {Salgado}, \&
  {Valero-Mart{\'\i}n}}]{Buckneretal19}
{Buckner}, A. S.~M., {Khorrami}, Z., {Khalaj}, P., {et~al.} 2019, \aap, 622,
  A184

\bibitem[{{C{\'a}novas} {et~al.}(2019){C{\'a}novas}, {Cantero}, {Cieza},
  {Bombrun}, {Lammers}, {Mer{\'\i}n}, {Mora}, {Ribas}, \&
  {Ru{\'\i}z-Rodr{\'\i}guez}}]{Canovasetal19}
{C{\'a}novas}, H., {Cantero}, C., {Cieza}, L., {et~al.} 2019, \aap, 626, A80

\bibitem[{{Cantat-Gaudin} {et~al.}(2018){Cantat-Gaudin}, {Jordi}, {Vallenari},
  {Bragaglia}, {Balaguer-N{\'u}{\~n}ez}, {Soubiran}, {Bossini}, {Moitinho},
  {Castro-Ginard}, {Krone-Martins}, {Casamiquela}, {Sordo}, \&
  {Carrera}}]{CantatGaudinetal18}
{Cantat-Gaudin}, T., {Jordi}, C., {Vallenari}, A., {et~al.} 2018, \aap, 618,
  A93

\bibitem[{{Cartwright}(2009)}]{Cartwright09}
{Cartwright}, A. 2009, \mnras, 400, 1427

\bibitem[{{Cartwright} \& {Whitworth}(2004)}]{CartwrightWhitworth04}
{Cartwright}, A. \& {Whitworth}, A.~P. 2004, \mnras, 348, 589

\bibitem[{{Cartwright} \& {Whitworth}(2009)}]{CartwrightWhitworth09}
{Cartwright}, A. \& {Whitworth}, A.~P. 2009, \mnras, 392, 341

\bibitem[{{Casertano} \& {Hut}(1985)}]{CasertanoHut85}
{Casertano}, S. \& {Hut}, P. 1985, \apj, 298, 80

\bibitem[{{Clarke}(2010)}]{Clarke10}
{Clarke}, C. 2010, Philosophical Transactions of the Royal Society of London
  Series A, 368, 733

\bibitem[{{Costado} {et~al.}(2017){Costado}, {Alfaro}, {Gonz{\'a}lez}, \&
  {Sampedro}}]{Costadoetal17}
{Costado}, M.~T., {Alfaro}, E.~J., {Gonz{\'a}lez}, M., \& {Sampedro}, L. 2017,
  \mnras, 465, 3879

\bibitem[{{Daffern-Powell} \& {Parker}(2020)}]{DaffernPowellParker20}
{Daffern-Powell}, E.~C. \& {Parker}, R.~J. 2020, \mnras, 493, 4925

\bibitem[{{Dib} {et~al.}(2018){Dib}, {Schmeja}, \& {Parker}}]{Dibetal18}
{Dib}, S., {Schmeja}, S., \& {Parker}, R.~J. 2018, \mnras, 473, 849

\bibitem[{{Diggle}(2003)}]{Diggle03}
{Diggle}, P.~J. 2003, Statistical Analysis of spatial point patterns (Wiley and
  sons)

\bibitem[{Ester {et~al.}(1996)Ester, Kriegel, Sander, \& Xu}]{Esteretal96}
Ester, M., Kriegel, H.-P., Sander, J., \& Xu, X. 1996, in Proceedings of the
  Second International Conference on Knowledge Discovery and Data Mining,
  KDD'96 (AAAI Press), 226--231

\bibitem[{{Feigelson} \& {Babu}(2012)}]{FeigelsonBabu12}
{Feigelson}, E.~D. \& {Babu}, G.~J. 2012, {Modern Statistical Methods for
  Astronomy}

\bibitem[{{F{\H{u}}r{\'e}sz} {et~al.}(2006){F{\H{u}}r{\'e}sz}, {Hartmann},
  {Szentgyorgyi}, {Ridge}, {Rebull}, {Stauffer}, {Latham}, {Conroy},
  {Fabricant}, \& {Roll}}]{Fureszetal06}
{F{\H{u}}r{\'e}sz}, G., {Hartmann}, L.~W., {Szentgyorgyi}, A.~H., {et~al.}
  2006, \apj, 648, 1090

\bibitem[{Fujii \& Portegies-Zwart(2016)}]{FujiiZwart16}
Fujii, M.~S. \& Portegies-Zwart, S. 2016, The Astrophysical Journal, 817, 4

\bibitem[{{Gaia Collaboration} {et~al.}(2016){Gaia Collaboration}, {Prusti},
  {de Bruijne}, {Brown}, {Vallenari}, {Babusiaux}, {Bailer-Jones}, {Bastian},
  {Biermann}, {Evans}, {Eyer}, {Jansen}, {Jordi}, {Klioner}, {Lammers},
  {Lindegren}, {Luri}, {Mignard}, {Milligan}, {Panem}, {Poinsignon},
  {Pourbaix}, {Randich}, {Sarri}, {Sartoretti}, {Siddiqui}, {Soubiran},
  {Valette}, {van Leeuwen}, {Walton}, {Aerts}, {Arenou}, {Cropper}, {Drimmel},
  {H{\o}g}, {Katz}, {Lattanzi}, {O'Mullane}, {Grebel}, {Holland}, {Huc},
  {Passot}, {Bramante}, {Cacciari}, {Casta{\~n}eda}, {Chaoul}, {Cheek}, {De
  Angeli}, {Fabricius}, {Guerra}, {Hern{\'a}ndez}, {Jean-Antoine-Piccolo},
  {Masana}, {Messineo}, {Mowlavi}, {Nienartowicz}, {Ord{\'o}{\~n}ez-Blanco},
  {Panuzzo}, {Portell}, {Richards}, {Riello}, {Seabroke}, {Tanga},
  {Th{\'e}venin}, {Torra}, {Els}, {Gracia-Abril}, {Comoretto},
  {Garcia-Reinaldos}, {Lock}, {Mercier}, {Altmann}, {Andrae}, {Astraatmadja},
  {Bellas-Velidis}, {Benson}, {Berthier}, {Blomme}, {Busso}, {Carry},
  {Cellino}, {Clementini}, {Cowell}, {Creevey}, {Cuypers}, {Davidson}, {De
  Ridder}, {de Torres}, {Delchambre}, {Dell'Oro}, {Ducourant}, {Fr{\'e}mat},
  {Garc{\'\i}a-Torres}, {Gosset}, {Halbwachs}, {Hambly}, {Harrison}, {Hauser},
  {Hestroffer}, {Hodgkin}, {Huckle}, {Hutton}, {Jasniewicz}, {Jordan},
  {Kontizas}, {Korn}, {Lanzafame}, {Manteiga}, {Moitinho}, {Muinonen},
  {Osinde}, {Pancino}, {Pauwels}, {Petit}, {Recio-Blanco}, {Robin}, {Sarro},
  {Siopis}, {Smith}, {Smith}, {Sozzetti}, {Thuillot}, {van Reeven}, {Viala},
  {Abbas}, {Abreu Aramburu}, {Accart}, {Aguado}, {Allan}, {Allasia},
  {Altavilla}, {{\'A}lvarez}, {Alves}, {Anderson}, {Andrei}, {Anglada Varela},
  {Antiche}, {Antoja}, {Ant{\'o}n}, {Arcay}, {Atzei}, {Ayache}, {Bach},
  {Baker}, {Balaguer-N{\'u}{\~n}ez}, {Barache}, {Barata}, {Barbier}, {Barblan},
  {Baroni}, {Barrado y Navascu{\'e}s}, {Barros}, {Barstow}, {Becciani},
  {Bellazzini}, {Bellei}, {Bello Garc{\'\i}a}, {Belokurov}, {Bendjoya},
  {Berihuete}, {Bianchi}, {Bienaym{\'e}}, {Billebaud}, {Blagorodnova},
  {Blanco-Cuaresma}, {Boch}, {Bombrun}, {Borrachero}, {Bouquillon}, {Bourda},
  {Bouy}, {Bragaglia}, {Breddels}, {Brouillet}, {Br{\"u}semeister},
  {Bucciarelli}, {Budnik}, {Burgess}, {Burgon}, {Burlacu}, {Busonero}, {Buzzi},
  {Caffau}, {Cambras}, {Campbell}, {Cancelliere}, {Cantat-Gaudin}, {Carlucci},
  {Carrasco}, {Castellani}, {Charlot}, {Charnas}, {Charvet}, {Chassat},
  {Chiavassa}, {Clotet}, {Cocozza}, {Collins}, {Collins}, {Costigan}, {Crifo},
  {Cross}, {Crosta}, {Crowley}, {Dafonte}, {Damerdji}, {Dapergolas}, {David},
  {David}, {De Cat}, {de Felice}, {de Laverny}, {De Luise}, {De March}, {de
  Martino}, {de Souza}, {Debosscher}, {del Pozo}, {Delbo}, {Delgado},
  {Delgado}, {di Marco}, {Di Matteo}, {Diakite}, {Distefano}, {Dolding}, {Dos
  Anjos}, {Drazinos}, {Dur{\'a}n}, {Dzigan}, {Ecale}, {Edvardsson}, {Enke},
  {Erdmann}, {Escolar}, {Espina}, {Evans}, {Eynard Bontemps}, {Fabre},
  {Fabrizio}, {Faigler}, {Falc{\~a}o}, {Farr{\`a}s Casas}, {Faye}, {Federici},
  {Fedorets}, {Fern{\'a}ndez-Hern{\'a}ndez}, {Fernique}, {Fienga}, {Figueras},
  {Filippi}, {Findeisen}, {Fonti}, {Fouesneau}, {Fraile}, {Fraser}, {Fuchs},
  {Furnell}, {Gai}, {Galleti}, {Galluccio}, {Garabato}, {Garc{\'\i}a-Sedano},
  {Gar{\'e}}, {Garofalo}, {Garralda}, {Gavras}, {Gerssen}, {Geyer}, {Gilmore},
  {Girona}, {Giuffrida}, {Gomes}, {Gonz{\'a}lez-Marcos},
  {Gonz{\'a}lez-N{\'u}{\~n}ez}, {Gonz{\'a}lez-Vidal}, {Granvik}, {Guerrier},
  {Guillout}, {Guiraud}, {G{\'u}rpide}, {Guti{\'e}rrez-S{\'a}nchez}, {Guy},
  {Haigron}, {Hatzidimitriou}, {Haywood}, {Heiter}, {Helmi}, {Hobbs},
  {Hofmann}, {Holl}, {Holland }, {Hunt}, {Hypki}, {Icardi}, {Irwin}, {Jevardat
  de Fombelle}, {Jofr{\'e}}, {Jonker}, {Jorissen}, {Julbe}, {Karampelas},
  {Kochoska}, {Kohley}, {Kolenberg}, {Kontizas}, {Koposov}, {Kordopatis},
  {Koubsky}, {Kowalczyk}, {Krone-Martins}, {Kudryashova}, {Kull}, {Bachchan},
  {Lacoste-Seris}, {Lanza}, {Lavigne}, {Le Poncin-Lafitte}, {Lebreton},
  {Lebzelter}, {Leccia}, {Leclerc}, {Lecoeur-Taibi}, {Lemaitre}, {Lenhardt},
  {Leroux}, {Liao}, {Licata}, {Lindstr{\o}m}, {Lister}, {Livanou}, {Lobel},
  {L{\"o}ffler}, {L{\'o}pez}, {Lopez-Lozano}, {Lorenz}, {Loureiro},
  {MacDonald}, {Magalh{\~a}es Fernandes}, {Managau}, {Mann}, {Mantelet},
  {Marchal}, {Marchant}, {Marconi}, {Marie}, {Marinoni}, {Marrese},
  {Marschalk{\'o}}, {Marshall}, {Mart{\'\i}n-Fleitas}, {Martino}, {Mary},
  {Matijevi{\v{c}}}, {Mazeh}, {McMillan}, {Messina}, {Mestre}, {Michalik},
  {Millar}, {Miranda}, {Molina}, {Molinaro}, {Molinaro}, {Moln{\'a}r},
  {Moniez}, {Montegriffo}, {Monteiro}, {Mor}, {Mora}, {Morbidelli}, {Morel},
  {Morgenthaler}, {Morley}, {Morris}, {Mulone}, {Muraveva}, {Musella},
  {Narbonne}, {Nelemans}, {Nicastro}, {Noval}, {Ord{\'e}novic},
  {Ordieres-Mer{\'e}}, {Osborne}, {Pagani}, {Pagano}, {Pailler}, {Palacin},
  {Palaversa}, {Parsons}, {Paulsen}, {Pecoraro}, {Pedrosa}, {Pentik{\"a}inen},
  {Pereira}, {Pichon}, {Piersimoni}, {Pineau}, {Plachy}, {Plum}, {Poujoulet},
  {Pr{\v{s}}a}, {Pulone}, {Ragaini}, {Rago}, {Rambaux}, {Ramos-Lerate},
  {Ranalli}, {Rauw}, {Read}, {Regibo}, {Renk}, {Reyl{\'e}}, {Ribeiro},
  {Rimoldini}, {Ripepi}, {Riva}, {Rixon}, {Roelens}, {Romero-G{\'o}mez},
  {Rowell}, {Royer}, {Rudolph}, {Ruiz-Dern}, {Sadowski}, {Sagrist{\`a}
  Sell{\'e}s}, {Sahlmann}, {Salgado}, {Salguero}, {Sarasso}, {Savietto},
  {Schnorhk}, {Schultheis}, {Sciacca}, {Segol}, {Segovia}, {Segransan},
  {Serpell}, {Shih}, {Smareglia}, {Smart}, {Smith}, {Solano}, {Solitro},
  {Sordo}, {Soria Nieto}, {Souchay}, {Spagna}, {Spoto}, {Stampa}, {Steele},
  {Steidelm{\"u}ller}, {Stephenson}, {Stoev}, {Suess}, {S{\"u}veges}, {Surdej},
  {Szabados}, {Szegedi-Elek}, {Tapiador}, {Taris}, {Tauran}, {Taylor},
  {Teixeira}, {Terrett}, {Tingley}, {Trager}, {Turon}, {Ulla}, {Utrilla},
  {Valentini}, {van Elteren}, {Van Hemelryck}, {van Leeuwen}, {Varadi},
  {Vecchiato}, {Veljanoski}, {Via}, {Vicente}, {Vogt}, {Voss}, {Votruba},
  {Voutsinas}, {Walmsley}, {Weiler}, {Weingrill}, {Werner}, {Wevers},
  {Whitehead}, {Wyrzykowski}, {Yoldas}, {{\v{Z}}erjal}, {Zucker}, {Zurbach},
  {Zwitter}, {Alecu}, {Allen}, {Allende Prieto}, {Amorim},
  {Anglada-Escud{\'e}}, {Arsenijevic}, {Azaz}, {Balm}, {Beck}, {Bernstein},
  {Bigot}, {Bijaoui}, {Blasco}, {Bonfigli}, {Bono}, {Boudreault}, {Bressan},
  {Brown}, {Brunet}, {Bunclark}, {Buonanno}, {Butkevich}, {Carret}, {Carrion},
  {Chemin}, {Ch{\'e}reau}, {Corcione}, {Darmigny}, {de Boer}, {de Teodoro}, {de
  Zeeuw}, {Delle Luche}, {Domingues}, {Dubath}, {Fodor}, {Fr{\'e}zouls},
  {Fries}, {Fustes}, {Fyfe}, {Gallardo}, {Gallegos}, {Gardiol}, {Gebran},
  {Gomboc}, {G{\'o}mez}, {Grux}, {Gueguen}, {Heyrovsky}, {Hoar}, {Iannicola},
  {Isasi Parache}, {Janotto}, {Joliet}, {Jonckheere}, {Keil}, {Kim},
  {Klagyivik}, {Klar}, {Knude}, {Kochukhov}, {Kolka}, {Kos}, {Kutka}, {Lainey},
  {LeBouquin}, {Liu}, {Loreggia}, {Makarov}, {Marseille}, {Martayan},
  {Martinez-Rubi}, {Massart}, {Meynadier}, {Mignot}, {Munari}, {Nguyen},
  {Nordlander}, {Ocvirk}, {O'Flaherty}, {Olias Sanz}, {Ortiz}, {Osorio},
  {Oszkiewicz}, {Ouzounis}, {Palmer}, {Park}, {Pasquato}, {Peltzer}, {Peralta},
  {P{\'e}turaud}, {Pieniluoma}, {Pigozzi}, {Poels}, {Prat}, {Prod'homme},
  {Raison}, {Rebordao}, {Risquez}, {Rocca-Volmerange}, {Rosen}, {Ruiz-Fuertes},
  {Russo}, {Sembay}, {Serraller Vizcaino}, {Short}, {Siebert}, {Silva},
  {Sinachopoulos}, {Slezak}, {Soffel}, {Sosnowska}, {Strai{\v{z}}ys}, {ter
  Linden}, {Terrell}, {Theil}, {Tiede}, {Troisi}, {Tsalmantza}, {Tur},
  {Vaccari}, {Vachier}, {Valles}, {Van Hamme}, {Veltz}, {Virtanen}, {Wallut},
  {Wichmann}, {Wilkinson}, {Ziaeepour}, \& {Zschocke}}]{Gaia2016}
{Gaia Collaboration}, {Prusti}, T., {de Bruijne}, J.~H.~J., {et~al.} 2016,
  \aap, 595, A1

\bibitem[{{Gomez} {et~al.}(1993){Gomez}, {Hartmann}, {Kenyon}, \&
  {Hewett}}]{Gomezetal93}
{Gomez}, M., {Hartmann}, L., {Kenyon}, S.~J., \& {Hewett}, R. 1993, \aj, 105,
  1927

\bibitem[{{Gonz{\'a}lez} \& {Alfaro}(2017)}]{GonzalezAlfaro17}
{Gonz{\'a}lez}, M. \& {Alfaro}, E.~J. 2017, \mnras, 465, 1889

\bibitem[{{Gutermuth} {et~al.}(2009){Gutermuth}, {Megeath}, {Myers}, {Allen},
  {Pipher}, \& {Fazio}}]{Gutermuthetal09}
{Gutermuth}, R.~A., {Megeath}, S.~T., {Myers}, P.~C., {et~al.} 2009, \apjs,
  184, 18

\bibitem[{{Hacar} {et~al.}(2017){Hacar}, {Tafalla}, \& {Alves}}]{Hacaretal17}
{Hacar}, A., {Tafalla}, M., \& {Alves}, J. 2017, \aap, 606, A123

\bibitem[{{Hetem} \& {Gregorio-Hetem}(2019)}]{HetemGregorioHetem19}
{Hetem}, A. \& {Gregorio-Hetem}, J. 2019, \mnras, 490, 2521

\bibitem[{{Illian} {et~al.}(2008){Illian}, {Penttinen}, {Stoyan}, \&
  {Stoyan}}]{Illianetal08}
{Illian}, J., {Penttinen}, A., {Stoyan}, H., \& {Stoyan}, D. 2008, Statistical
  Analysis and Modelling of spatial point patterns (John Wiley and Sons, Ltd)

\bibitem[{{Jaffa} {et~al.}(2017){Jaffa}, {Whitworth}, \& {Lomax}}]{Jaffaetal17}
{Jaffa}, S.~E., {Whitworth}, A.~P., \& {Lomax}, O. 2017, \mnras, 466, 1082

\bibitem[{{Joncour}(2019)}]{Joncour19}
{Joncour}, I. 2019, Astronomical Society of the Pacific Conference Series, Vol.
  523, {Multiscale Spatial Analysis of Young Stars Complex using the dbscan
  Clustering Algorithm}, ed. P.~J. {Teuben}, M.~W. {Pound}, B.~A. {Thomas}, \&
  E.~M. {Warner}, 87

\bibitem[{{Joncour} {et~al.}(2020){Joncour}, {Buckner}, {Khalaj}, {Moraux}, \&
  {Motte}}]{JoncourReportClustering}
{Joncour}, I., {Buckner}, A., {Khalaj}, P., {Moraux}, E., \& {Motte}, F. 2020,
  arXiv e-prints, arXiv:2006.07830

\bibitem[{{Joncour} {et~al.}(2017){Joncour}, {Duch{\^e}ne}, \&
  {Moraux}}]{Joncouretal17}
{Joncour}, I., {Duch{\^e}ne}, G., \& {Moraux}, E. 2017, \aap, 599, A14

\bibitem[{{Joncour} {et~al.}(2018){Joncour}, {Duch{\^e}ne}, {Moraux}, \&
  {Motte}}]{Joncouretal18}
{Joncour}, I., {Duch{\^e}ne}, G., {Moraux}, E., \& {Motte}, F. 2018, \aap, 620,
  A27

\bibitem[{{Kenyon} {et~al.}(2008){Kenyon}, {G{\'o}mez}, \&
  {Whitney}}]{Kenyonetal08}
{Kenyon}, S.~J., {G{\'o}mez}, M., \& {Whitney}, B.~A. 2008, {Low Mass Star
  Formation in the Taurus-Auriga Clouds}, ed. B.~{Reipurth}, Vol.~4, 405

\bibitem[{{Kirk} {et~al.}(2016){Kirk}, {Johnstone}, {Di Francesco}, {Lane},
  {Buckle}, {Berry}, {Broekhoven-Fiene}, {Currie}, {Fich}, {Hatchell},
  {Jenness}, {Mottram}, {Nutter}, {Pattle}, {Pineda}, {Quinn}, {Salji}, {Tisi},
  {Hogerheijde}, {Ward-Thompson}, \& {JCMT Gould Belt Survey
  Team}}]{Kirketal16}
{Kirk}, H., {Johnstone}, D., {Di Francesco}, J., {et~al.} 2016, \apj, 821, 98

\bibitem[{{Kirk} \& {Myers}(2011)}]{KirkMyers11}
{Kirk}, H. \& {Myers}, P.~C. 2011, \apj, 727, 64

\bibitem[{{Kounkel} \& {Covey}(2019)}]{KounkelCovey19}
{Kounkel}, M. \& {Covey}, K. 2019, \aj, 158, 122

\bibitem[{{Kraus} \& {Hillenbrand}(2008)}]{KrausHillenbrand08}
{Kraus}, A.~L. \& {Hillenbrand}, L.~A. 2008, \apjl, 686, L111

\bibitem[{{Kuhn} {et~al.}(2014){Kuhn}, {Feigelson}, {Getman}, {Baddeley},
  {Broos}, {Sills}, {Bate}, {Povich}, {Luhman}, {Busk}, {Naylor}, \&
  {King}}]{Kuhnetal14}
{Kuhn}, M.~A., {Feigelson}, E.~D., {Getman}, K.~V., {et~al.} 2014, \apj, 787,
  107

\bibitem[{{K{\"u}pper} {et~al.}(2011){K{\"u}pper}, {Maschberger}, {Kroupa}, \&
  {Baumgardt}}]{Kuepperetal11}
{K{\"u}pper}, A. H.~W., {Maschberger}, T., {Kroupa}, P., \& {Baumgardt}, H.
  2011, \mnras, 417, 2300

\bibitem[{{Larson}(1995)}]{Larson95}
{Larson}, R.~B. 1995, \mnras, 272, 213

\bibitem[{{Larson}(2007)}]{Larson07}
{Larson}, R.~B. 2007, Reports on Progress in Physics, 70, 337

\bibitem[{{Lomax} {et~al.}(2018){Lomax}, {Bates}, \& {Whitworth}}]{Lomaxetal18}
{Lomax}, O., {Bates}, M.~L., \& {Whitworth}, A.~P. 2018, \mnras, 480, 371

\bibitem[{{Luhman}(2018)}]{Luhman18}
{Luhman}, K.~L. 2018, \aj, 156, 271

\bibitem[{{Luhman} {et~al.}(2010){Luhman}, {Allen}, {Espaillat}, {Hartmann}, \&
  {Calvet}}]{Luhmanetal10}
{Luhman}, K.~L., {Allen}, P.~R., {Espaillat}, C., {Hartmann}, L., \& {Calvet},
  N. 2010, \apjs, 186, 111

\bibitem[{{Luhman} {et~al.}(2016){Luhman}, {Esplin}, \&
  {Loutrel}}]{Luhmanetal16}
{Luhman}, K.~L., {Esplin}, T.~L., \& {Loutrel}, N.~P. 2016, \apj, 827, 52

\bibitem[{{Luhman} {et~al.}(2018){Luhman}, {Herrmann}, {Mamajek}, {Esplin}, \&
  {Pecaut}}]{Luhmanetal18}
{Luhman}, K.~L., {Herrmann}, K.~A., {Mamajek}, E.~E., {Esplin}, T.~L., \&
  {Pecaut}, M.~J. 2018, \aj, 156, 76

\bibitem[{{Ma{\'\i}z-Apell{\'a}niz} {et~al.}(2004){Ma{\'\i}z-Apell{\'a}niz},
  {P{\'e}rez}, \& {Mas-Hesse}}]{MaizApellanizetal04}
{Ma{\'\i}z-Apell{\'a}niz}, J., {P{\'e}rez}, E., \& {Mas-Hesse}, J.~M. 2004,
  \aj, 128, 1196

\bibitem[{{Maschberger} \& {Clarke}(2011)}]{MaschbergerClarke11}
{Maschberger}, T. \& {Clarke}, C.~J. 2011, \mnras, 416, 541

\bibitem[{{Parker}(2018)}]{Parker18}
{Parker}, R.~J. 2018, \mnras, 476, 617

\bibitem[{{Parker} \& {Dale}(2013)}]{ParkerDale13}
{Parker}, R.~J. \& {Dale}, J.~E. 2013, \mnras, 432, 986

\bibitem[{{Parker} \& {Goodwin}(2015)}]{ParkerGoodwin15}
{Parker}, R.~J. \& {Goodwin}, S.~P. 2015, \mnras, 449, 3381

\bibitem[{{Parker} \& {Meyer}(2012)}]{ParkerMeyer12}
{Parker}, R.~J. \& {Meyer}, M.~R. 2012, \mnras, 427, 637

\bibitem[{{Parker} \& {Wright}(2018)}]{ParkerWright18}
{Parker}, R.~J. \& {Wright}, N.~J. 2018, \mnras, 481, 1679

\bibitem[{{Parker} {et~al.}(2014){Parker}, {Wright}, {Goodwin}, \&
  {Meyer}}]{Parkeretal14}
{Parker}, R.~J., {Wright}, N.~J., {Goodwin}, S.~P., \& {Meyer}, M.~R. 2014,
  \mnras, 438, 620

\bibitem[{Pedregosa {et~al.}(2011)Pedregosa, Varoquaux, Gramfort, Michel,
  Thirion, Grisel, Blondel, Prettenhofer, Weiss, Dubourg, Vanderplas, Passos,
  Cournapeau, Brucher, Perrot, \& Duchesnay}]{scikit-learn}
Pedregosa, F., Varoquaux, G., Gramfort, A., {et~al.} 2011, Journal of Machine
  Learning Research, 12, 2825

\bibitem[{{Peebles}(1980)}]{Peebles80}
{Peebles}, P.~J.~E. 1980, in Ninth Texas Symposium on Relativistic
  Astrophysics, Vol. 336, 161--171

\bibitem[{{Pfalzner} {et~al.}(2012){Pfalzner}, {Kaczmarek}, \&
  {Olczak}}]{Pfalzneretal12}
{Pfalzner}, S., {Kaczmarek}, T., \& {Olczak}, C. 2012, \aap, 545, A122

\bibitem[{Pflamm-Altenburg \& Kroupa(2008)}]{PflammAltenburgKroupa08}
Pflamm-Altenburg, J. \& Kroupa, P. 2008, Nature, 455, 641

\bibitem[{{Pilbratt}(2010)}]{Herschel}
{Pilbratt}, G. 2010, in JENAM 2010, Joint European and National Astronomy
  Meeting, 149

\bibitem[{{Reiter} \& {Parker}(2019)}]{ReiterParker19}
{Reiter}, M. \& {Parker}, R.~J. 2019, \mnras, 486, 4354

\bibitem[{{Retter} {et~al.}(2019){Retter}, {Hatchell}, \&
  {Naylor}}]{Retteretal19}
{Retter}, B., {Hatchell}, J., \& {Naylor}, T. 2019, \mnras, 487, 887

\bibitem[{{Robitaille} {et~al.}(2020){Robitaille}, {Abdeldayem}, {Joncour},
  {Moraux}, {Motte}, {Lesaffre}, \& {Khalil}}]{Robitailleetal20}
{Robitaille}, J.-F., {Abdeldayem}, A., {Joncour}, I., {et~al.} 2020, Astronomy
  \& Astrophysics, in press

\bibitem[{{Robitaille} {et~al.}(2019){Robitaille}, {Motte}, {Schneider},
  {Elia}, \& {Bontemps}}]{Robitailleetal19}
{Robitaille}, J.~F., {Motte}, F., {Schneider}, N., {Elia}, D., \& {Bontemps},
  S. 2019, \aap, 628, A33

\bibitem[{{Sacco} {et~al.}(2017){Sacco}, {Spina}, {Randich}, {Palla}, {Parker},
  {Jeffries}, {Jackson}, {Meyer}, {Mapelli}, {Lanzafame}, {Bonito}, {Damiani},
  {Franciosini}, {Frasca}, {Klutsch}, {Prisinzano}, {Tognelli},
  {Degl'Innocenti}, {Prada Moroni}, {Alfaro}, {Micela}, {Prusti}, {Barrado},
  {Biazzo}, {Bouy}, {Bravi}, {Lopez-Santiago}, {Wright}, {Bayo}, {Gilmore},
  {Bragaglia}, {Flaccomio}, {Koposov}, {Pancino}, {Casey}, {Costado}, {Donati},
  {Hourihane}, {Jofr{\'e}}, {Lardo}, {Lewis}, {Magrini}, {Monaco},
  {Morbidelli}, {Sousa}, {Worley}, \& {Zaggia}}]{Saccoetal17}
{Sacco}, G.~G., {Spina}, L., {Randich}, S., {et~al.} 2017, \aap, 601, A97

\bibitem[{{S{\'a}nchez} \& {Alfaro}(2009)}]{SanchezAlfaro09}
{S{\'a}nchez}, N. \& {Alfaro}, E.~J. 2009, \apj, 696, 2086

\bibitem[{{Schmeja} \& {Klessen}(2006)}]{SchmejaKlessen06}
{Schmeja}, S. \& {Klessen}, R.~S. 2006, \aap, 449, 151

\bibitem[{{Siemiginowska} {et~al.}(2019){Siemiginowska}, {Eadie}, {Czekala},
  {Feigelson}, {Ford}, {Kashyap}, {Kuhn}, {Loredo}, {Ntampaka}, {Stevens},
  {Avelino}, {Borne}, {Budavari}, {Burkhart}, {Cisewski-Kehe}, {Civano},
  {Chilingarian}, {van Dyk}, {Fabbiano}, {Finkbeiner}, {Foreman-Mackey},
  {Freeman}, {Fruscione}, {Goodman}, {Graham}, {Guenther}, {Hakkila},
  {Hernquist}, {Huppenkothen}, {James}, {Law}, {Lazio}, {Lee},
  {L{\'o}pez-Morales}, {Mahabal}, {Mandel}, {Meng}, {Moustakas}, {Muna},
  {Peek}, {Richards}, {Portillo}, {Scargle}, {de Souza}, {Speagle}, {Stassun},
  {Stenning}, {Taylor}, {Tremblay}, {Trimble}, {Yanamand ra-Fisher}, \&
  {Young}}]{Siemiginowskaetal19}
{Siemiginowska}, A., {Eadie}, G., {Czekala}, I., {et~al.} 2019, \baas, 51, 355

\bibitem[{{Sills} {et~al.}(2018){Sills}, {Rieder}, {Scora}, {McCloskey}, \&
  {Jaffa}}]{Sillsetal18}
{Sills}, A., {Rieder}, S., {Scora}, J., {McCloskey}, J., \& {Jaffa}, S. 2018,
  \mnras, 477, 1903

\bibitem[{{Silverman}(1986)}]{Silverman86}
{Silverman}, B.~W. 1986, {Density estimation for statistics and data analysis}

\bibitem[{Vermeesch {et~al.}(2016)Vermeesch, Resentini, \&
  Garzanti}]{ProvenanceRpackage}
Vermeesch, P., Resentini, A., \& Garzanti, E. 2016, Sedimentary Geology, 336,
  14 , sediment generation and provenance: processes and pathways

\bibitem[{{Wright} {et~al.}(2014){Wright}, {Parker}, {Goodwin}, \&
  {Drake}}]{Wrightetal14}
{Wright}, N.~J., {Parker}, R.~J., {Goodwin}, S.~P., \& {Drake}, J.~J. 2014,
  \mnras, 438, 639

\end{thebibliography}

\begin{appendix}
\section{Public implementations of S2D2}\label{implementations}

The development of this work is part of the StarFormMapper EU project, which will provide the community with different tools and materials to study the spatial distribution of youg stars in SFRs and thus constrain the star formation process. 

In the following, we present the implementations of the procedure that are available for the community. These make use of different software packages: scikit.learn and astropy in python \citep{scikit-learn, Astropy}, and spatstat in R \citep{Baddeleyetal15}. In the StarFormMapper web page (\url{https://starformmapper.org}), updated links to the implementations of the procedure will be available, each with its own user guide.

\subsection{Basic users: DEAVI}

One of the milestones of StarFormMapper is DEAVI, \citep{Bainesetal19}, an added value interface to manage and exploit data from the GAIA and Herschel missions, which is in final development and will be presented and publicly available during 2020 in \url{https://sfmdeavi.quasarsr.com/}. In addition to Gaia and Herschel data, DEAVI will provide simulations of the stellar and gas component of SFRs, and several tools of analysis, including S2D2 and INDICATE (B19).

The S2D2 interface in DEAVI is very simple, allowing any user to apply S2D2 without the need to compile or install anything. 
Due to its simplicity, the default procedure will be very conservative so relevance of the structures detected is ensured.
By default, the confidence level for structure detection will be fixed at 99.85 \% and the complete procedure (calculating $\epsilon$ and $N_{min}$ and retrieving the structures) will only be applied to regions where we can guarantee there is global structure. This will be done by requiring that the $Q$ structure parameter (CW04) is below 0.8. Not only that, if $Q$ is larger than 0.7, the program output will include a warning so the user is cautious with the interpretation of the structures retrieved. The motivation behind these limitations is explained in the text, particularly in section \ref{results}, and summarised in section \ref{conclusions}.

We are aware that these conditions can be too strict for some applications, so we also allow the users to manually introduce $\epsilon$ and $N_{min}$ values to apply DBSCAN and analyse regions with larger $Q$ values, either to retrieve structures of different sizes or with lower levels of confidence.
 
\subsection{Advanced usage: GitHub}

We also provide the community with GitHub repositories with python (\url{https://github.com/martaGG/S2D2}) and R (\url{https://github.com/martaGG/S2D2_R}.) complete implementations of the procedure for more advanced users, that want to have more control on the procedure.

In these implementations the user will be able to modify a parameter file containing all the options available to DEAVI users and, additionally, the $Q$ limit for performing the analysis, and the significance limit for structure detection. 

In the GitHub repositories, the whole code will be available, so further modifications of the procedure will be possible. 

\section{Measuring the structure of stellar clusters: limits and extensions of the $Q$ parameter.}\label{AppendixQ}
 Even though defining and quantifying the level and nature of structure of a region is beyond the main objectives of this work, it is clearly related. In this appendix we review the currently most popular method to evaluate the structure of star-forming regions: the $Q$ parameter, used in this work. We will discuss its underlying model, assumption and limits, along with two recent extensions/alternatives to $Q$, namely $Q^{+}$ by \citet{Jaffaetal17}  (henceforth JF17),  and \textit{ClusterFrac}, by \citet{Lomaxetal18} (henceforth LX18).
 \subsection{$Q$ parameter}
 
 The $Q$ parameter was introduced in CW04, and utilises central values of the edge length distribution of the complete graph (CG) and minimum spanning tree (MST) to describe the spatial structure of a point distribution, allowing to distinguish between substructured, homogeneous and concentrated distributions. This was a feat, since CG and MST distributions separately do give a different signal for clustered, homogeneous and inhibited distributions (where these definitions are in comparison to homogeneous, as explained in the main text, in section \ref{epsSelect}), but they cannot distinguish if a clustered distribution has only one mode, like in a concentrated distribution, or several. 
 
We remind the reader that $Q=\bar{m}/\bar{s}$ is the ratio between the normalised mean length of the MST $\bar{m}$, and that of the complete graph, $\bar{s}$, making it a scale-free parameter. 
In the original work by CW04, the $Q$ parameter was calibrated with synthetic clusters from the box-fractal and radial families, as the ones used in this work, ranging from very substructured to very concentrated regimes and testing both the fractal and radial approximation to a homogeneous distribution. They found that homogeneous distributions had $Q \sim 0.8$,  fractals were characterised by lower $Q$ values, and radials by larger $Q$. In a later work, \citet{Cartwright09} searched for a similar pattern for clusters taking into account kinematical information, but found that including velocity in the calculations did not improve the discriminant power of $Q$. 
\begin{figure}[h!]
\begin{center}
\resizebox{\hsize}{!}{\includegraphics{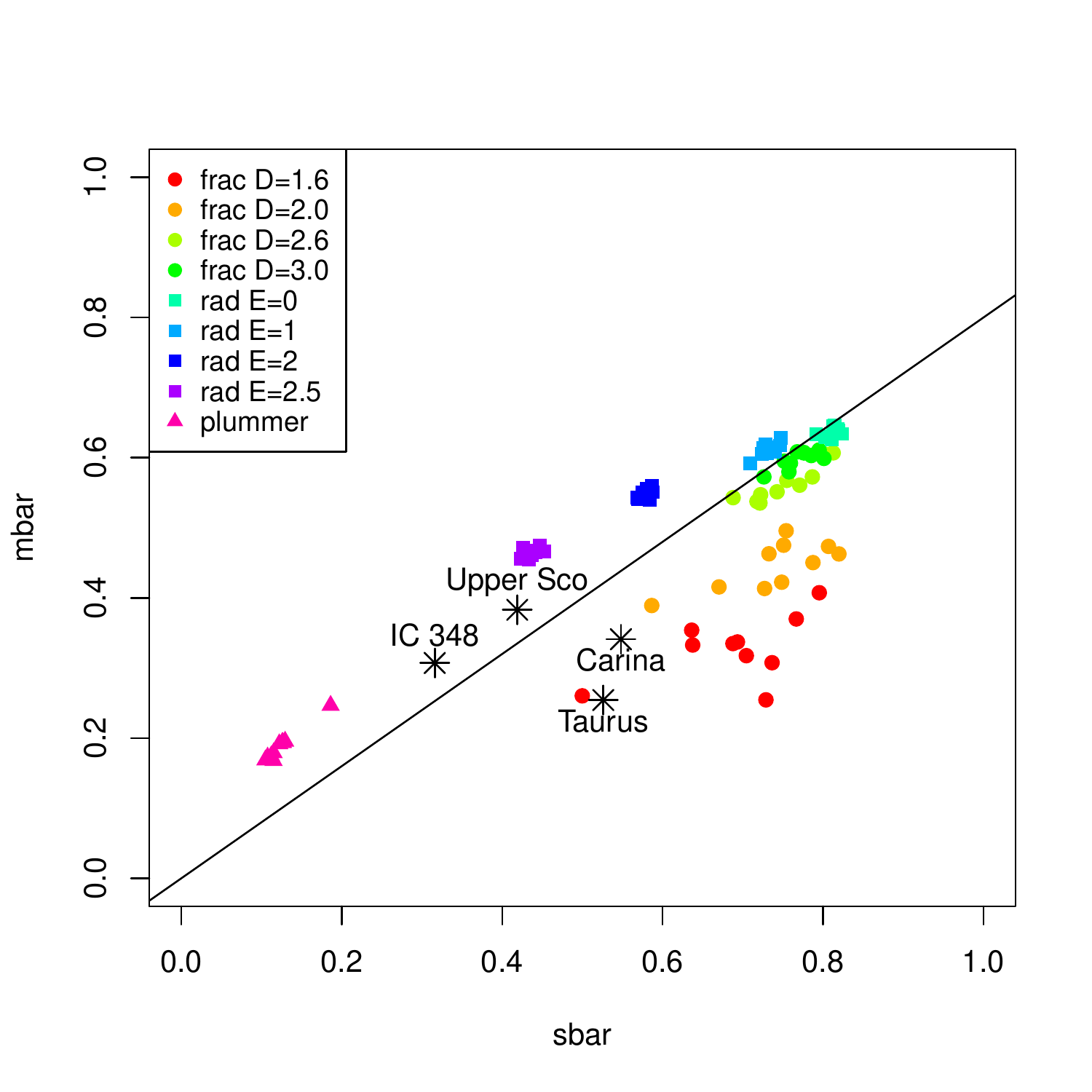}}
\resizebox{\hsize}{!}{\includegraphics{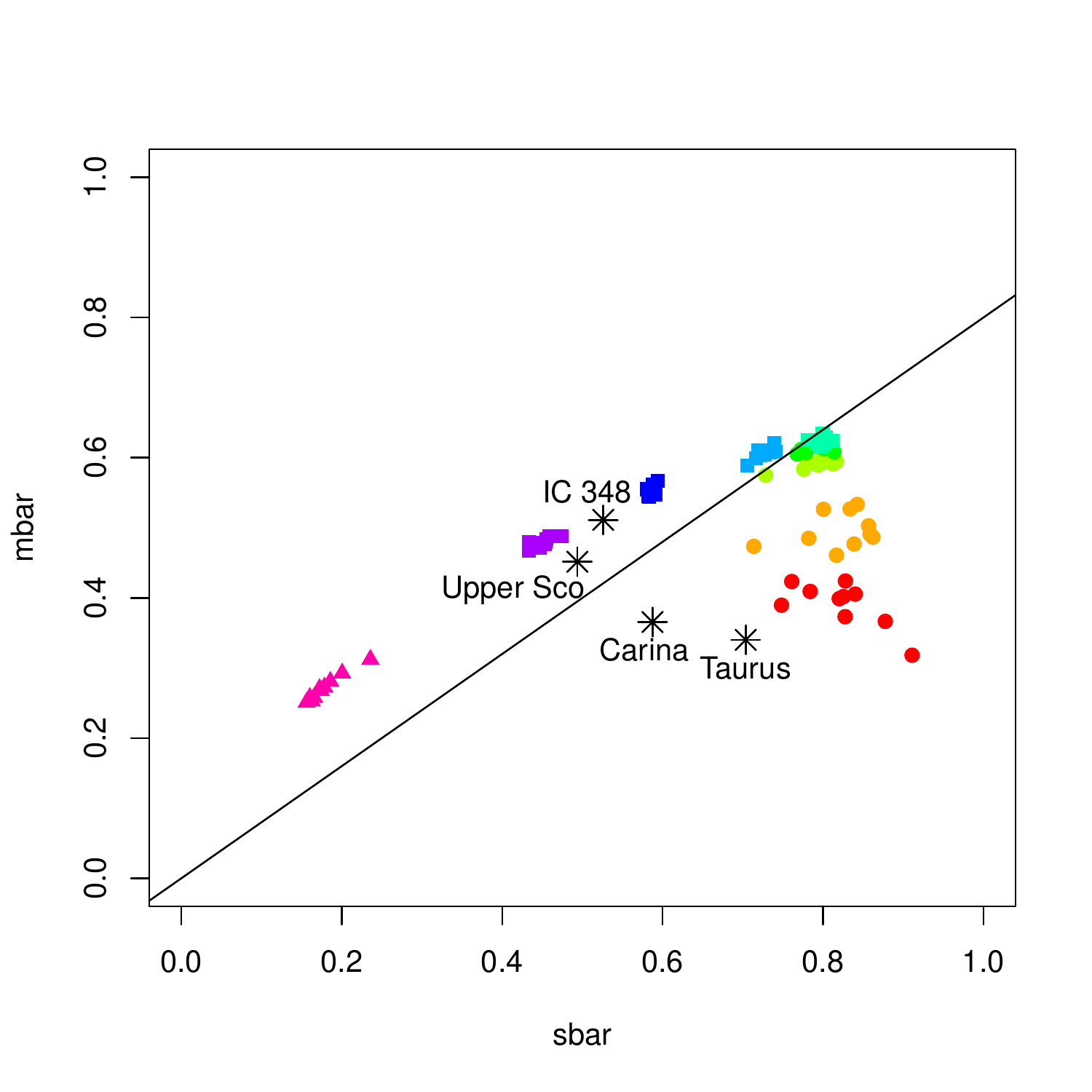}}

\end{center}

\caption{ $\bar{s},  \bar{m}$ plots of the simulated clusters in this work, described in \ref{tablaDist}. The upper panel uses the original normalisation in CW04, while the lower plot uses the convex hull, proposed by \citet{SchmejaKlessen06}. Coloured circles, squares and triangles represent fractal, radial, and Plummer distributions. Black stars represent the real clusters, each of them tagged appropriately. The black line represents the limit $Q=0.8$ traditionally separating structured from concentrated clusters.}
\label{smPlots}
\end{figure}

A constant value of $Q=Q_{0}$, such as the threshold 0.8 found to separate fractal from radial distributions, corresponds to a line in the  $(\bar{s},  \bar{m})$ 2D space passing by the origin with slope $Q_{0}$. The position of clusters in this space has also been used to characterize the structure of clusters, in a variety of works, like e.g. \citet{Cartwright09,ParkerDale13}; LX18; \citet{DaffernPowellParker20}, and references therein. In Figure \ref{smPlots}, the positions of the distributions analyised in this work are shown in the $(\bar{s},  \bar{m})$ space. In the upper panel of Fig. \ref{smPlots} the normalisation used is the standard one, from CW04, that uses the maximum radius and the area of the circle corresponding to that radius to normalize the MST and CG mean values. The lower panel shows the results of the \cite{SchmejaKlessen06} normalisation, that uses the area of the convex hull of the point distribution and the radius of the equivalent circle to normalise. We note that both of these normalisations produce the same $Q$ values, although the differences in the $(\bar{s},  \bar{m})$ space are obvious from the plot. This was explored in \citet{Parker18}, along with a different normalisation proposed by \cite{Kirketal16}, where the area of the convex hull was used but its radius was calculated as the maximum distance from the mean position to the most distant point in the hull. \citet{Parker18} concluded that this distorted the $\bar{s},  \bar{m}$, and $Q$ values to the point that the $Q=0.8$ limit was not valid anymore.
The \cite{SchmejaKlessen06} normalisation is more robust to the presence of outliers in the $(\bar{s},  \bar{m})$ space, effectively confining the fractal distributions, that naturally have strong dispersion to smaller regions of the $(\bar{s},  \bar{m})$ plane. This was shown in the appendix of \cite{ParkerGoodwin15}. This highlights the differences between the substructured real regions (Carina and Taurus), and the fractal distributions, that were not that obvious with the original CW04 normalisation.

% The $(\bar{s},  \bar{m})$ plane was also used by LX18 to show that pure box-fractal/radial models do not accurately capture the behaviour of real regions, while the Fractal Brownian Model they propose covers the apropriate $(\bar{s},  \bar{m})$ region.

The original $Q$ definition assumes the spherical symmetry of the clusters. \citet{CartwrightWhitworth09} explore the effect of elongation, that they calculate as the ratio of the extent of the data in the direction maximizing that extent and the perpendicular one. \cite{SchmejaKlessen06} propose an alternate definition for the elongation based on the ratio between the areas of the maximum distance circle and that of the convex hull.  \citet{CartwrightWhitworth09} concluded that even though elongation modified the $Q$ values, the effect for moderate elongations (of the order or less than 3) was within the uncertainty ranges of $Q$ for spherically symmetrical clusters. For larger elongations, they calculated a table of corrections. 

\citet{CartwrightWhitworth09} warned that the values of both $Q$ and the elongation would be affected by the presence of outliers. This was explored in Chamaleon I by \citet{Saccoetal17}, where they not only use jackknife resampling to estimate the uncertainty of the $Q$ value, but they also calculate the $Q$ parameter of different subsamples restricted to  the central areas of the cluster. They showed that, even though the $Q$ of the central subsample indicated strong substructure, including the outer part increased it to values typical of homogeneous distributions.

One of the main advantages of the $Q$ parameter, already highlighted in CW04, is that it can trace to a fractal dimension or a radial exponent, in the box-fractal/radial paradigm, and that this estimate is robust even for small samples ($N_{star}\sim 100$). This was confirmed with further numerical experiments by \citep{SanchezAlfaro09}, who found that, even if the calculation of the fractal dimension using the two point correlation function (TPCF, as defined in the text, in section \ref{epsSelect}) was in general more exact, it was not reliable for $N_{star}<200$. They also found that the $Q$ estimate varied with the number of points, lowering the threshold separating substructured from concentrated regions from 0.8 to 0.785. \citet{ParkerDale13} expanded the study of the variation of $Q$ with the number of members in a region, and also tested a variety of synthetic and more complex cases, like e.g. masking a region of a sample to produce a hole or combining different synthetic regions of different caracteristics in a single sample. The general behaviour of the $Q$ becomes, in those cases, very hard to predict.
  
\subsection{Expanding the box-fractal model: $Q^{+}$}

JF17 expand the simple box-fractal model used in CW04 to one with two additional parameters (the length scale $L$ and the volume-density exponent $C$) to generate more realistic fractal models. The model is still box fractal, since it is based on the subsequent subdivision of a box, the root cube, where only a fixed proportion of the new smaller boxes are fertile and keep being divided. The star distribution is finally created by generating points in the fertile areas of the original box. 
\begin{itemize}
\item{The fractal Dimension D, with the same meaning as in CW04, where D is related indicates to the probability of cubes in each division to be fertile $P_{fertile}=2^{\rm{D}-Dim}$, where Dim is the dimension of the space. Increasing D increases the filling factor of a cluster, having more fertile cubes, and the number of stars. }
\item{The lenght scale $L$ is associated to the number of levels $l$, which indicates the finite spatial range of scales associated to the fractal through the relationship $2^{L}=R$, where $R$ is the ratio between the size of cubes in the last level and the root cube. Increasing $L$ increases $R$ the relative size of the smallest separations compared to the size of the cluster, and the number of stars.}
\item{The volume density scaling exponent $C$ is associated to the degree of concentration in the smallest scales. The additional volume density in level $l$ is $\delta n_{l}=n_{0}2^{C}$, with $n_{0}$ the density of the root cube. Increasing $C$ concentrates stars in the smaller structures, decreasing the number of stars.} 
\end{itemize}

Within that paradigm the CW04 fractal clusters as the ones used in this work have $\rm{D=D}$, $L=log_{2}(N_{star}^{1/\rm{D}})$, and $C=\infty$. When generating a JF17 fractal cluster with a specific set of $\rm{D}, L, C$, the number of members cannot be prefixed.

\begin{figure}[h!]
\begin{center}

\resizebox{\hsize}{!}{\includegraphics{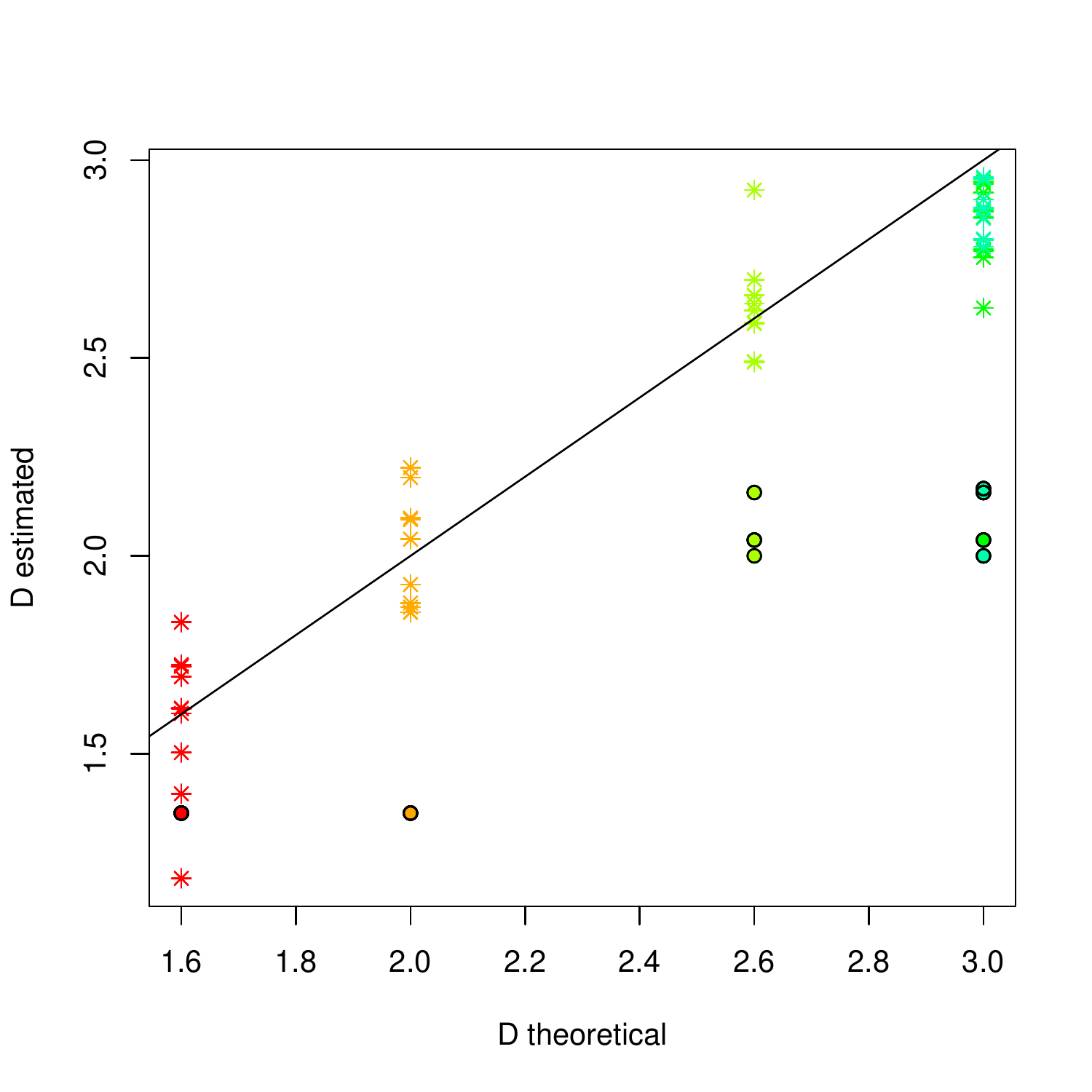}}
\resizebox{\hsize}{!}{\includegraphics{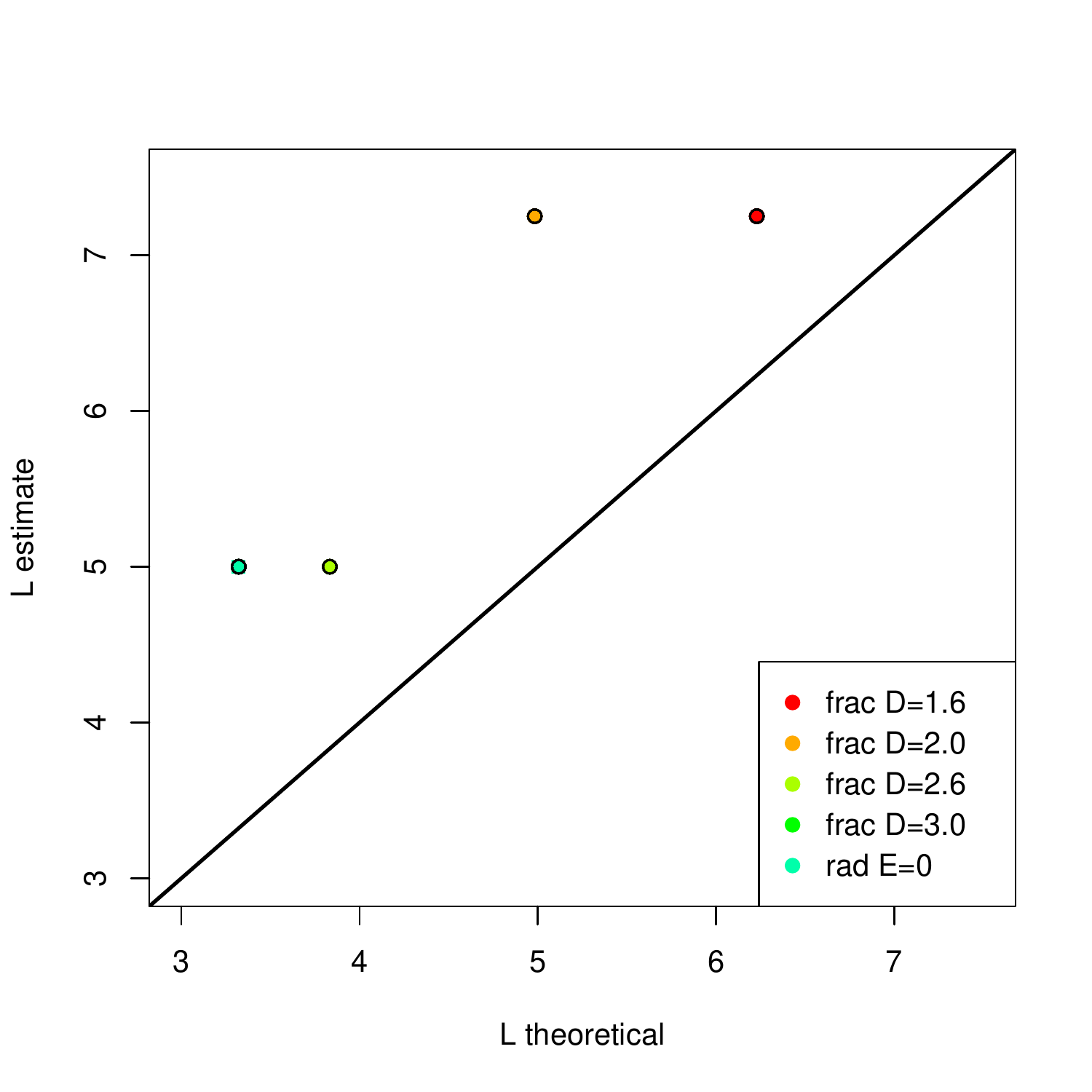}}

\end{center}

\caption{
Fractal dimension estimated using the standard $Q$ approach from CW04 (stars), and the $Q^{+}$ from JF17 (black rimmed coloured dots) against the fractal dimension used to generate the clusters. Bottom: Estimated against theoretical value of $L$ for the simulated clusters considered in this work. Each colour respresents a different true fractal dimension. The black line is the identity function, where the theoretical and estimated values are equal. Please note that the Fractal with D=3 and the radial with E=0 both represent a homogeneous distribution, and that some of the $D$ estimates and all the $L$ estimates given by $Q^{+}$ overlap.}
\label{DimFrac}
\end{figure}

\begin{figure}[h!]
\begin{center}

\resizebox{\hsize}{!}{\includegraphics{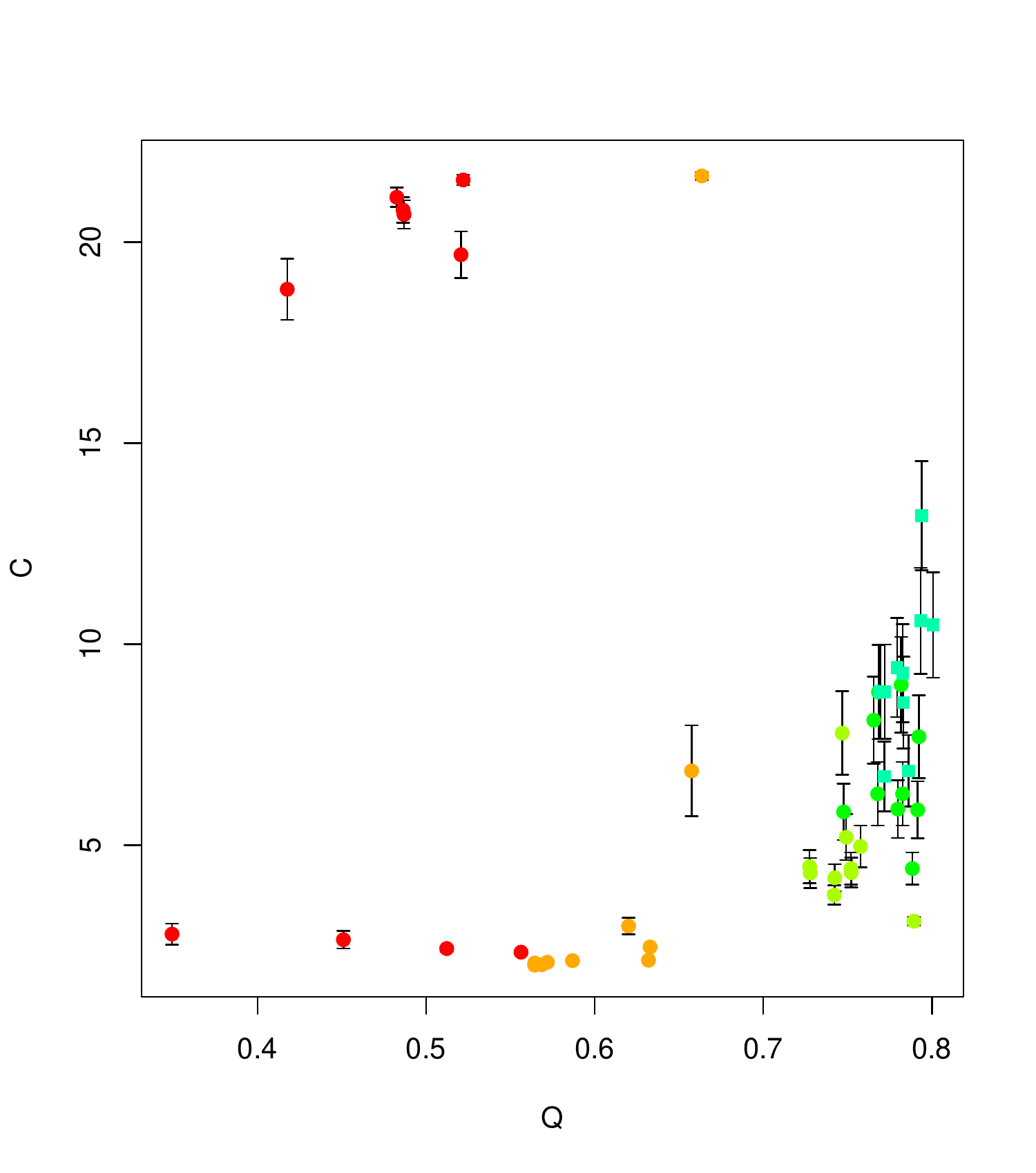}}

\end{center}

\caption{estimated value of $C$ for the simulated clusters. The colour code is the same as in Fig. \ref{DimFrac}}
\label{QplusLc}
\end{figure}

JF17 calculate, for each synthetic cluster, a set of 7 features describing the characteristics of the cluster and based on the MST and CG of the distribution.
%\begin{itemize}
%\item{The logarithm of the members in the cluster, log($N_{star}$).}
%\item{The scale ratio log($R$)=$s_{max}/s_{min,5}$, where $s_{max}$ and $s_{min,5}$ are the largest and fifth shortest edges of the CG. \footnote{The fifth shortest edge is chosen for robustness with respect to chance alignments.}}
%\item{The normalised mean edge length of the CG $\bar{s}$}
%\item{The normalised mean edge length of the MST $\bar{m}$}
%\item{The  mean edge length of the MST $\mu_{m}$}
%\item{The standard deviation of the edge length distribution of the MST $\sigma_{m}$}
%\item{The normalised area above the cumulative distribution of the MST edge lenght, $\mathcal{A}$}
%
%\end{itemize}
%
%Simulations in J17 varying D, $L$ and $C$ show that from this set of features, only $\mathcal{A}$ varies with $C$.
Based on these features, they perform simulations varying  D, $L$ and $C$ to which they apply principal component analysis (PCA), obtaining a 2D principal component (PC) space on which they estimate the values for D and $L$. $C$ is estimated independently using a bayesian approach. 
The results obtained by JF17 in simulations of their fractal models are very good, except for $C=\infty$, which is not well constrained. Even in the PC space, the effect of $C$ is hard to disentangle from that of D and $L$.  In real regions, they obtain estimates of the fractal dimension which are systematically lower, but follow the same trend as previous estimates using the standard $Q$ parameter. JF17 also report that clusters with D$>2.32$ will appear smooth.  

Figures \ref{DimFrac} and \ref{QplusLc} show the results of the $Q^{+}$ algorithm (from the author's github python implementation, provided in JF17) on our simulated clusters. We have only applied the code to the fractal distributions and the homogeneous radial approximation, since concentrated distributions are beyond their scope, and the user is warned when introducing a distribution with $Q>=0.9$. From the top panel in Fig. \ref{DimFrac} it is clear that the standard $Q$ estimates for the fractal dimension are much better than those provided by $Q^{+}$, which is not surprising, since the clusters were generated as simple box-fractal models following CW04's recipe. As for the $Q^{+}$ estimates in our synthetic clusters, they are always lower than the values obtained from the simple $Q$, and they do not help in distinguishig fractals with dimension 1.6 from fractals with dimension 2, or fractals with D=2.6 from homogeneous distributions.
 
The bottom plot in Fig. \ref{DimFrac} shows the $L$ estimates, which have similar problems as those of D: they all overlap, and they cannot discriminate amongst distributions. In addition they are systematically overestimated.

Figure  \ref{QplusLc} shows the values of $C$ estimated with $Q^{+}$ vs their standard $Q$ parameter. As already indicated in JF17, the values are not well constrained, and in fact, a significant amount of the distributions tested have $C<5$, which are no good estimates considering that the theoretical value is $=\infty$.

It must be noted that JF17 remark that the $Q{+}$ estimates will improve as they increase the number of simulations for the PCA and bayesian calculations, better sampling the (D, $L$, $C$) space. We conclude that, at least in its current state, a $Q^{+}$ analysis does not improve the simple $Q$ analysis for the box-fractal models, as those in this work. 

\subsection{The Fractal Brownian model: ClusFrac}

LX18 extend the complete box-fractal/radial paradigm (not just the fractal part of it, as in JF17), to cover a much larger part of the  $(\bar{s},  \bar{m})$ space, filling in particular the areas usually occupied by real clusters, as shown in the lower plot of Fig. \ref{smPlots}.

The fractal brownian model clusters start from a fractal brownian model (FBM), usually used for describing the structrure of clouds, with fixed drift exponent $H$, which is related to the fractal dimension of the cloud and the slope of its fourier power spectrum. The cloud field $f(r, H)$ is obtained as the inverse fourier transform of its power spectrum. After filtering for the smallest spatial scales of self similarity, the cloud is then exponentiated with parameter $\sigma$, transforming the Gaussian distribution into a lognormal:
\begin{equation}
g(r, H, \sigma)=\rm{exp}\left(\frac{\sigma f(r, H)}{\sqrt[2]{\langle f(r, H) \rangle^{2}}}  \right)
\end{equation}

This lognormal field will be used as the probability density function to sample stars and build clusters, and $H, \sigma$ are its parameters.
For fixed $H$, varying  $\sigma$ changes the dynamic range of the cloud. This lognormal cloud fulfils the conditions to be a probability density function, from which the stars are sampled. 

LX18 simulated a variety of FBM clusters with different random seeds and values of $H$ and $\sigma$, finding that there is an anticorrelation between  $\bar{m}$ and $\sigma$, and that clusters with $H=1$ fulfil the same role as radials.  

To estimate $H$ and $\sigma$ LX18 fit artificial neural network regressors (ANNs, specifically multilayer perceptrons with one hidden layer) using as features the MST and CG edge length distribution moments (from the first to the fourth) after whitening the clusters (rescaling and eliminating elongation). They trained 3 ANNs, each of which fit clusters with different population ranges (r $N_{star}<99$, $100<N_{star}<315$, and $316<N_{star}<999$). Their test results are very good in artificial FBM clusters, particularly when they are well populated. The results in observed clusters are hard to evaluate: we get some values for $H$ and $\sigma$, but their meaning is not clear: looking at them we cannot say whether the distribution is clumpy or not. This means that we cannot rule out the possibility that FBM clusters are overfitted by the ANNs, in which case results of clusters that do not follow a FBM could be unreliable. Overfitting can be of importance since the exponentiated fractal Brownian model is not a complete descriptor of the interstellar medium \citep[see e.g.][and references therein]{Robitailleetal20}.

We cannot directly calculate and compare the values of $H$ and $\sigma$ that would correspond to our clusters, but LX18 warn that $H$, even though related to the fractal dimension, must be taken with caution, since with the effect of sigma does not necessarily correspond with what we consider clumpy. Despite the great advantage that is describing clusters with parameters and models that can be applied to the gas component of SFRs,  $H$ and $\sigma$ are still a bit obscure, and somehow lack intuitive descriptive power. 
This can be due simply to their novelty, so more theoretical and applied studies of the FBM should be done to deepen our understanding of  $H$ and $\sigma$ and better interpret them.

\section{One Point Correlation Function and simulations}\label{AppendixOPCF}
\begin{figure}[h!]
\begin{center}
\resizebox{\hsize}{!}{\includegraphics{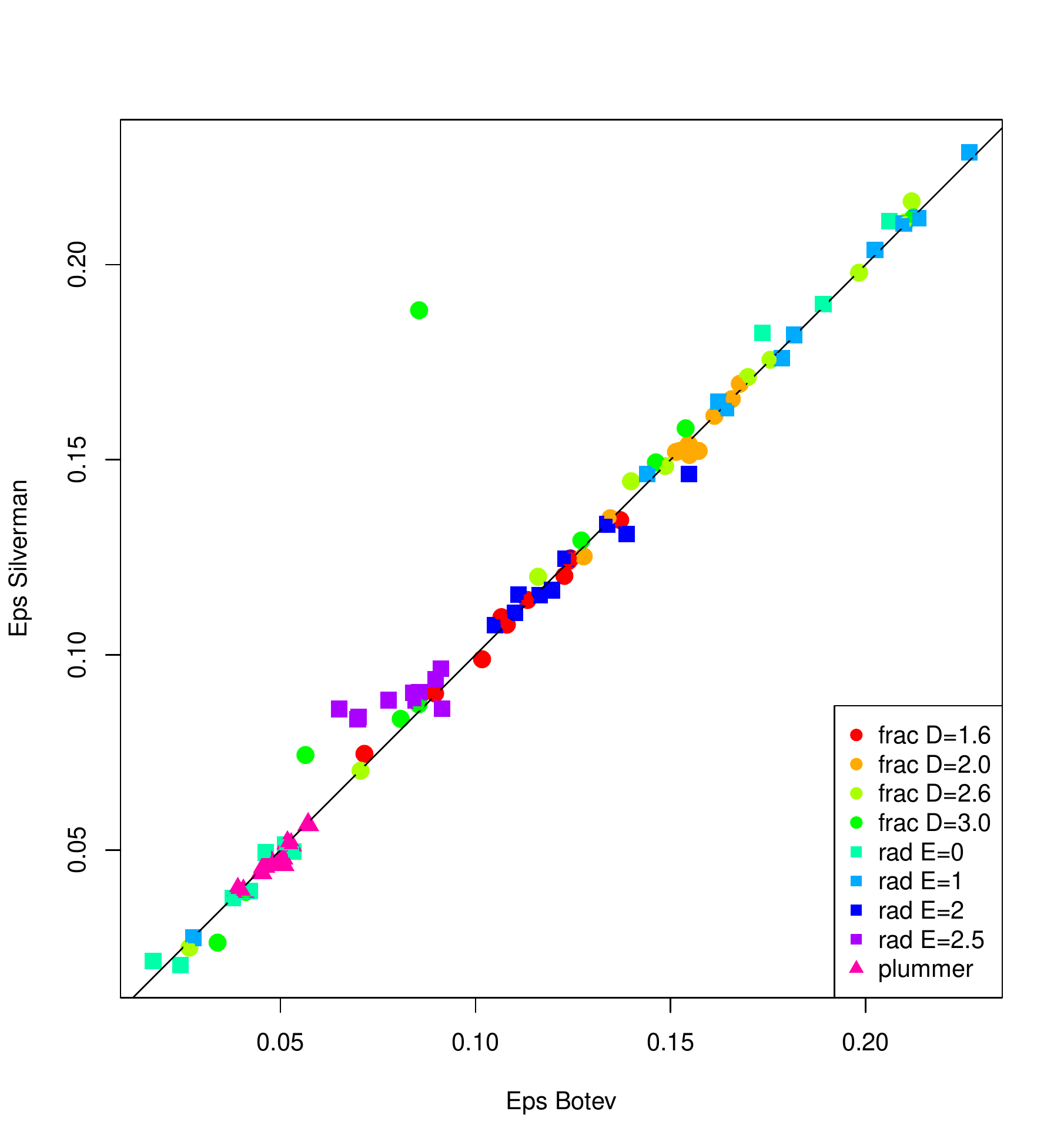}}

\end{center}

\caption{$\epsilon$ value obtained by S2D2 when the OPCF is generated using Botev's algorithm \citep{Botev10} instead of Silverman's formula \citep{Silverman86} for the bandwidth.}
\label{Botev}
\end{figure}

We show the general behaviour of the OPCF in simulated clusters, as support to the results shown in the main text. We remind the reader that OPCF stands for the One Point Correlation Function $\Psi(r)$ introduced in J17, that compares the first nearest neighbour distance distribution of the sample with that of a theoretical homogeneous random distribution. The OPCF is used by S2D2 to calculate $\epsilon$, the relevant scale to search for small, significant substructure, as described in section \ref{epsSelect} in the main text. $\epsilon$ is the smallest distance where the OPCF crosses the 1 boundary decreasingly, separating excess (with respect to random) of stars with nearest neighbour at smaller $r$ and depletion at larger. 

\begin{figure}[h!]
\begin{center}
\resizebox{\hsize}{!}{\includegraphics{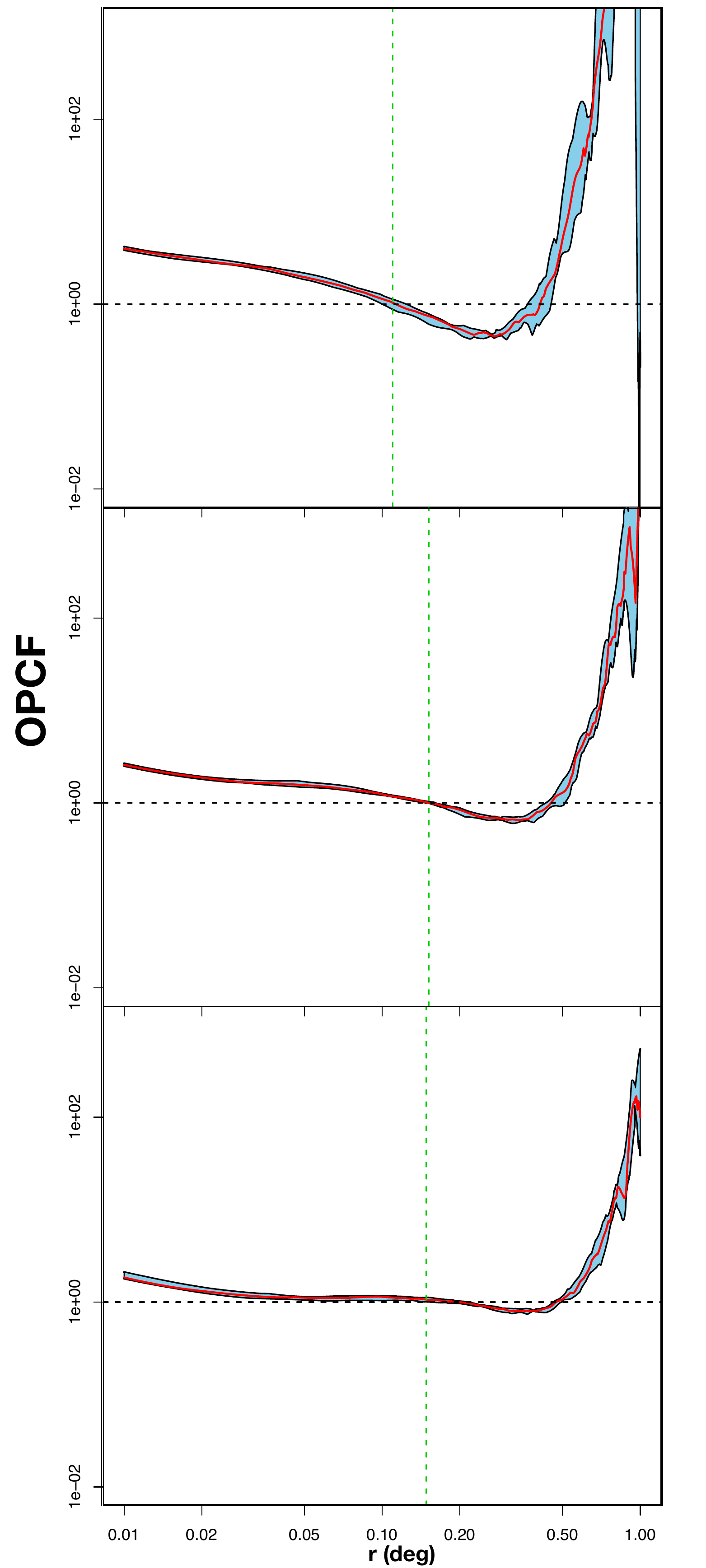}}

\end{center}

\caption{Top: In red, median value of the OPCF at each distance for 10 realisations of a fractal distribution with D=1.6. Black solid lines represent the quartiles $Q_{1}$ and $Q_{3}$, with the inter-quartile range area shaded in blue. The horizontal dashed black line at constant value 1 represents the theoretical value for CSR, and the vertical dashed green line shows the average value of the $\epsilon$ obtained by S2D2 (see Table \ref{tableSynth}). Middle: same for a fractal distribution with D=2.0. Bottom: same for a fractal distribution with D=2.6.}

\label{OPCFStruct}
\end{figure}

First of all, in Figure \ref{Botev} we show the $\epsilon$ values obtained using Silverman's formula to calculate the bandwidth for the Gaussian Kernel against the $\epsilon$ obtained with Botev's algorithm \citep{Botev10}. As described in section \ref{epsSelect}, this Gaussian Kernel is used to calculate the distribution of the first nearest neighbour distance in the sample.
Botev's algortithm is complex, and based on diffusion equations.  We use the R package provenance \citep{ProvenanceRpackage} to calculate the bandwidth.

\begin{figure}[h!]
\begin{center}
\resizebox{\hsize}{!}{\includegraphics{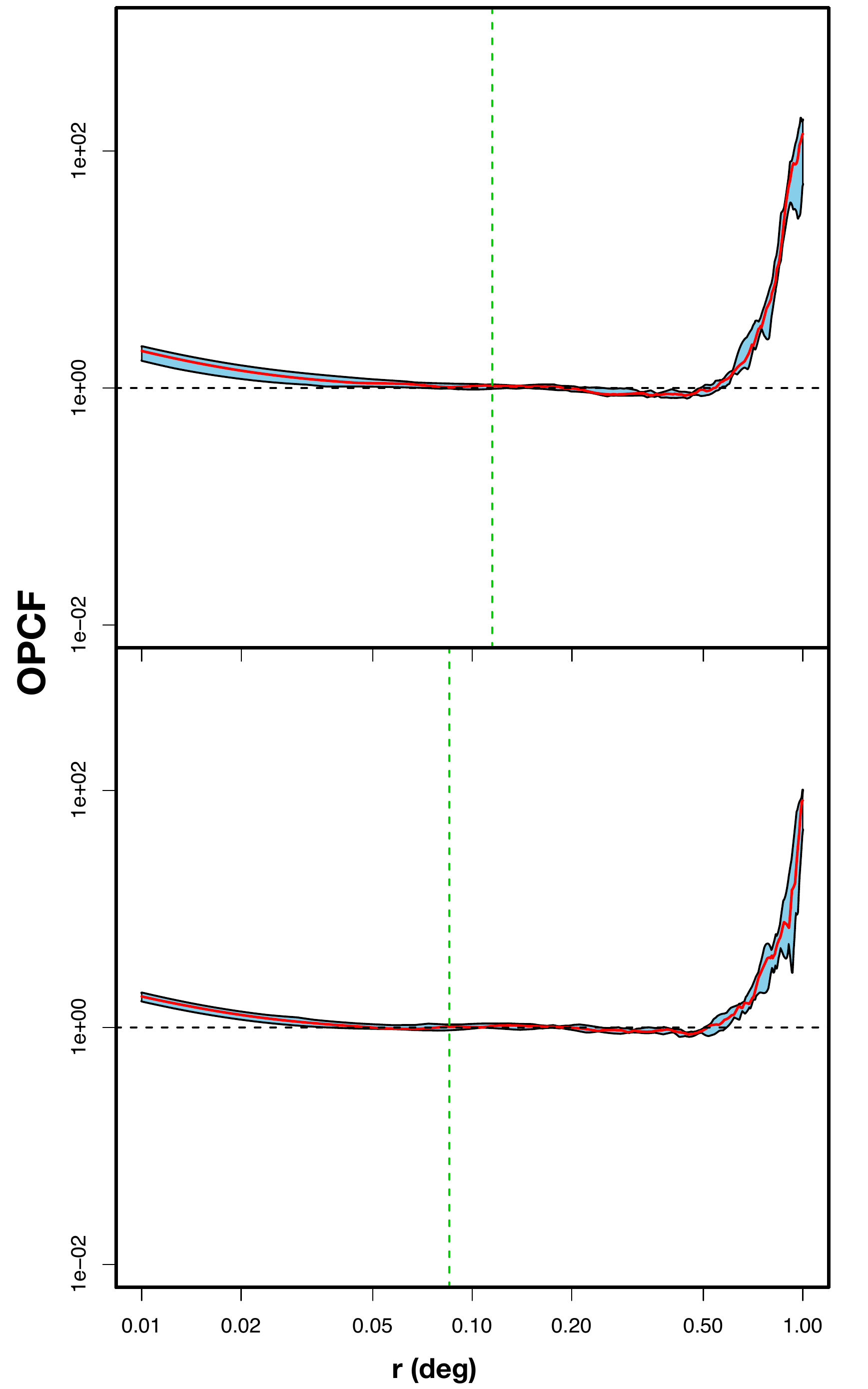}}

\end{center}

\caption{Top: Plot analogous to those in Figure \ref{OPCFStruct} summarising the OPCF in a fractal with D=3.0. Bottom: Same for a Radial with E=0.0}
\label{OPCFHomog}
\end{figure}

Both approaches give similar results, with a very significant difference only in one realisation of a fractal approximation to a homogeneous distribution (D=3), where the dispersion of the $\epsilon$ values is large due to the closeness of the first nearest neighbour distribution to the theoretical homogeneous one. The shape of the OPCF in this case is shown in Fig \ref{OPCFHomog}.

\begin{figure*}[h!]
\begin{center}
\includegraphics[width=17cm]{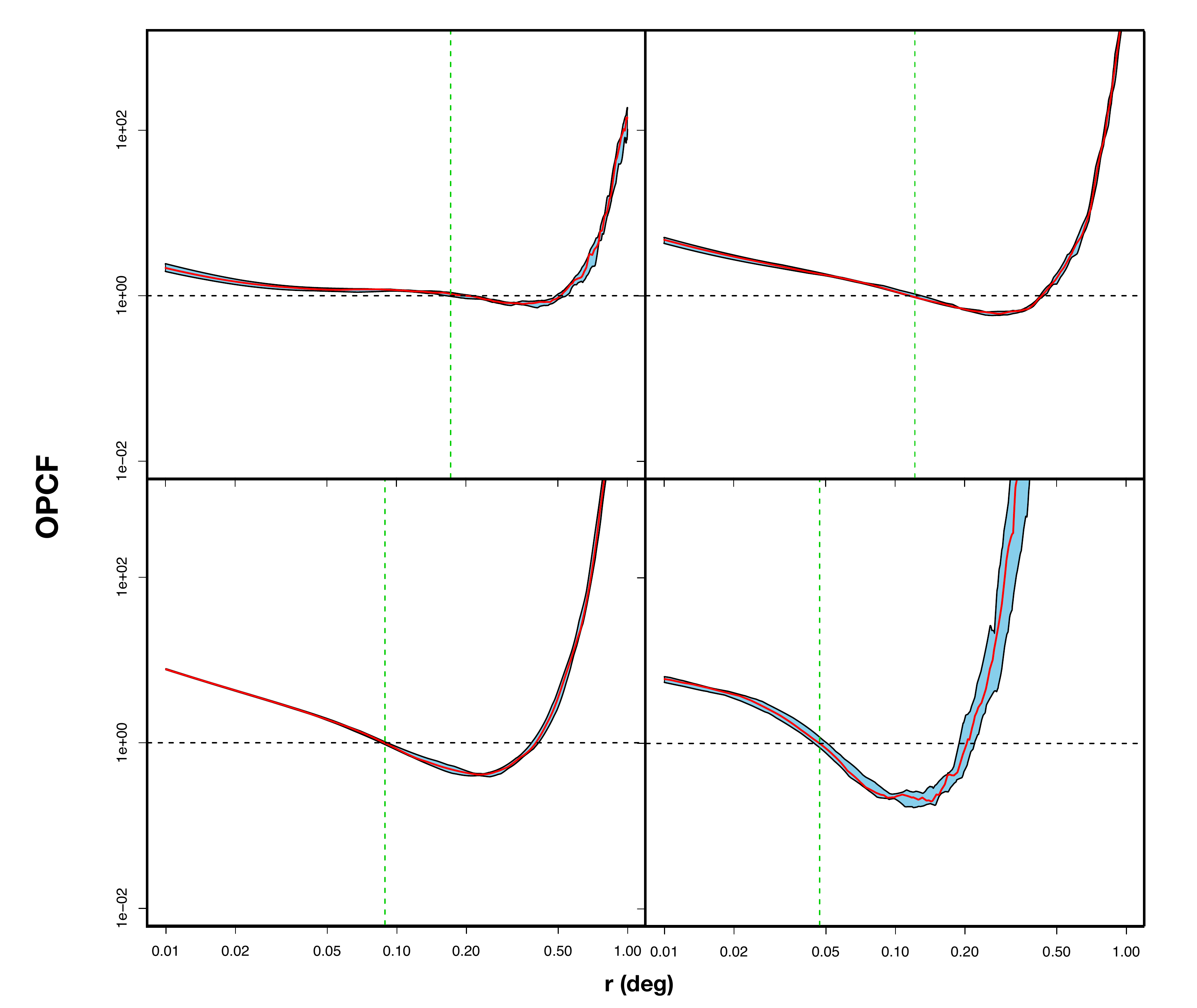}

\end{center}
\caption{Plot analogous to those in Figure \ref{OPCFStruct} summarising the OPCF for concentrated distributions. From left to right and top to bottom: radial distribution with exponent E=1, radial distribution with exponent E=2, radial distribution with exponent E=2.5, and Plummer distribution.}
\label{OPCFConcentrated}
\end{figure*}

Each panel in Figures \ref{OPCFStruct}, \ref{OPCFHomog}, and \ref{OPCFConcentrated} summarises the OPCF values for the fractal, homogeneous and concentrated distributions in this work (see Table \ref{tablaDist}). For each distribution, the black lines show the quartiles $Q_{1}$ and $Q_{3}$  of the OPCF at each distance, the interquartile area is shaded in blue, and the red line shows its median value. The theoretical value of the OPCF at a random homogeneous distribution (which would be a function with constant value 1) is shown as a black horizontal dashed line, and the mean value of $\epsilon$ calculated by S2D2 (also shown in Table \ref{tableSynth}) is shown as a green vertical dashed line.
 
Even though there are differences amongst distributions, the use of the OPCF by itself to determine the structure of a point pattern is limited. This is partially due to the fact that it is based on the first nearest neighbour and only measures first order effects. This is similar to the lack of discriminant power between substructured and concentrated distributions associated to the MST distribution, as described in Appendix B from CW4. In their work, they include second order effects through the distance distribution of the sample, defining the $Q$ parameter. The $Q$ parameter has been thoroughly described in Appendix \ref{AppendixQ}, but we remind the reader that it is a global quantity, normalized to account for the size of the region.

 Both substructured and concentrated distributions are clustered compared to CSR. This is evident in Figures \ref{OPCFStruct} and \ref{OPCFConcentrated}, where a pattern on the characteristics of stars with respect to random associated to distance can be seen (as in e.g J17). This pattern consists in an excess of close stars compared to random expectation at small distances (shown by an OPCF larger than one), followed by depletion of stars with a first nearest neighbour at intermediate distances and an excess at larger distances. In both cases, the closer a distribution is to CSR, the closer the OPCF values get to 1, with both the fractal and radial distributions converging towards the theoretical homogeneous case of $\Psi(r)=1$, as shown in Figure \ref{OPCFHomog}. We note that for homogeneous distribution, the values of the OPCF are close to unity in a significant part of its domain. As explained in the text, this results in the $\epsilon$ obtained by S2D2, having very different values based on random sampling effects.

\end{appendix}

\end{document}